Title: “How should parallel cluster randomized trials with a baseline period be analyzed? A survey of estimands and common estimators”

Short Title: “How to analyze PB-CRTs”

Author List: Kenneth Menglin Lee,[1*] Fan Li[2,3]

[1] Centre for Quantitative Medicine, Duke-NUS Medical School, Singapore 169857

[2] Department of Biostatistics, Yale School of Public Health, New Haven, CT, USA

[3] Center for Methods in Implementation and Prevention Science, Yale School of Public Health, New Haven, CT, USA

*Corresponding author. Center for Quantitative Medicine, Duke-NUS Medical School, Singapore, 20 College Road, Singapore 169857

E-mail: klee@u.duke.nus.edu.

**Abstract**

The parallel cluster randomized trial with baseline (PB-CRT) is a common variant of the standard parallel cluster randomized trial (P-CRT). We define two natural estimands in the context of PB-CRTs with informative cluster sizes, the participant-average treatment effect (pATE) and cluster-average treatment effect (cATE), to address participant and cluster-level hypotheses. In this work, we theoretically derive the convergence of the unweighted and inverse cluster-period size weighted (i.) independence estimating equation, (ii.) fixed-effects model, (iii.) exchangeable mixed-effects model, and (iv.) nested-exchangeable mixed-effects model treatment effect estimators in a PB-CRT with continuous outcomes. Overall, we theoretically show that the unweighted and weighted independence estimating equation and fixed-effects model yield consistent estimators for the pATE and cATE estimands. Although mixed-effects models yield inconsistent estimators to these two natural estimands under informative cluster sizes, we empirically demonstrate that the exchangeable mixed-effects model is surprisingly robust to bias. This is in sharp contrast to the corresponding analyses in P-CRTs and the nested-exchangeable mixed-effects model in PB-CRTs, and may carry implications for practice. We report a simulation study and conclude with a re-analysis of a PB-CRT examining the effects of community youth teams on improving mental health among adolescent girls in rural eastern India.

# 1. Introduction

Cluster randomized trials (CRTs) refer to the collection of study designs where randomization is carried out at the group level (such as a hospital, clinic, or worksite level), and often includes outcome measurements collected at the participant level (Turner et al., 2017). The standard parallel cluster randomized trial (P-CRT) with data collected over a single time period has two estimands with particularly natural interpretations, the cluster-average treatment effect (cATE) and the participant-average treatment effect (pATE) (Kahan, Blette, et al., 2023; Kahan, Li, Copas, et al., 2023). Briefly, the cATE is the average treatment effect over the cluster level and is commonly of interest when studying interventions designed for implementation at the cluster level. The pATE is the average treatment effect over the individual participant level, is often the target estimand had individual randomization been feasible and may often be of relevance when studying individual-level interventions (which may be cluster randomized as the only feasible randomization scheme). These two estimands represent different units of inference and can differ in magnitude when cluster sizes vary.

To account for within-cluster correlation, P-CRTs are often analyzed with an exchangeable correlation structure specified through a mixed-effects model or generalized estimating equation (GEE) framework. When P-CRTs have variable cluster sizes, inverse cluster size weights can additionally be specified to ensure that all clusters contribute equally regardless of cluster size (Williamson et al., 2003). This is useful in the presence of heterogeneous treatment effects according to cluster size, also referred to as "informative cluster sizes" (Bugni et al., 2024; Kahan, Li, Blette, et al., 2023; B. Wang, Park, et al., 2024; X. Wang et al., 2022; Williamson et al., 2003). It can then be of interest to combine the specified exchangeable correlation structure with inverse cluster size weights for estimation of the average treatment effect. However, previous work by Wang et al. has indicated that specifying both an inverse cluster size weight and an exchangeable correlation structure in the analysis of P-CRTs can potentially lead to estimation of an uninterpretable treatment effect estimand that is neither the cATE (sometimes also referred to as the "unit average treatment effect" or "UATE" (Imai et al., 2009)) nor pATE under informative cluster sizes, and accordingly observe that "Two weights make a wrong" (X. Wang et al., 2022).

A common extension to the P-CRT design is the parallel CRT with a baseline period (PB-CRT). Such a design maintains parallel randomization of clusters across two sequences, with one sequence of clusters receiving the treatment following the initial baseline period. This additionally allows for both within-cluster and between-cluster comparisons (Hooper et al., 2018). Accordingly, the PB-CRT design allows for more analytic options (Hooper et al., 2018) and is typically more efficient than a standard P-CRT (Lee & Cheung, 2024).

In this article, our goal is to explore if "two weights make a wrong" still holds in PB-CRT designs with continuous outcomes, cross-sectional measurements, and variable cluster-period sizes between clusters. We start by focusing on two natural estimands of interest in Section 2 before introducing standard estimators for PB-CRTs in Section 3. We then derive the convergence probability limits of the point estimators from the following set of commonly used models for PB-CRTs (Section 4):

i. Independence estimating equation (IEE);
ii. Independence estimating equation with inverse cluster-period size weighting (IEEw);
iii. Fixed-effects model (FE);
iv. Fixed-effects model with inverse cluster-period size weighting (FEw);
v. Exchangeable mixed-effects model (EME);
vi. Exchangeable mixed-effects model with inverse cluster-period size weighting (EMEw);
vii. Nested-exchangeable mixed-effects model (NEME);
viii. Nested-exchangeable mixed-effects model with inverse cluster-period size weighting (NEMEw).

We further characterize the bias of the EMEw and NEMEw estimators for the cATE estimand in Section 5.

In particular, with an identity link, the generalized estimating equation (GEE) and mixed-effects model (as estimated by generalized least squares (GLS)) treatment effect estimators are expected to coincide (Gardiner et al., 2009; Hubbard et al., 2010), and therefore we only focus on the latter under a mixed-effects setup in this article. Furthermore, we often assume in this paper that cluster-period sizes only vary between clusters, but not between periods within clusters, allowing us to define inverse cluster-period size weights at the cluster-level in the style of Williamson et al. (Williamson et al., 2003) and Wang et al. (X. Wang et al., 2022).

In this paper, the overarching goal is to determine which point estimators (from the collection of commonly-used models) are theoretically consistent for the pATE and cATE estimands in the context of PB-CRTs. In addition, we report a simulation study with informative cluster sizes to compare the performance of the different estimators in terms of bias and efficiency with model-based and jackknife variance estimators (Section 6). We then re-analyze a case study of a PB-CRT examining the effects of community youth teams on improving mental health among adolescent girls in rural eastern India (Section 7), and end with concluding remarks (Section 8).

## 2. Potential Outcomes framework & Estimands of interest

In this article, we focus on the basic PB-CRT design with 2 periods, $I$ clusters, and $K_{ij}$ individuals in each cluster-period cell. Data from each cluster are collected in the 2 discrete, equally-spaced periods indexed by $j = 0, 1$, with period $j = 0$ being the "pre-rollout" baseline period where all clusters are in the control group and period $j = 1$ being the "rollout" follow-up period where $I/2$ clusters randomized to the treatment sequence $S_i = 1$ receive the treatment. Accordingly, the total number of individuals across all clusters and periods is $n = \sum_{i=1}^{I}(K_{i0} + K_{i1})$. In this article, we focus on continuous outcomes $Y_{ijk}$ for each individual $k$ in period $j$ of cluster $i$.

For illustrative purposes, we provide an example 4 cluster, 2 period PB-CRT in Figure 1. This $I = 4$ cluster, 2 period PB-CRT with continuous outcomes $Y_{ijk}$ has cluster-period sizes of $K_{ij}$. Half of the clusters are randomized into sequence $S_i = 1$ where they receive the treatment in period $j = 1$. Cluster-period cells that receive the treatment are highlighted in gray (Figure 1).

Figure 1. An example of a 4 cluster, 2 period PB-CRT with continuous outcomes $Y_{ijk}$ and cluster period sizes $K_{ij}$. Cluster-period cells receiving the treatment are highlighted in gray.

| | Baseline Period $j = 0$ | Follow-up Period $j = 1$ |
|---|---|---|
| Cluster $i = 1$, Sequence $S_1 = 1$ | $(Y_{10k}, K_{10})$ | $(Y_{11k}, K_{11})$ |
| Cluster $i = 2$, Sequence $S_2 = 1$ | $(Y_{20k}, K_{20})$ | $(Y_{21k}, K_{21})$ |
| Cluster $i = 3$, Sequence $S_3 = 0$ | $(Y_{30k}, K_{30})$ | $(Y_{31k}, K_{31})$ |
| Cluster $i = 4$, Sequence $S_4 = 0$ | $(Y_{40k}, K_{40})$ | $(Y_{41k}, K_{41})$ |

Previous work in stepped-wedge cluster randomized trials (SW-CRT) has defined estimands with potential outcomes by breaking the SW-CRT design into a “pre-rollout” period $j = 0$ where all clusters receive the control, “rollout” periods $j = 1, \ldots, J$ during which some clusters are receiving treatment and some are receiving the control, and a “post-rollout” period $j = J + 1$ where all clusters receive the treatment (Chen & Li, 2024; B. Wang, Wang, et al., 2024). Estimands are then defined with potential outcomes during the “rollout” periods where there is within-period variation in treatment (Chen & Li, 2024; B. Wang, Wang, et al., 2024).

Analogously, we can use the potential outcomes framework to define the treatment effect estimands in a PB-CRT during the “rollout” periods. We denote $Y_{ijk}(z)$ as the potential outcome for individual $k$ in period $j$ of cluster $i$ receiving treatment $z = 1$ (treatment) or 0 (control). Notably, like the SW-CRT, the PB-CRT by design induces treatment non-positivity in the “pre-rollout” baseline period $j = 0$ (Chen & Li, 2024; B. Wang, Wang, et al., 2024), where potential outcomes under treatment are never observable by design and $Y_{i0k} = Y_{i0k}(0)\ \forall\ i, k$. Therefore, focusing only on the “rollout” periods (the follow-up period in a PB-CRT), we can connect the observed outcome $Y_{i1k}$ to the potential outcomes via:

$$Y_{i1k} = S_i Y_{i1k}(1) + (1 - S_i) Y_{i1k}(0)\,.$$

We assume a cluster superpopulation framework, where sampled clusters are independent and identically distributed draws from an infinite superpopulation of clusters. Under such a framework, randomness is introduced in the sampling of clusters, followed by the randomization of half the sampled clusters to the treatment sequence. Participants are assumed to be observed and fixed within each cluster-period cell, as opposed to further sampled. For simplicity, we assume a cross-sectional design in this article.

Under the potential outcomes framework, since the rollout period is precisely the follow-up period in a PB-CRT, the participant-average treatment effect (pATE) and cluster-average treatment effect (cATE) in the presence of informative cluster sizes can be defined as the average treatment effect across participants (where participants are given equal weight) and across clusters (where clusters are given equal weight) during the follow-up period, respectively (Kahan, Li, Copas, et al., 2023). Accordingly, their mathematical expressions are:

a. **participant-Average Treatment Effect (pATE):**

$$pATE = E\left[\frac{1}{E[K_{i1}]}\sum_{k=1}^{K_{i1}}[Y_{i1k}(1) - Y_{i1k}(0)]\right]$$

b. **cluster-Average Treatment Effect (cATE):**

$$cATE = E\left[\frac{1}{K_{i1}}\sum_{k=1}^{K_{i1}}[Y_{i1k}(1) - Y_{i1k}(0)]\right]$$

with more details included in the Appendix (A.1). Notably, when cluster sizes in the follow-up period are independent of the potential outcomes ($K_{i1} \perp\!\!\!\perp \{Y_{i1k}(1), Y_{i1k}(0)\}$), as is the case when there are non-informative cluster sizes, both the pATE and cATE estimands can be further simplified to $E[Y_{i1k}(1) - Y_{i1k}(0)]$ (Appendix A.1).

We focus on these two estimands due to their natural interpretations, and by no means indicate that these are the only possible estimands of interest in PB-CRTs. Importantly, these two estimands address different levels of hypothesis. The pATE addresses a participant-level hypothesis whereas the cATE addresses a cluster-level hypothesis; see Kahan et al. (Kahan, Li, Copas, et al., 2023) for a more detailed discussion. When all clusters have the same cluster-period size or there is a homogeneous treatment effect across clusters, the two estimands will coincide, otherwise they generally differ in magnitude.

Notably, these estimand definitions are identical to those defined in P-CRTs without a baseline period (X. Wang et al., 2022) due to the lack of treatment positivity in the baseline period. However, the presence of a baseline period allows for more analytic options. Our central goal is thus to explore whether some of the analytic options previously proposed for PB-CRTs actually target the pATE or cATE estimands in the presence of informative cluster sizes.

# 3. Analytic Models

There are several common analytic models for PB-CRTs (Hooper et al., 2018). First, PB-CRTs can be simply analyzed with an independence estimating equation with a treatment indicator and period fixed effects. Such a model strictly makes "vertical" within-period comparisons, therefore the baseline period does not contribute any information to the treatment effect estimator. Accordingly, with continuous outcomes, the independence estimating equation yields an equivalent estimator to the ordinary least squares (OLS) estimator applied to only the follow-up period data as if it were a P-CRT. Throughout this article, we will simply refer to this as the "independence estimating equation" (IEE) model.

Alternatively, we can also use information from the baseline period to analyze PB-CRT data using a "two-way fixed-effects model" which includes fixed effects for both the clusters and periods. With a PB-CRT design, the two-way fixed-effects model estimator makes both between and within-cluster comparisons and closely resembles a standard difference-in-differences (DiD) estimator (Goodman-Bacon, 2021). Throughout this article, we will simply refer to this as the "fixed-effects" (FE) model.

Cluster effects in such multi-period CRT designs can also be accounted for with a mixed-effects model. In contrast to the fixed-effects model, mixed-effects models specify cluster intercepts (among other terms) as random. Therefore, a major distinction between modelling clusters as random or fixed depends on whether such confounders may exist (Gardiner et al., 2009). In this article, we derive an "exchangeable mixed-effects" (EME) model, which specifies a cluster random effect to induce an exchangeable correlation structure. Such a model has been previously referred to as a "constrained baseline analysis with a less flexible correlation structure" in the analysis of PB-CRTs (Hooper et al., 2018). We also derive a "nested exchangeable mixed-effects" (NEME) model, which specifies both a cluster random effect and a cluster-period random interaction to induce a nested exchangeable correlation structure between outcomes within the same cluster. Such a model has been previously referred to as a "constrained baseline analysis with a realistic correlation structure" in the analysis of PB-CRTs (Hooper et al., 2018).

## 3.1 IEE model

We can analyze the treatment effects in a PB-CRT by specifying treatment and period fixed effects in an independence estimating equation (IEE) model, which is equivalent to OLS with a treatment indicator and period fixed effects. As mentioned previously, such a model yields an equivalent estimator to the ordinary least squares (OLS) estimator applied to only the follow-up period data as if it were a P-CRT. Accordingly, this estimating equation has an independence correlation structure within and between clusters:

*Equation 3.1*

$$Y_{ijk} = \mu + X_{ij}\delta + \Phi_j + e_{ijk}$$

$$e_{ijk} \overset{iid}{\sim} N(0, \sigma_w^2)$$

where $\mu$ is the grand mean, $X_{ij}$ and $\delta$ are the indicator variable and fixed effect for the treatment effect, $\Phi_j$ is the period fixed effect for period $j$ ($\Phi_0 = 0$ for identifiability), and $e_{ijk}$ is the residual error.

The treatment effect in the IEE model can be estimated using ordinary least squares (OLS), with Equation 3.1 rewritten as:

*Equation 3.2*

$$Y = Z\theta_{IEE} + \epsilon$$

$$\epsilon \overset{iid}{\sim} N(0, \dot{V})$$

with $Y$ being the $n$ by 1 vector of individual level outcomes $Y_{ijk}$. $Z$ is the conventional $n$ by $(2+1)$ design matrix and $\theta_{IEE} = (\mu, \delta, \phi_1)'$ is the $J+1$ by 1 vector of parameters. $\dot{V} = \sigma_w^2 \mathrm{I}_n$ (where $\mathrm{I}_n$ is an $n$ by $n$ dimension identity matrix) denotes the variance-covariance matrix of $Y$. Recall that $n = \sum_{i=1}^{I} \sum_{j=0}^{1} K_{ij}$ for the total sample size.

The resulting IEE parameter point estimator is then:

*Equation 3.3*

$$\theta_{IEE} = (Z'Z)^{-1}Z'Y\,.$$

## 3.2 Fixed-effects model

We can define the fixed-effects (FE) model in the analysis of a PB-CRT with a period fixed effect and cluster fixed effects, shown below as Equation 3.4:

*Equation 3.4*

$$Y_{ijk} = \mu + X_{ij}\delta + \Phi_j + \alpha_i + e_{ijk}$$

$$e_{ijk} \overset{iid}{\sim} N(0, \sigma_w^2)$$

where $\alpha_i$ are the $I-1$ fixed cluster deviations from $\mu$ ($\alpha_1 = 0$ for identifiability).

The treatment effect in the FE model can be estimated using OLS, with Equation 3.4 rewritten as:

*Equation 3.5*

$$Y = \check{Z}\theta_{FE} + \epsilon$$

$$\epsilon \overset{iid}{\sim} N\big(0, \dot{V}\big)\,.$$

This resembles the IEE model (Equation 3.2), but is instead defined with $\check{Z}$ as the $n$ by $(I+2)$ design matrix and $\theta_{FE} = (\mu, \delta, \phi_1, \alpha_2, \dots, \alpha_I)'$ is the $(I+2)$ by 1 vector of parameters.

The resulting fixed-effects parameter point estimator is then:

*Equation 3.6*

$$\theta_{FE} = \big(\check{Z}'\check{Z}\big)^{-1}\check{Z}'Y\,.$$

## 3.3 Exchangeable mixed-effects model

We can define the exchangeable mixed-effects (EME) model in the analysis of a PB-CRT by including an additional cluster random effect in Equation 3.1, shown below as Equation 3.7:

*Equation 3.7*

$$Y_{ijk} = \mu + X_{ij}\delta + \Phi_j + \alpha_i + e_{ijk}$$

$$\alpha_i \overset{iid}{\sim} N(0, \tau_\alpha^2)$$

$$e_{ijk} \overset{iid}{\sim} N(0, \sigma_w^2)\,.$$

In contrast to the FE model (Equation 3.4), here $\alpha_i \overset{iid}{\sim} N(0, \tau_\alpha^2)$ is defined as the random (instead of fixed) cluster deviations from $\mu$.

The treatment effect in the EME model can be estimated using generalized least squares (GLS), with Equation 3.7 rewritten as:

*Equation 3.8*

$$Y = Z\theta_{EME} + \tilde{\epsilon}$$

$$\tilde{\epsilon} \sim N(0, \tilde{V})$$

which resembles the IEE model (Equation 3.2) but is instead defined with $\tilde{V}$ denoting the variance-covariance matrix of $Y$ as a $n$ by $n$ block diagonal matrix. $\tilde{V} = \mathrm{I}_I \otimes R_i^{EME}$ (where $\mathrm{I}_I$ is an $I$ by $I$ dimension identity matrix):

$$\tilde{V}_{ijk} = \begin{pmatrix} R_1^{EME} & 0 & \cdots & 0 \\ 0 & R_i^{EME} & \cdots & 0 \\ \vdots & \vdots & \ddots & \vdots \\ 0 & 0 & \cdots & R_I^{EME} \end{pmatrix}$$

and each block $R_i^{EME}$ is a $\sum_{j=0}^{1} K_{ij}$ by $\sum_{j=0}^{1} K_{ij}$ symmetric matrix:

$$R_i^{EME} = \mathrm{I}_{\sum_{j=0}^{1} K_{ij}} \sigma_w^2 + \mathrm{J}_{\sum_{j=0}^{1} K_{ij}} \tau_\alpha^2 = \begin{pmatrix} \sigma_w^2 + \tau_\alpha^2 & \tau_\alpha^2 & \cdots & \tau_\alpha^2 \\ \tau_\alpha^2 & \sigma_w^2 + \tau_\alpha^2 & \cdots & \tau_\alpha^2 \\ \vdots & \vdots & \ddots & \vdots \\ \tau_\alpha^2 & \tau_\alpha^2 & \cdots & \sigma_w^2 + \tau_\alpha^2 \end{pmatrix}$$

where $\mathrm{I}_{\sum_{j=0}^{1} K_{ij}}$ and $\mathrm{J}_{\sum_{j=0}^{1} K_{ij}}$ are a $\sum_{j=0}^{1} K_{ij}$ by $\sum_{j=0}^{1} K_{ij}$ dimension identity matrix and matrix of ones, respectively.

The resulting exchangeable mixed-effects parameter point estimator is then:

*Equation 3.9*

$$\theta_{EME} = \left(Z'\tilde{V}^{-1}Z\right)^{-1} Z'\tilde{V}^{-1}Y \,.$$

## 3.4 Nested exchangeable mixed-effects model

We can further define the nested exchangeable mixed-effects (NEME) model in the analysis of a PB-CRT by specifying a cluster random effect and an additional cluster-period random interaction, shown below as Equation 3.10:

*Equation 3.10*

$$Y_{ijk} = \mu + X_{ij}\delta + \Phi_j + \alpha_i + \gamma_{ij} + e_{ijk}$$

$$\alpha_i \overset{iid}{\sim} N(0, \tau_\alpha^2)$$

$$\gamma_{ij} \overset{iid}{\sim} N\left(0, \tau_\gamma^2\right)$$

$$e_{ijk} \overset{iid}{\sim} N(0, \sigma_w^2) \,.$$

The NEME model resembles the EME model (Equation 3.7), but with an additionally specified cluster-period random interaction term $\gamma_{ij} \sim N(0, \tau_\gamma^2)$.

The treatment effect in the NEME model can then be estimated using GLS, with Equation 3.10 rewritten as:

*Equation 3.11*

$$Y = Z\theta_{NEME} + \check{\epsilon}$$

$$\check{\epsilon} \sim N(0, \check{V})$$

which is defined with $\check{V}$, denoting the variance-covariance matrix of $Y$, as an $n$ by $n$ block diagonal matrix. $\check{V} = \mathrm{I}_I \otimes R_i^{NEME}$ (where $\mathrm{I}_I$ is an $I$ by $I$ dimension identity matrix):

$$\check{V} = \begin{pmatrix} R_1^{NEME} & 0 & \cdots & 0 \\ 0 & R_i^{NEME} & \cdots & 0 \\ \vdots & \vdots & \ddots & \vdots \\ 0 & 0 & \cdots & R_I^{NEME} \end{pmatrix}$$

and each block $R_i^{NEME}$ is a $\sum_{j=0}^{1} K_{ij}$ by $\sum_{j=0}^{1} K_{ij}$ symmetric matrix that can be written as the following block matrices:

$$R_i = \begin{pmatrix} R_{i1}^{NEME} & R_{i2}^{NEME} \\ R_{i3}^{NEME} & R_{i4}^{NEME} \end{pmatrix}.$$

Assuming equal cluster-period cell sizes between periods within clusters, $K_{i0} = K_{i1} = K_{i-}$, for simplicity, the components of the correlation matrix are:

$$R_{i1}^{NEME} = R_{i4}^{NEME} = \left(\mathrm{I}_{K_{i-}}\sigma_w^2 + \mathrm{J}_{K_{i-}}(\tau_\alpha^2 + \tau_\gamma^2)\right)$$
$$= \begin{pmatrix} \sigma_w^2 + \tau_\alpha^2 + \tau_\gamma^2 & \tau_\alpha^2 + \tau_\gamma^2 & \cdots & \tau_\alpha^2 + \tau_\gamma^2 \\ \tau_\alpha^2 + \tau_\gamma^2 & \sigma_w^2 + \tau_\alpha^2 + \tau_\gamma^2 & \cdots & \tau_\alpha^2 + \tau_\gamma^2 \\ \vdots & \vdots & \ddots & \vdots \\ \tau_\alpha^2 + \tau_\gamma^2 & \tau_\alpha^2 + \tau_\gamma^2 & \cdots & \sigma_w^2 + \tau_\alpha^2 + \tau_\gamma^2 \end{pmatrix},$$

$$R_{i2}^{NEME} = R_{i3}^{NEME} = \left(\mathrm{J}_{K_{i-}}\tau_\alpha^2\right) = \begin{pmatrix} \tau_\alpha^2 & \tau_\alpha^2 & \cdots & \tau_\alpha^2 \\ \tau_\alpha^2 & \tau_\alpha^2 & \cdots & \tau_\alpha^2 \\ \vdots & \vdots & \ddots & \vdots \\ \tau_\alpha^2 & \tau_\alpha^2 & \cdots & \tau_\alpha^2 \end{pmatrix}$$

where $\mathrm{I}_{K_{i-}}$ and $\mathrm{J}_{K_{i-}}$ are a $K_{i-}$ by $K_{i-}$ dimension identity matrix and matrix of ones, respectively.

The resulting nested-exchangeable mixed-effects parameter point estimator is then:

*Equation 3.12*

$$\theta_{NEME} = \left(Z'\check{V}^{-1}Z\right)^{-1} Z'\check{V}^{-1}Y.$$

## 4. Convergence of different estimators

In this section, we derive different estimators and their corresponding convergence probability limits for estimating the pATE and cATE; full derivations of the estimators are included in the Appendix (A.2). The properties of these estimators are summarized in Table 1.

The cATE estimand can be estimated by additionally utilizing inverse cluster size weights in the analysis of P-CRTs. In a simple P-CRT, the weighted treatment effect estimator, as described in Williamson et al. (Williamson et al., 2003), can be obtained by solving for the weighted estimating equations:

$$\sum_{i=1}^{I} \frac{D_i' V_i^{-1}(Y_i - \mu_i)}{w_i} = 0$$

where $D_i = \frac{d\mu_i}{d\theta}$. Subsequently, $Y_i = \left(Y_{i1}, \dots, Y_{iK_i}\right)'$ and $\mu_i = (\mu_i, \dots, \mu_i)'$ are the $K_i$ by 1 vector of outcomes and marginal means $\mu_i = E[Y_{ik}|Z_i] = Z_i\theta$, respectively. Here, $Z_i$ describes the P-CRT design matrix for observations in cluster $i$. Subsequently, $w_i$ is the cluster $i$-specific weight equal to 1 for equal participant-level weights or cluster size $K_i$ for equal cluster-level weights. This weighted estimating equation can be re-written as solving for:

$$\sum_{i=1}^{I} D_i' \boldsymbol{W}_i^{-1}(Y_i - \mu_i) = 0$$

where $\boldsymbol{W}_i = \mathrm{I}_I \otimes Q_i$ (where $\mathrm{I}_I$ is an $I$ by $I$ dimension identity matrix) for $I$ total clusters. With an independence and exchangeable correlation structure, we define the model-specific values of $Q_i$ as:

$$Q_i^{IEE} = w_i\left[(\sigma_e^2) I_{K_i}\right],$$

$$Q_i^{EME} = w_i\left[(\sigma_e^2)\mathrm{I}_{K_i} + (\tau_\alpha^2)\mathrm{J}_{K_i}\right],$$

respectively, where $\mathrm{I}_{K_i}$ is the $K_i \times K_i$ identity matrix and $\mathrm{J}_{K_i}$ is the $K_i \times K_i$ matrix of ones. Recall that $\sigma_e^2$ and $\tau_\alpha^2$ are the residual and cluster random effect variances. Altogether, in a P-CRT, the vector of coefficients is estimated by:

$$\hat{\theta} = \left(\sum_{i=1}^{I} Z_i' \boldsymbol{W}_i^{-1} Z_i\right)^{-1} \left(\sum_{i=1}^{I} Z_i' \boldsymbol{W}_i^{-1} Y_i\right) = (Z'\boldsymbol{W}^{-1}Z)^{-1} Z\boldsymbol{W}^{-1}Y\,.$$

Accordingly, the cATE estimand can be estimated by additionally utilizing inverse cluster-period size weights in the analysis of PB-CRTs. We can extend this weighted treatment effect estimator to PB-CRTs using inverse cluster-period size weights. In realistic scenarios where cluster-period cell sizes vary between periods within cluster, similar weighting can be implemented in analyses with uncorrelated errors (IEEw and FEw) by performing the corresponding analyses with cluster-period cell means (Kahan, Blette, et al., 2023). However, to our knowledge, the extension of variable cluster-period size weighting to analyses with correlated errors (EMEw and NEMEw) is not as clear in the existing literature, nor is it obvious if such weighted mixed-effects analyses are even desirable.

We assume for simplicity that cluster-period sizes vary between clusters but not between periods within clusters in a PB-CRT, $K_{i0} = K_{i1} = K_{i-}$. Therefore, the weighted treatment effect estimator in a PB-CRT can be obtained by solving for:

$$\sum_{i=1}^{I} \frac{D_i' V_i^{-1}(Y_i - \mu_i)}{w_i} = 0$$

where $D_i = \frac{d\mu_i}{d\theta}$. Subsequently, $Y_i = \left(Y_{i01}, \dots, Y_{i0K_{i-}}, Y_{i11}, \dots, Y_{i1K_{i-}}\right)'$ and $\mu_i = (\mu_{i0}, \dots, \mu_{i0}, \mu_{i1}, \dots, \mu_{i1})'$ are the $(2K_{i-})$ by 1 vector of outcomes and marginal means, respectively, with $\mu_i = E\left[Y_{ijk} | Z_i\right] = Z_i\theta$ ($\check{Z}$ in a fixed-effects model). The quantity $w_i$ is the cluster-specific weight equal to 1 for equal participant-level weights or cluster-period size $K_{i-}$. As in a P-CRT, this can be re-written in a PB-CRT as solving for:

$$\sum_{i=1}^{I} D_i' \boldsymbol{W}_i^{-1}(Y_i - \mu_i) = 0$$

where $\boldsymbol{W}_i = \mathrm{I}_I \otimes Q_i$ (where $\mathrm{I}_I$ is an $I$ by $I$ dimension identity matrix) for $I$ total clusters. With an IEE, FE, EME, and NEME model, we define the corresponding model-specific values of $Q_i$ as :

$$Q_i^{IEE} = Q_i^{FE} = w_i \sigma_e^2,$$

$$Q_i^{EME} = w_i R_i^{EME},$$

$$Q_i^{NEME} = w_i R_i^{NEME},$$

respectively. Altogether, in a PB-CRT:

$$\hat{\theta} = \left(\sum\nolimits_{i=1}^{I} Z_i' \boldsymbol{W}_i^{-1} Z_i\right)^{-1} \left(\sum\nolimits_{i=1}^{I} Z_i' \boldsymbol{W}_i^{-1} Y_i\right) = (Z' \boldsymbol{W}^{-1} Z)^{-1} Z \boldsymbol{W}^{-1} Y$$

($\check{Z}$ in a fixed-effects model).

Table 1. The convergence probability limits of the different treatment effect estimators $\hat{\delta}$ described here for continuous outcomes, alongside the sufficient conditions under which each method converges to the pATE or cATE estimands. Conditions with non-informative cluster sizes ("No ICS") can occur when $K_{i-} = K \ \forall \ i$ or with homogeneous treatment effects. The ICC is denoted in the EME and EMEw models with $\rho = \frac{\tau_\alpha^2}{\tau_\alpha^2+\sigma_w^2}$. The within-period ICC and between-period ICC in the NEME and NEMEw models are accordingly $\rho_{wp} = \frac{\tau_\alpha^2+\tau_\gamma^2}{\tau_\alpha^2+\tau_\gamma^2+\sigma_w^2}$ and $\rho_{bp} = \frac{\tau_\alpha^2}{\tau_\alpha^2+\tau_\gamma^2+\sigma_w^2}$. (* We assume that cluster-period cell sizes do not vary within clusters $K_{i0} = K_{i1} = K_{i-}$ in the derivation of these estimators.)

| **Method** | $\hat{\delta} \xrightarrow{P}$ | **Sufficient conditions for convergence to the:** | |
|---|---|---|---|
| | | **pATE** | **cATE** |
| IEE | $E\left[\frac{1}{E[K_{i1}]}\sum_{k=1}^{K_{i1}}[Y_{i1k}(1)-Y_{i1k}(0)]\right]$ | Always | No ICS |
| IEEw | $E\left[\frac{1}{K_{i1}}\sum_{k=1}^{K_{i1}}[Y_{i1k}(1)-Y_{i1k}(0)]\right]$ | No ICS | Always |
| FE | $E\left[\frac{\frac{K_{i0}}{\sum_{j=0}^{1}K_{ij}}\sum_{k=1}^{K_{i1}}[Y_{i1k}(1)-Y_{i1k}(0)]}{E\left[\frac{\prod_{j=0}^{1}K_{ij}}{\sum_{j=0}^{1}K_{ij}}\right]}\right]$ | When $K_{i0} = K_{i1}$ or no ICS | No ICS |
| FEw | $E\left[\frac{1}{K_{i1}}\sum_{k=1}^{K_{i1}}[Y_{i1k}(1)-Y_{i1k}(0)]\right]$ | No ICS | Always |
| EME* | $E\left[\frac{\left(\frac{1+(K_{i-}-1)\rho}{1+(2K_{i-}-1)\rho}\right)}{E\left[\left(\frac{1+(K_{i-}-1)\rho}{1+(2K_{i-}-1)\rho}\right)K_{i-}\right]}\sum_{k=1}^{K_{i-}}[Y_{i1k}(1)-Y_{i1k}(0)]\right]$ | When $\rho = 0$ or 1, or no ICS | No ICS |
| EMEw* | $E\left[\frac{\left(\frac{1+(K_{i-}-1)\rho}{1+(2K_{i-}-1)\rho}\right)}{E\left[\left(\frac{1+(K_{i-}-1)\rho}{1+(2K_{i-}-1)\rho}\right)\right]}\left(\frac{1}{K_{i-}}\sum_{k=1}^{K_{i-}}[Y_{i1k}(1)-Y_{i1k}(0)]\right)\right]$ | No ICS | When $\rho = 0$ or 1, or no ICS |
| NEME* | $E\left[\frac{\left(\frac{1+(K_{i-}-1)\rho_{wp}}{\left(1+(K_{i-}-1)\rho_{wp}\right)^2-(K_{i-})^2\rho_{bp}^2}\right)}{E\left[\left(\frac{1+(K_{i-}-1)\rho_{wp}}{\left(1+(K_{i-}-1)\rho_{wp}\right)^2-(K_{i-})^2\rho_{bp}^2}\right)K_{i-}\right]}\sum_{k=1}^{K_{i-}}[Y_{i1k}(1)-Y_{i1k}(0)]\right]$ | When $\rho_{wp} = \rho_{bp} = 0$, or no ICS | No ICS |
| NEMEw* | $E\left[\frac{\left(\frac{1+(K_{i-}-1)\rho_{wp}}{\left(1+(K_{i-}-1)\rho_{wp}\right)^2-(K_{i-})^2\rho_{bp}^2}\right)}{E\left[\frac{1+(K_{i-}-1)\rho_{wp}}{\left(1+(K_{i-}-1)\rho_{wp}\right)^2-(K_{i-})^2\rho_{bp}^2}\right]}\left(\frac{1}{K_{i-}}\sum_{k=1}^{K_{i-}}[Y_{i1k}(1)-Y_{i1k}(0)]\right)\right]$ | No ICS | When $\rho_{wp} = \rho_{bp} = 0$, or no ICS |

## 4.1 Independence estimating equation (IEE)

Earlier, we mentioned how the independence estimating equation (IEE) treatment effect estimator only uses information from the follow-up period of the PB-CRT, and can be written with potential outcomes $\left(Y_{i1k}(0), Y_{i1k}(1)\right)$ for participant $k \in (1, \dots, K_{i1})$ in period $j = 1$ of cluster $i \in (1, \dots, I)$. Let $S_i$ be an indicator for whether individuals are assigned to cluster sequence $S_i = 1$, the IEE treatment effect estimator is then:

<u>*Equation 4.1*</u>

$$\hat{\delta}_{IEE} = \left[\frac{\sum_{i=1}^{I} S_i \sum_{k=1}^{K_{i1}} Y_{i1k}(1)}{\sum_{i=1}^{I} S_i K_{i1}}\right] - \left[\frac{\sum_{i=1}^{I}(1-S_i)\sum_{k=1}^{K_{i1}} Y_{i1k}(0)}{\sum_{i=1}^{K_{i1}}(1-S_i)K_{i1}}\right].$$

We can demonstrate that this estimator is consistent and asymptotically unbiased for the pATE:

*Equation 4.2*

$$\hat{\delta}_{IEE} \xrightarrow{P} E\left[\frac{1}{E[K_{i1}]}\sum_{k=1}^{K_{i1}}[Y_{i1k}(1) - Y_{i1k}(0)]\right].$$

More information about the derivation of these estimators is included in the Appendix (A.2.1).

## 4.2 Independence estimating equation with inverse cluster-period size weighting (IEEw)

The independence estimating equation with inverse cluster-period size weighting (IEEw) treatment effect estimator can then be written as:

$$\hat{\delta}_{IEEw} = \left[\frac{\sum_{i=1}^{I} S_i \frac{1}{K_{i1}} \sum_{k=1}^{K_{i1}} Y_{i1k}(1)}{\sum_{i=1}^{I} S_i}\right] - \left[\frac{\sum_{i=1}^{I}(1 - S_i)\frac{1}{K_{i1}}\sum_{k=1}^{K_{i1}} Y_{i1k}(0)}{\sum_{i=1}^{I}(1 - S_i)}\right]$$

We can demonstrate that this estimator is consistent and asymptotically unbiased for the cATE:

*Equation 4.3*

$$\hat{\delta}_{IEEw} \xrightarrow{P} E\left[\frac{1}{K_{i1}}\sum_{k=1}^{K_{i1}}[Y_{i1k}(1) - Y_{i1k}(0)]\right].$$

Additionally, the IEEw estimator is also unbiased for the cATE in expectation over the sampling distribution (A.2.2). This is a stronger result than consistency.

## 4.3 Fixed-effects model (FE)

The fixed-effects (FE) treatment effect estimator can be written with potential outcomes $\left(Y_{ijk}(0), Y_{i1k}(1)\right)$ for participant $k \in (1, \ldots, K_{i-})$ in period $j \in (0,1)$ of cluster $i \in (1, \ldots, I)$. Let $S_i$ be an indicator for whether individuals are assigned to cluster sequence $S_i = 1$:

*Equation 4.4*

$$\hat{\delta}_{FE} = \left[\frac{\left(\sum_{i=1}^{I} S_i \frac{K_{i0}}{\sum_{j=0}^{1} K_{ij}} \sum_{k=1}^{K_{i1}} Y_{i1k}(1)\right) - \left(\sum_{i=1}^{I} S_i \frac{K_{i1}}{\sum_{j=0}^{1} K_{ij}} \sum_{k} Y_{i0k}(0)\right)}{\sum_{i=1}^{I} S_i \frac{\prod_{j=0}^{1} K_{ij}}{\sum_{j=0}^{1} K_{ij}}}\right] - \left[\frac{\left(\sum_{i=1}^{I}(1 - S_i) \frac{K_{i0}}{\sum_{j=0}^{1} K_{ij}} \sum_{k} Y_{i1k}(0)\right) - \left(\sum_{i=1}^{I}(1 - S_i) \frac{K_{i1}}{\sum_{j=0}^{1} K_{ij}} \sum_{k} Y_{i0k}(0)\right)}{\sum_{i=1}^{I}(1 - S_i) \frac{\prod_{j=0}^{1} K_{ij}}{\sum_{j=0}^{1} K_{ij}}}\right].$$

Accordingly, we have the following convergence in probability:

*Equation 4.5*

$$\hat{\delta}_{FE} \overset{P}{\to} E\left[\frac{\frac{K_{i0}}{\sum_{j=0}^{1} K_{ij}} \sum_{k=1}^{K_{i1}} [Y_{i1k}(1) - Y_{i1k}(0)]}{E\left[\frac{\prod_{j=0}^{1} K_{ij}}{\sum_{j=0}^{1} K_{ij}}\right]}\right]$$

when cluster-period sizes vary between periods within clusters, $K_{i0} \neq K_{i1}$. Notably, this FE estimator is not a theoretically consistent estimator for the pATE estimand when cluster-period sizes vary within clusters. More information is included in the Appendix (A.2.3).

However, if we assume that cluster-period sizes vary between clusters but not between periods within clusters, $K_{i0} = K_{i1} = K_{i-}$, Equation 4.4 simplifies to:

*Equation 4.6*

$$\hat{\delta}_{FE} = \left[\frac{\left(\sum_{i=1}^{I} S_i \sum_{k=1}^{K_{i-}} Y_{i1k}(1)\right) - \left(\sum_{i=1}^{I} S_i \sum_{k=1}^{K_{i-}} Y_{i0k}(0)\right)}{\sum_{i=1}^{I} S_i K_{i-}}\right] - \left[\frac{\left(\sum_{i=1}^{I} (1 - S_i) \sum_{k=1}^{K_{i-}} Y_{i1k}(0)\right) - \left(\sum_{i=1}^{I} (1 - S_i) \sum_{k=1}^{K_{i-}} Y_{i0k}(0)\right)}{\sum_{i=1}^{I} (1 - S_i) K_{i-}}\right].$$

We can demonstrate that this estimator is consistent and asymptotically unbiased for the pATE when cluster-period sizes do not vary within clusters:

*Equation 4.7*

$$\hat{\delta}_{FE} \overset{P}{\to} E\left[\frac{1}{E[K_{i1}]} \sum_{k=1}^{K_{i1}} [Y_{i1k}(1) - Y_{i1k}(0)]\right]$$

More information is included in the Appendix (A.2.3).

## 4.4 Fixed-effects model with inverse cluster-period size weighting (FEw)

When cluster-period sizes vary between periods within clusters, $K_{i0} \neq K_{i1}$, we can still easily specify inverse cluster-period size weights with the fixed-effects (FEw) estimator. Accordingly, this gives the following FEw treatment effect estimator:

*Equation 4.8*

$$\hat{\delta}_{FEw} = \left[\frac{\left(\sum_{i=1}^{I} S_i \frac{1}{K_{i1}} \sum_{k=1}^{K_{i1}} Y_{i1k}(1)\right) - \left(\sum_{i=1}^{I} S_i \frac{1}{K_{i0}} \sum_{k=1}^{K_{i0}} Y_{i0k}(0)\right)}{\sum_{i=1}^{I} S_i}\right] - \left[\frac{\left(\sum_{i=1}^{I} (1 - S_i) \frac{1}{K_{i1}} \sum_{k=1}^{K_{i1}} Y_{i1k}(0)\right) - \left(\sum_{i=1}^{I} (1 - S_i) \frac{1}{K_{i0}} \sum_{k=1}^{K_{i0}} Y_{i0k}(0)\right)}{\sum_{i=1}^{I} (1 - S_i)}\right].$$

We can demonstrate that this estimator is consistent and asymptotically unbiased for the cATE:

*Equation 4.9*

$$\hat{\delta}_{FEw} \xrightarrow{P} E\left[\frac{1}{K_{i1}}\sum_{k=1}^{K_{i1}}[Y_{i1k}(1) - Y_{i1k}(0)]\right]$$

Additionally, the FEw estimator is also unbiased for the cATE in expectation over the sampling distribution (A.2.4), similar to the property of IEEw described in Section 4.2.

Notably, such weights deviate from Williamson's derivation with cluster-specific weights and are instead specified as cluster-period-specific weights. Such inverse cluster-period size weights can be easily defined in analyses with an independence correlation structure (IEEw and FEw) by specifying analyses with cluster-period cell means.

The FEw estimator is equivalent to a standard difference-in-differences (DiD) model with the cluster-period means. Without randomization, the estimator in expectation is still a theoretically consistent estimator for the cluster-average treatment effect on the treated (cATT) estimand when assuming parallel trends. More information is included in the Appendix (A.2.4).

## 4.5 Mixed-effects models

In the subsequent sections (4.6-4.9), we describe the different mixed-effects models (weighted and unweighted), assuming $K_{i0} = K_{i1} = K_{i-}$ and randomization. We will generally demonstrate that the mixed-effects treatment effect estimators are of the form:

*Equation 4.10*

$$\hat{\delta}_{ME} \xrightarrow{P} E\left[\frac{A_i/K_{i-}}{E[A_i]}\sum_{k=1}^{K_{i-}}[Y_{i1k}(1) - Y_{i1k}(0)]\right]$$

with mixed-effects model-specific values for $A_i$. Therefore, unless $A_i \propto 1$, $\hat{\delta}_{ME}$ generally converges to a weighted average treatment effect estimand with data-dependent model-specific weights that is difficult to interpret. This issue was discussed by Kahan et al. (Kahan, Blette, et al., 2023) in the context of P-CRTs without a baseline period, and is generalized here in the context of PB-CRTs. As a result, unweighted and weighted mixed-effects models are not theoretically consistent estimators for the pATE nor cATE estimands, unless under some extreme conditions (Table 1).

## 4.6 Exchangeable Mixed-effects model (EME)

The exchangeable mixed-effects (EME) treatment effect estimator can be specified based on Equation 3.9 where the diagonal terms $(D_i)$ and off-diagonal terms $(F_i)$ in the block matrix corresponding to the observations within cluster $i$ in $\tilde{V}_{ijk}^{-1}$ is:

$$D_i = \frac{1}{\sigma_w^2}\left(\frac{\sigma_w^2 + (2K_{i-} - 1)\tau_\alpha^2}{\sigma_w^2 + 2K_{i-}\tau_\alpha^2}\right),$$

$$F_i = -\frac{1}{\sigma_w^2}\left(\frac{\tau_\alpha^2}{\sigma_w^2 + 2K_{i-}\tau_\alpha^2}\right),$$

assuming that cluster-period sizes vary between clusters but not between periods within clusters, $K_{i0} = K_{i1} = K_{i-}$, with $2K_{i-}$ observations within each cluster.

Accordingly, using Equation 4.10 where:

$$A_i = K_{i-}[D_i + (K_{i-} - 1)F_i] = \frac{1}{\sigma_w^2}\left(\frac{1 + (K_{i-} - 1)\rho}{1 + (2K_{i-} - 1)\rho}\right)K_{i-}$$

with intracluster correlation coefficient (ICC) $\rho = \frac{\tau_\alpha^2}{\tau_\alpha^2 + \sigma_w^2}$, we can demonstrate that the EME treatment effect estimator converges in probability to:

*Equation 4.11*

$$\hat{\delta}_{EME} \overset{P}{\to} E\left[\frac{\left(\frac{1 + (K_{i-} - 1)\rho}{1 + (2K_{i-} - 1)\rho}\right)}{E\left[\left(\frac{1 + (K_{i-} - 1)\rho}{1 + (2K_{i-} - 1)\rho}\right)K_{i-}\right]}\sum_{k=1}^{K_{i-}}[Y_{i1k}(1) - Y_{i1k}(0)]\right].$$

This estimand includes a cluster-specific weight that depends on both the cluster-period size as well as the probability limit of the ICC $\rho$. Therefore, in general, the ICC will dictate the estimands and therefore the analysis of a different outcome may inadvertently change the estimand, which can be undesirable.

There are several special cases where this estimand reduces to familiar estimands. First, when $\rho = 0$ and there is simply no ICC, then $E[\hat{\delta}_{EME}] \overset{P}{\to} E\left[\frac{1}{E[K_{i-}]}\sum_{k=1}^{K_{i-}}[Y_{i1k}(1) - Y_{i1k}(0)]\right]$, which is the pATE estimand. Second, when $\rho = 1$ and the outcomes are perfectly correlated, $\hat{\delta}_{EME}$ also converges to the pATE estimand. However, both conditions are unlikely to be realistic for PB-CRTs. More information is included in the Appendix (A.2.5).

## 4.7 Exchangeable Mixed-effects model with inverse cluster-period size weighting (EMEw)

The weighted EMEw treatment effect estimator can be specified similarly to the unweighted EME estimator, but with the diagonal terms $(D_i)$ and off-diagonal terms $(F_i)$ in the block matrices of the inverted weighted correlation structure $\boldsymbol{W}_i^{-1}$ (as described in Section 4) corresponding to the observations within cluster $i$ specified as:

$$D_i = \frac{1}{(K_{i-})\sigma_w^2}\left(\frac{\sigma_w^2 + (2K_{i-} - 1)\tau_\alpha^2}{\sigma_w^2 + 2K_{i-}\tau_\alpha^2}\right),$$

$$F_i = -\frac{1}{(K_{i-})\sigma_w^2}\left(\frac{\tau_\alpha^2}{\sigma_w^2 + 2K_{i-}\tau_\alpha^2}\right),$$

assuming that cluster-period sizes vary between clusters but not between periods within clusters, $K_{i0} = K_{i1} = K_{i-}$.

Accordingly, using Equation 4.10 where:

$$A_i = K_{i-}[D_i + (K_{i-} - 1)F_i] = \frac{1}{\sigma_w^2}\left(\frac{1 + (K_{i-} - 1)\rho}{1 + (2K_{i-} - 1)\rho}\right),$$

with intracluster correlation coefficient (ICC) $\rho = \frac{\tau_\alpha^2}{\tau_\alpha^2+\sigma_w^2}$, we can demonstrate that the EMEw treatment effect estimator converges in probability to:

*Equation 4.12*

$$\hat{\delta}_{EMEw} \xrightarrow{P} E\left[\frac{\left(\frac{1+(K_{i-}-1)\rho}{1+(2K_{i-}-1)\rho}\right)}{E\left[\left(\frac{1+(K_{i-}-1)\rho}{1+(2K_{i-}-1)\rho}\right)\right]}\left(\frac{1}{K_{i-}}\sum_{k=1}^{K_{i-}}[Y_{i1k}(1)-Y_{i1k}(0)]\right)\right]$$

More information is included in the Appendix (A.2.6).

Like the EME estimator (Equation 4.11), this estimand includes a cluster-specific weight that depends on both the cluster-period size as well as the probability limit of the ICC $\rho$. Notably, for $\rho \rightarrow 0$ or 1, the $E[\hat{\delta}_{EMEw}] \rightarrow cATE$.

Surprisingly, unlike the results in (X. Wang et al., 2022) with a P-CRT, we will further show in subsequent sections (Sections 5 and 6) that the expected value of the treatment effect with an exchangeable correlation structure and inverse cluster-period size weights can yield mostly unbiased estimates for the cATE in a PB-CRT.

## 4.8 Nested Exchangeable Mixed-effects model (NEME)

The nested exchangeable mixed-effects (NEME) treatment effect estimator can be specified based on Equation 3.12. Assuming that cluster-period sizes vary between clusters but not between periods within clusters, $K_{i0} = K_{i1} = K_{i-}$ with $2K_{i-}$ observations within each cluster, the specified correlation structure is then $\check{V}_{ijk} = \mathrm{I}_I \otimes R_i$, where:

$$R_i = \begin{pmatrix} R_{i1}^{NME} & R_{i2}^{NME} \\ R_{i3}^{NME} & R_{i4}^{NME} \end{pmatrix},$$

with block matrices:

$$R_{i1}^{NME} = R_{i4}^{NME} = \left(\mathrm{I}_{K_{i-}}\sigma_w^2 + \mathrm{J}_{K_{i-}}(\tau_\alpha^2+\tau_\gamma^2)\right) = \begin{pmatrix} \sigma_w^2+\tau_\alpha^2+\tau_\gamma^2 & \tau_\alpha^2+\tau_\gamma^2 & \cdots & \tau_\alpha^2+\tau_\gamma^2 \\ \tau_\alpha^2+\tau_\gamma^2 & \sigma_w^2+\tau_\alpha^2+\tau_\gamma^2 & \cdots & \tau_\alpha^2+\tau_\gamma^2 \\ \vdots & \vdots & \ddots & \vdots \\ \tau_\alpha^2+\tau_\gamma^2 & \tau_\alpha^2+\tau_\gamma^2 & \cdots & \sigma_w^2+\tau_\alpha^2+\tau_\gamma^2 \end{pmatrix},$$

$$R_{i2}^{NME} = R_{i3}^{NME} = \left(\mathrm{J}_{K_{i-}}\tau_\alpha^2\right) = \begin{pmatrix} \tau_\alpha^2 & \tau_\alpha^2 & \cdots & \tau_\alpha^2 \\ \tau_\alpha^2 & \tau_\alpha^2 & \cdots & \tau_\alpha^2 \\ \vdots & \vdots & \ddots & \vdots \\ \tau_\alpha^2 & \tau_\alpha^2 & \cdots & \tau_\alpha^2 \end{pmatrix},$$

where $\mathrm{I}_{K_{i-}}$ is a $K_{i-}$ by $K_{i-}$ dimension identity matrix and $\mathrm{J}_{K_{i-}}$ is a $K_{i-}$ by $K_{i-}$ dimension matrix of ones. Accordingly, $\tilde{V}_{ijk}^{-1} = \mathrm{I}_I \otimes R_i^{-1}$, where:

$$R_i^{-1} = \begin{pmatrix} [R_{i1}^{NME} - R_{i2}^{NME}(R_{i1}^{NME})^{-1}R_{i2}^{NME}]^{-1} & -[R_{i1}^{NME} - R_{i2}^{NME}(R_{i1}^{NME})^{-1}R_{i2}^{NME}]^{-1}R_{i2}^{NME}(R_{i1}^{NME})^{-1} \\ -[R_{i1}^{NME} - R_{i2}^{NME}(R_{i1}^{NME})^{-1}R_{i2}^{NME}]^{-1}R_{i2}^{NME}(R_{i1}^{NME})^{-1} & [R_{i1}^{NME} - R_{i2}^{NME}(R_{i1}^{NME})^{-1}R_{i2}^{NME}]^{-1} \end{pmatrix}.$$

We define:

$$[R_{i1}^{NME} - R_{i2}^{NME}(R_{i1}^{NME})^{-1}R_{i2}^{NME}]^{-1} = \left(\mathrm{I}_{K_{i-}}(D_i - F_i) + \mathrm{J}_{K_{i-}}(F_i)\right)$$

where the diagonal terms $(D_i)$ and off-diagonal terms $(F_i)$ are:

$$D_i = \frac{1}{\sigma_w^2}\left(\frac{\sigma_w^2 + (K_{i-} - 1)\left[\left(\tau_\alpha^2 + \tau_\gamma^2\right) - \frac{(K_{i-})(\tau_\alpha^2)^2}{\sigma_w^2 + (K_{i-})\left(\tau_\alpha^2 + \tau_\gamma^2\right)}\right]}{\sigma_w^2 + (K_{i-})\left[\left(\tau_\alpha^2 + \tau_\gamma^2\right) - \frac{(K_{i-})(\tau_\alpha^2)^2}{\sigma_w^2 + (K_{i-})\left(\tau_\alpha^2 + \tau_\gamma^2\right)}\right]}\right),$$

$$F_i = -\frac{1}{\sigma_w^2}\left(\frac{\left(\tau_\alpha^2 + \tau_\gamma^2\right) - \frac{(K_{i-})(\tau_\alpha^2)^2}{\sigma_w^2 + (K_{i-})\left(\tau_\alpha^2 + \tau_\gamma^2\right)}}{\sigma_w^2 + (K_{i-})\left[\left(\tau_\alpha^2 + \tau_\gamma^2\right) - \frac{(K_{i-})(\tau_\alpha^2)^2}{\sigma_w^2 + (K_{i-})\left(\tau_\alpha^2 + \tau_\gamma^2\right)}\right]}\right).$$

Accordingly, using Equation 4.10 where:

$$A_i = K_{i-}[D_i + (K_{i-} - 1)F_i]$$

$$= K_{i-}\left(\frac{1 + (K_{i-} - 1)\rho_{wp}}{\left[\left(1 + (K_{i-} - 1)\rho_{wp}\right)^2 - (K_{i-})^2\rho_{bp}^2\right]\left(\tau_\alpha^2 + \tau_\gamma^2 + \sigma_w^2\right)}\right)$$

with within-period ICC $\left(\rho_{wp} = \frac{\tau_\alpha^2 + \tau_\gamma^2}{\tau_\alpha^2 + \tau_\gamma^2 + \sigma_w^2}\right)$ and between-period ICC $\left(\rho_{bp} = \frac{\tau_\alpha^2}{\tau_\alpha^2 + \tau_\gamma^2 + \sigma_w^2}\right)$, we can demonstrate that the NEME treatment effect estimator converges in probability to:

_Equation 4.13_

$$\hat{\delta}_{NEME} \xrightarrow{P} E\left[\frac{\left(\frac{1 + (K_{i-} - 1)\rho_{wp}}{\left(1 + (K_{i-} - 1)\rho_{wp}\right)^2 - (K_{i-})^2\rho_{bp}^2}\right)}{E\left[\left(\frac{1 + (K_{i-} - 1)\rho_{wp}}{\left(1 + (K_{i-} - 1)\rho_{wp}\right)^2 - (K_{i-})^2\rho_{bp}^2}\right)K_{i-}\right]}\sum_{k=1}^{K_{i-}}[Y_{i1k}(1) - Y_{i1k}(0)]\right].$$

More information is included in the Appendix (A.2.7).

The NEME estimand includes a cluster-specific weight that depends on both the cluster-period size as well as the probability limit of the within-period and between-period ICC, $\rho_{wp}$ and $\rho_{bp}$, respectively. Notably, for $\rho_{wp} = \rho_{bp} \to 0$, the $E[\hat{\delta}_{NEME}] \to pATE$. However, this condition is unlikely to be realistic for PB-CRTs.

## 4.9 Nested Exchangeable Mixed-effects model with inverse cluster-period size weighting (NEMEw)

The weighted nested exchangeable mixed-effects model with inverse cluster-period size weighting (NEMEw) treatment effect estimator can be specified similarly to the unweighted NEME estimator, but with the following terms:

$$D_i = \frac{1}{(K_{i-})\sigma_w^2}\left(\frac{\sigma_w^2 + (K_{i-} - 1)\left[\left(\tau_\alpha^2 + \tau_\gamma^2\right) - \frac{(K_{i-})(\tau_\alpha^2)^2}{\sigma_w^2 + (K_{i-})\left(\tau_\alpha^2 + \tau_\gamma^2\right)}\right]}{\sigma_w^2 + (K_{i-})\left[\left(\tau_\alpha^2 + \tau_\gamma^2\right) - \frac{(K_{i-})(\tau_\alpha^2)^2}{\sigma_w^2 + (K_{i-})\left(\tau_\alpha^2 + \tau_\gamma^2\right)}\right]}\right),$$

$$F_i = -\frac{1}{(K_{i-})\sigma_w^2}\left(\frac{\left(\tau_\alpha^2 + \tau_\gamma^2\right) - \frac{(K_{i-})(\tau_\alpha^2)^2}{\sigma_w^2 + (K_{i-})\left(\tau_\alpha^2 + \tau_\gamma^2\right)}}{\sigma_w^2 + (K_{i-})\left[\left(\tau_\alpha^2 + \tau_\gamma^2\right) - \frac{(K_{i-})(\tau_\alpha^2)^2}{\sigma_w^2 + (K_{i-})\left(\tau_\alpha^2 + \tau_\gamma^2\right)}\right]}\right).$$

Accordingly, using Equation 4.10 where:

$$A_i = K_{i-}[D_i + (K_{i-} - 1)F_i]$$

$$= \left(\frac{1 + (K_{i-} - 1)\rho_{wp}}{\left[\left(1 + (K_{i-} - 1)\rho_{wp}\right)^2 - (K_{i-})^2\rho_{bp}^2\right]\left(\tau_\alpha^2 + \tau_\gamma^2 + \sigma_w^2\right)}\right)$$

with within-period ICC $\left(\rho_{wp} = \frac{\tau_\alpha^2+\tau_\gamma^2}{\tau_\alpha^2+\tau_\gamma^2+\sigma_w^2}\right)$ and between-period ICC $\left(\rho_{bp} = \frac{\tau_\alpha^2}{\tau_\alpha^2+\tau_\gamma^2+\sigma_w^2}\right)$, we can demonstrate that the NEMEw treatment effect estimator converges in probability to:

*<u>Equation 4.14</u>*

$$\hat{\delta}_{NEMEw} \overset{P}{\to} E\left[\frac{\left(\frac{1 + (K_{i-} - 1)\rho_{wp}}{\left(1 + (K_{i-} - 1)\rho_{wp}\right)^2 - (K_{i-})^2\rho_{bp}^2}\right)}{E\left[\frac{1 + (K_{i-} - 1)\rho_{wp}}{\left(1 + (K_{i-} - 1)\rho_{wp}\right)^2 - (K_{i-})^2\rho_{bp}^2}\right]}\left(\frac{1}{K_{i-}}\sum_{k=1}^{K_{i-}}[Y_{i1k}(1) - Y_{i1k}(0)]\right)\right]$$

More information is included in the Appendix (A.2.8).

The NEMEw estimand includes a cluster-specific weight that depends on both the cluster-period size as well as the probability limit of the within-period and between-period ICC, $\rho_{wp}$ and $\rho_{bp}$, respectively. Notably, for $\rho_{wp} = \rho_{bp} \to 0$, the $E\left[\hat{\delta}_{NEMEw}\right] \to cATE$. However, as mentioned previously, this condition is unlikely to be realistic for PB-CRTs.

# 5. Evaluation of EMEw and NEMEw estimators

As we previously demonstrated in Sections 4.5 - 4.9, the unweighted and weighted mixed-effects models are not theoretically consistent estimators for the pATE nor the cATE, and instead converge to weighted versions of these estimands.

In this section, we will explore the EMEw and NEMEw treatment effect estimators, characterize their estimand weights, and illustrate the pattern of the estimators' biases for the cATE estimand. We focus on the estimand weights in the EMEw and NEMEw estimators for the cATE estimand, and not in the EME and NEME estimators for the pATE estimand, due to the simpler interpretation.

In Equation 4.10, we demonstrated that the exchangeable and nested-exchangeable mixed-effects model estimators share a common general form in the analysis of PB-CRTs. Notably, the EMEw treatment effect estimator converges in probability to a weighted average of the cluster-specific cATE estimands:

$$\hat{\delta}_{EMEw} \xrightarrow{P} E\left[\lambda_{EMEw}\left(\frac{1}{K_{i-}}\sum_{k=1}^{K_{i-}}[Y_{i1k}(1) - Y_{i1k}(0)]\right)\right]$$

with cluster-specific estimand weights:

$$\lambda_{EMEw} = \frac{\left(\frac{1 + (K_{i-} - 1)\rho}{1 + (2K_{i-} - 1)\rho}\right)}{E\left[\left(\frac{1 + (K_{i-} - 1)\rho}{1 + (2K_{i-} - 1)\rho}\right)\right]}.$$

Similarly, the NEMEw treatment effect estimator converges in probability to a weighted average of the cluster-specific cATE estimands:

$$\hat{\delta}_{NEMEw} \xrightarrow{P} E\left[\lambda_{NEMEw}\left(\frac{1}{K_{i-}}\sum_{k=1}^{K_{i-}}[Y_{i1k}(1) - Y_{i1k}(0)]\right)\right]$$

with cluster-specific estimand weights:

$$\lambda_{NEMEw} = \frac{\left(\frac{1 + (K_{i-} - 1)\rho_{wp}}{\left(1 + (K_{i-} - 1)\rho_{wp}\right)^2 - (K_{i-})^2\rho_{bp}^2}\right)}{E\left[\frac{1 + (K_{i-} - 1)\rho_{wp}}{\left(1 + (K_{i-} - 1)\rho_{wp}\right)^2 - (K_{i-})^2\rho_{bp}^2}\right]}.$$

When the estimand weights $\lambda_{EMEw} = 1$ and $\lambda_{NEMEw} = 1$, the EMEw and NEMEw treatment effect estimators converge to the cATE estimand. We plot some numerical examples to explore when these estimand weights deviate from 1 and illustrate the pattern of the bias in the EMEw and NEMEw estimators for the cATE estimand in the presence of informative cluster sizes.

In Figure 2, we plot the values of the estimand weights from the EMEw estimator $\lambda_{EMEw,u}$ across different ICC values $\rho \in [0,1]$, in scenarios with two subpopulations $u = 1,2$ with fixed subpopulation-specific cluster-period sizes $K_{i-,1}$ and $K_{i-,2}$ that differ by a ratio of $\zeta = \frac{K_{i-,1}}{K_{i-,2}}$, and equal probabilities of being sampled $P(u = 1) = 0.5$ and $P(u = 2) = 1 - P(u = 1)$.

Figure 2. EMEw estimand weights $\lambda_{EMEw,u}$ are plotted across two subpopulations $u = 1, 2$ and different values of ICC, with subpopulation-specific cluster-period sizes $K_{i-,1}$ and $K_{i-,2}$ that differ by a ratio of $\zeta = \frac{K_{i-,1}}{K_{i-,2}}$, and equal probabilities of being sampled $P(u = 1) = P(u = 2) = 0.5$.

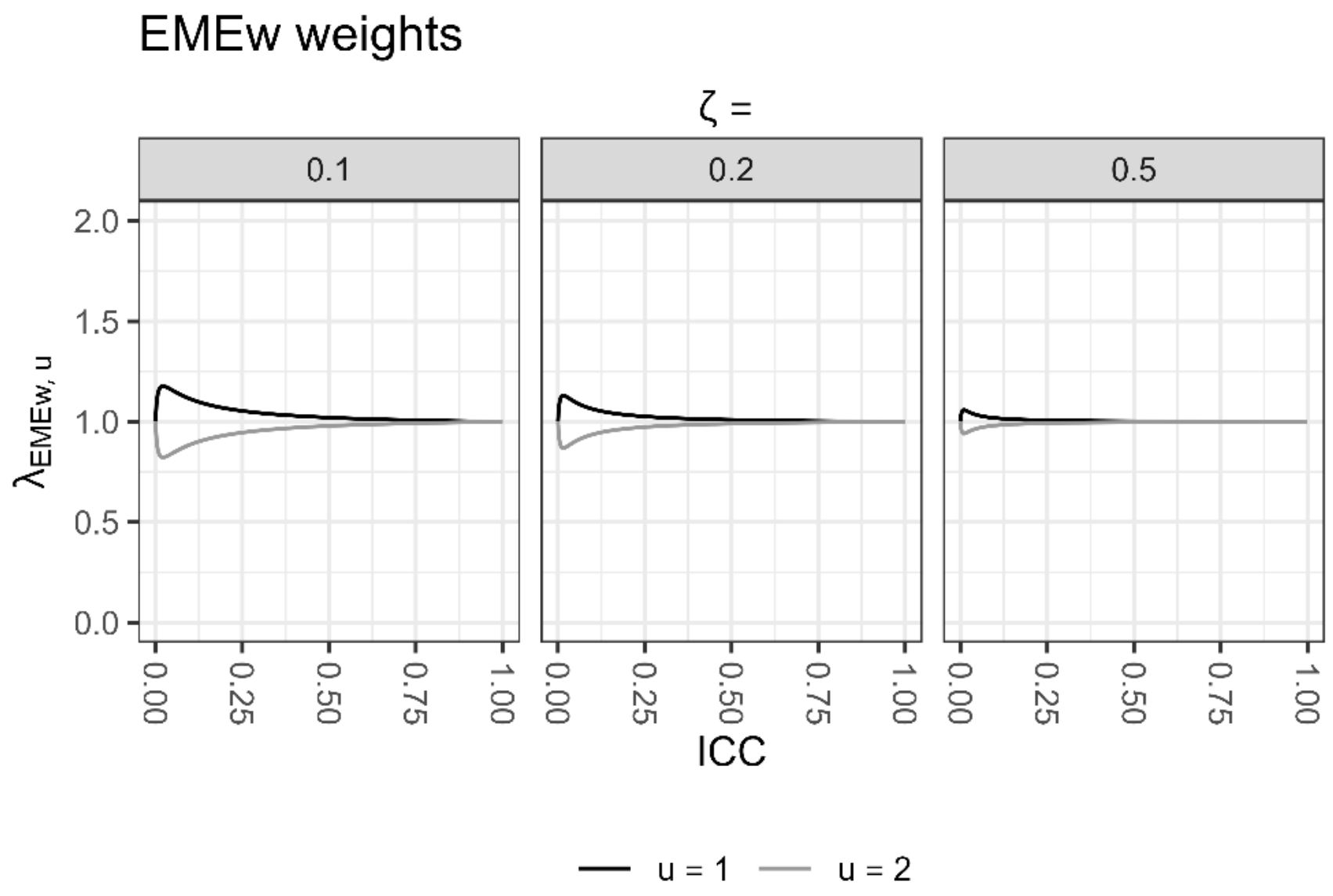

In Figure 3, we plot the values of the estimand weights from the NEMEw estimator $\lambda_{NEMEw,u}$ across different within-period ICC values $\rho_{wp} \in [0,1]$ with cluster auto-correlation $CAC = \frac{\rho_{bp}}{\rho_{wp}} = 0.8$, in scenarios with two subpopulations $u = 1,2$ with fixed subpopulation-specific cluster-period sizes $K_{i-,1}$ and $K_{i-,2}$ that differ by a ratio of $\zeta = \frac{K_{i-,1}}{K_{i-,2}}$. Again, these subpopulations have equal probabilities of being sampled from the overall population $P(u = 1) = 0.5$ and $P(u = 2) = 1 - P(u = 1)$.

Figure 3. NEMEw estimand weights $\lambda_{NEMEw,u}$ are plotted across two subpopulations $u = 1, 2$ and different values of within-period ICC and fixed cluster auto-correlation $CAC = 0.8$, with subpopulation-specific cluster-period sizes $K_{i-,1}$ and $K_{i-,2}$ that differ by a ratio of $\zeta = \frac{K_{i-,1}}{K_{i-,2}}$, and equal probabilities of being sampled $P(u = 1) = P(u = 2) = 0.5$.

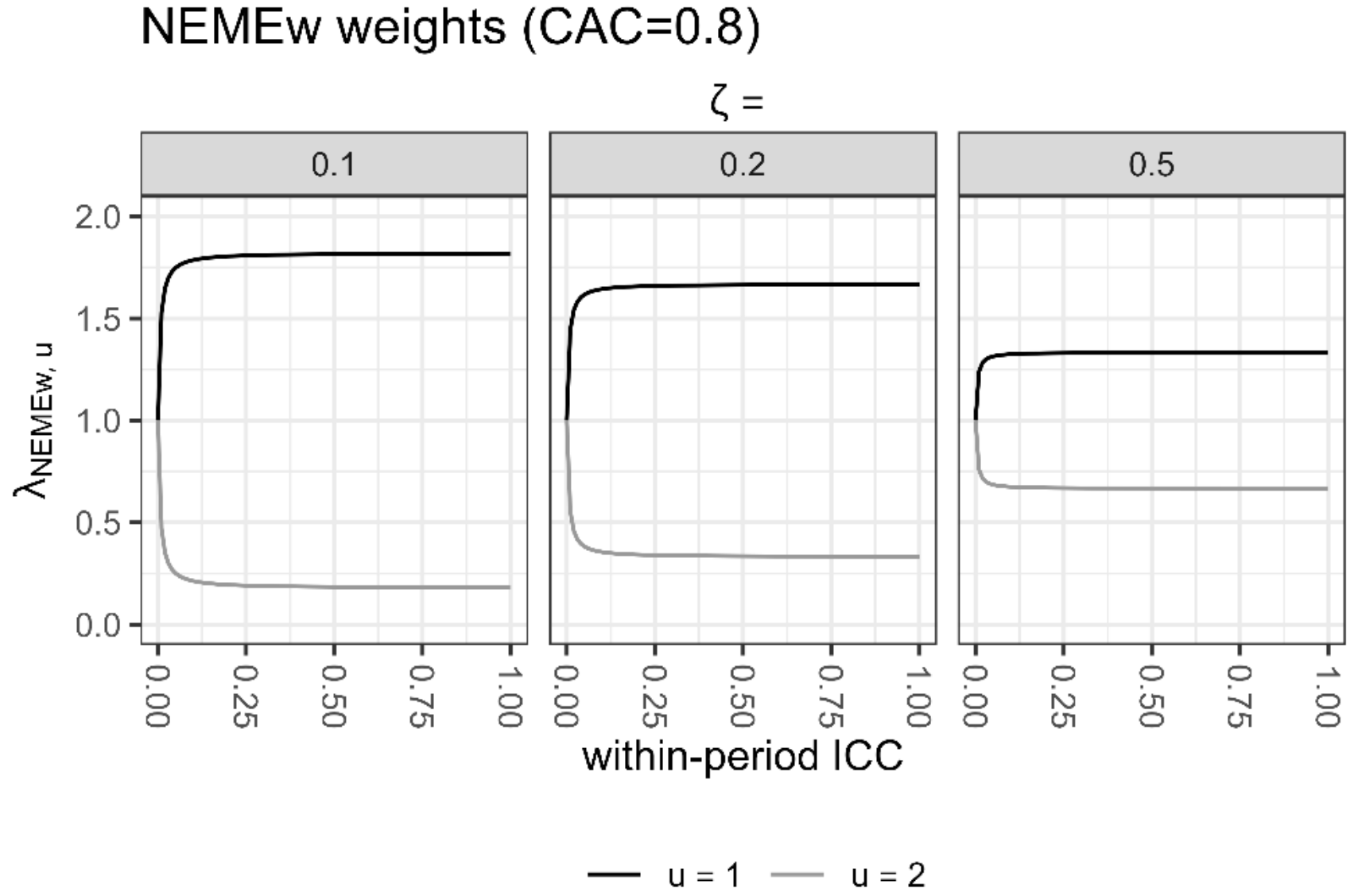

Overall, we observe that the estimand weights $\lambda_{EMEw}$ in the EMEw treatment effect estimator are slightly elevated in scenarios with low ICC, low cluster-period size ratios $\zeta$, and high sampling probabilities $P(u = 1)$ (Figure 2). Accordingly, the EMEw treatment effect estimates will be more biased for the cATE estimand in these scenarios. However, the weights are minimized for larger values of ICC (Figure 2) which would result in a less biased EMEw treatment effect estimate.

Notably, the maximum amount of difference between the estimand weights of the two subpopulations in the EMEw estimator $(\lambda_{EMEw,1} - \lambda_{EMEw,2})$ (Figure 2) is still considerably lower than that of the NEMEw estimator $(\lambda_{NEMEw,1} - \lambda_{NEMEw,2})$ (Figure 3). We observe that the estimand weights $\lambda_{NEMEw}$ in the NEMEw treatment effect estimator are in general severely unbalanced and remain so for nearly all values of within-period ICC (Figure 3). This indicates that the NEMEw treatment effect estimator will generally yield biased estimates for the cATE estimand across most scenarios.

We observe that the maximum difference between the EMEw estimand weights increases with larger values of $\zeta$. However, graphing up to an extreme value of $\zeta = 0.001$ with the optimal $P(u = 1)$ and $\rho$ to maximize the bias still yielded relatively balanced estimand weights with fairly small values for $\lambda_{EMEw,1} - \lambda_{EMEw,2}$ (Appendix A.3.3).

In the EMEw estimator, given two subpopulations each with fixed subpopulation-specific cluster-period sizes $K_{i-,1}$ and $K_{i-,2}$, we can prove that the maximum difference between the estimand weights, and accordingly the maximum amount of bias, occur when the ICC is equivalent to:

$$\rho = \frac{1}{1+\sqrt{2K_{i-,1}K_{i-,2}}}$$

as demonstrated in the Appendix (A.3.1). Subsequently, we also prove that the maximum amount of bias occurs when the probability of belonging to subpopulation $u = 1$ is equivalent to:

$$P(u = 1) = \frac{\sqrt{\lambda_{EMEw,2}}}{\sqrt{\lambda_{EMEw,1}} + \sqrt{\lambda_{EMEw,2}}}$$

which tends to be $\approx 0.5$, as demonstrated in the Appendix (A.3.2).

Overall, we demonstrate that in the analysis of a PB-CRT, the EMEw treatment effect estimator is more resistant to bias than the NEMEw treatment effect estimator in estimating the cATE estimand.

# 6. Simulation

Here, we simulated a scenario of PB-CRT data with continuous outcomes and homogeneous or heterogeneous treatment effects according to cluster size to empirically demonstrate the theoretical results derived in Section 4.

Potential outcomes were generated with the following data generating process (DGP):

$$Y_{i0k}(0) = \mu + \alpha_i + \gamma_{ij} + e_{ijk}$$

$$Y_{i1k}(0) = \mu + \Phi_1 + \alpha_i + \gamma_{ij} + e_{ijk}$$

$$Y_{i1k}(1) = \mu + X_{ij}\delta_u + \Phi_1 + \alpha_i + \gamma_{ij} + e_{ijk}$$

$$\alpha_i \overset{iid}{\sim} N(0, \tau_\alpha^2)$$

$$\gamma_{ij} \overset{iid}{\sim} N(0, \tau_\gamma^2)$$

$$e_{ijk} \overset{iid}{\sim} N(0, \sigma_w^2 = 1)$$

with $\delta_u$ being the $u$ subpopulation-specific heterogeneous treatment effect. We set $\mu = 1$ with a period effect $\Phi_1 = 0.2$. Cluster random intercepts and cluster-period random interaction terms are drawn from independent normal distributions with variances $\tau_\alpha^2 = 0.053$ and $\tau_\gamma^2 = 0.013$, respectively, to yield a within-period intracluster correlation coefficient of $\rho_{wp} = \frac{\tau_\alpha^2+\tau_\gamma^2}{\tau_\alpha^2+\tau_\gamma^2+\sigma_w^2} \approx 0.06$, between-period intracluster correlation coefficient of $\rho_{bp} = \frac{\tau_\alpha^2}{\tau_\alpha^2+\tau_\gamma^2+\sigma_w^2} \approx 0.05$, and a cluster auto-correlation of $CAC = \frac{\rho_{bp}}{\rho_{wp}} = \frac{\tau_\alpha^2}{\tau_\alpha^2+\tau_\gamma^2} = 0.8$.

For illustration, we simulated data from a 10 cluster, 2 period PB-CRT. Half of the clusters arose from subpopulation $u = 1$ with the other half from population $u = 2$ (accordingly, the cluster sampling probabilities are $P(u = 1) = P(u = 2) = 0.5$). Cluster sizes for subpopulations $u = 1$ and 2 were generated with $K_{i-,1} \sim Poisson(20)$ and $K_{i-,2} \sim Poisson(100)$, respectively. In our simulations, we fixed the cluster-period sizes to be the same between periods within clusters. Half of the clusters were randomized to receive the treatment in period $j = 1$, with the other half receiving the control throughout the trial.

In scenarios with noninformative cluster sizes, we simulated a homogeneous treatment effect $\delta_1 = \delta_2 = 0.35$. In scenarios with informative cluster sizes between subpopulations $u = 1$ and 2, we simulated heterogeneous treatment effects $\delta_1 = 0.2$ and $\delta_2 = 0.5$. Given the cluster-level treatment indicator $X_{ij}$, the observed outcome in follow-up period $j = 1$ was $Y_{i1k} = X_{i1}Y_{i1k}(1) + (1 - X_{i1})Y_{i1k}(0)$.

We analyzed the data with the IEE, FE, EME, NEME, and IEEw, FEw, EMEw, NEMEw models. With the described DGP, the true pATE estimand is:

$$pATE = E\left[\frac{1}{E[K_{i1}]}\sum_{k=1}^{K_{i1}}[Y_{i1k}(1) - Y_{i1k}(0)]\right]$$

$$= \frac{E\big[E[K_{i-}\delta_u|u]\big]}{E\big[E[K_{i-}|u]\big]} = \frac{P(u = 1)E[K_{i-}\delta_u|u = 1] + P(u = 2)E[K_{i-}\delta_u|u = 2]}{P(u = 1)E[K_{i-}|u = 1] + P(u = 2)E[K_{i-}|u = 2]}$$

$$= \left(\frac{0.5(20\delta_1) + 0.5(100\delta_2)}{0.5(20) + 0.5(100)}\right) = \frac{10\delta_1 + 50\delta_2}{60}$$

which is equal to 0.35 in scenarios with non-informative cluster sizes ($\delta_1 = \delta_2 = 0.35$) and 0.45 in scenarios with informative cluster sizes ($\delta_1 = 0.2$ and $\delta_2 = 0.5$). The true cATE estimand is then:

$$cATE = E\left[\frac{1}{K_{i1}}\sum_{k=1}^{K_{i1}}[Y_{i1k}(1) - Y_{i1k}(0)]\right]$$

$$= E\big[E[\delta_u|u]\big] = P(u = 1)E[\delta_u|u = 1] + P(u = 2)E[\delta_u|u = 2]$$

$$= 0.5(\delta_1) + 0.5(\delta_2) = 0.35$$

in simulation scenarios with non-informative cluster-sizes or informative cluster sizes.

For each scenario, we simulated 1000 PB-CRT data sets. Accordingly, we presented the results in terms of percent relative bias $\left(= \frac{\overline{\hat{\delta} - cATE}}{cATE} \times 100\right)$ and root mean square error $\left(RMSE = \sqrt{\overline{(\hat{\delta} - cATE)^2}}\right)$, with the bar denoting the average over the 1000 simulated data sets. Furthermore, we explored the efficiency $\left(= \overline{Var(\hat{\delta})}\right)$ of model-based and the leave-one-cluster-out jackknife variance estimators alongside the Monte Carlo variance (the variance of the 1000 simulated point estimates), and used the model-based and jackknife variance estimators to calculate the coverage probability (CP) of the 95% confidence interval and the power.

The weighted mixed-effects models were run using the *WeMix* package in R. The leave-one-cluster-out jackknife variance estimator was manually programmed in R version 4.3.2 following the description by Bell & McCaffrey (Bell & McCaffrey, 2002). A similar jackknife variance estimator has been previously demonstrated to yield robust inference with mixed-effects model misspecification in SW-CRTs (Ouyang et al., 2024). We additionally evaluated the bias-reduced linearization method described by Pustejovsky & Tipton (Pustejovsky & Tipton, 2018) and recommended for inference with two-way fixed-effects models. The bias-reduced linearization method was easily implemented with the *clubSandwich* package in R for the IEE, IEEw, FE, and FEw models. However, *clubSandwich* is not compatible with *WeMix*. Therefore, the bias-reduced-linearization method was not used with the weighted mixed-effects models; its results are reported in the Appendix (A.4.1).

## 6.1 Simulation results with non-informative cluster sizes

The simulation results in scenarios with homogeneous treatment effects (non-informative cluster sizes) are shown below. With such homogeneous treatment effects, the pATE and cATE estimands coincide. As expected, all the unweighted and weighted treatment effect estimators were unbiased for the true average treatment effect estimand (Figure 4). Comparing the weighted models against their unweighted counterparts, we observe that the IEEw model can yield lower RMSE than the IEE model, whereas the NEMEw model can yield higher RMSE than the NEME model (Figure 4).

Figure 4. Simulation bias and RMSE results in scenarios with homogeneous treatment effects (non-informative cluster sizes). Dashed lines show a relative bias (%) of 5% and -5%.

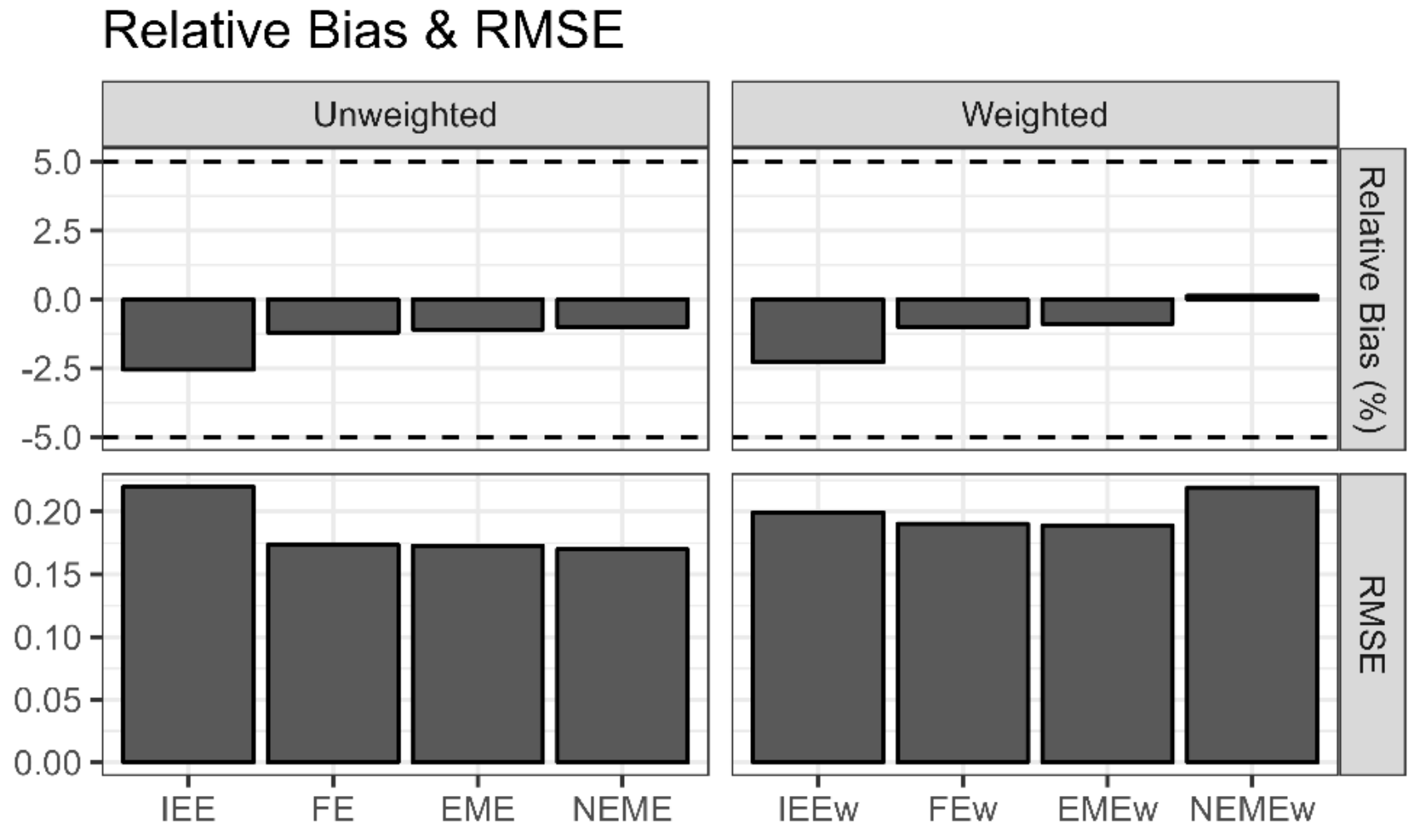

The averages of the model-based and leave-one-cluster-out jackknife variance estimates are plotted in Figure 5, alongside the Monte Carlo variance. The Monte Carlo variance has also been commonly referred to as the “empirical,” “observed,” or “sampling” variances of the point estimates over the simulation replicates (Morris et al., 2019). The model-based and jackknife variance estimators explicitly target the Monte Carlo variance, with systematic deviations representing a bias in the estimation of the variance (Morris et al., 2019). Comparing the variance estimates in Figure 5 where the true underlying DGP has a nested exchangeable correlation structure, we observe that the jackknife variance estimates roughly approximate the Monte Carlo variances, even in scenarios where the correlation structure is misspecified.

Figure 5. The efficiency of the different unweighted and weighted models as captured by the average of the model-based and leave-one-cluster jackknife variance estimates over the 1000 simulation replicates, graphed alongside the corresponding Monte Carlo variances.

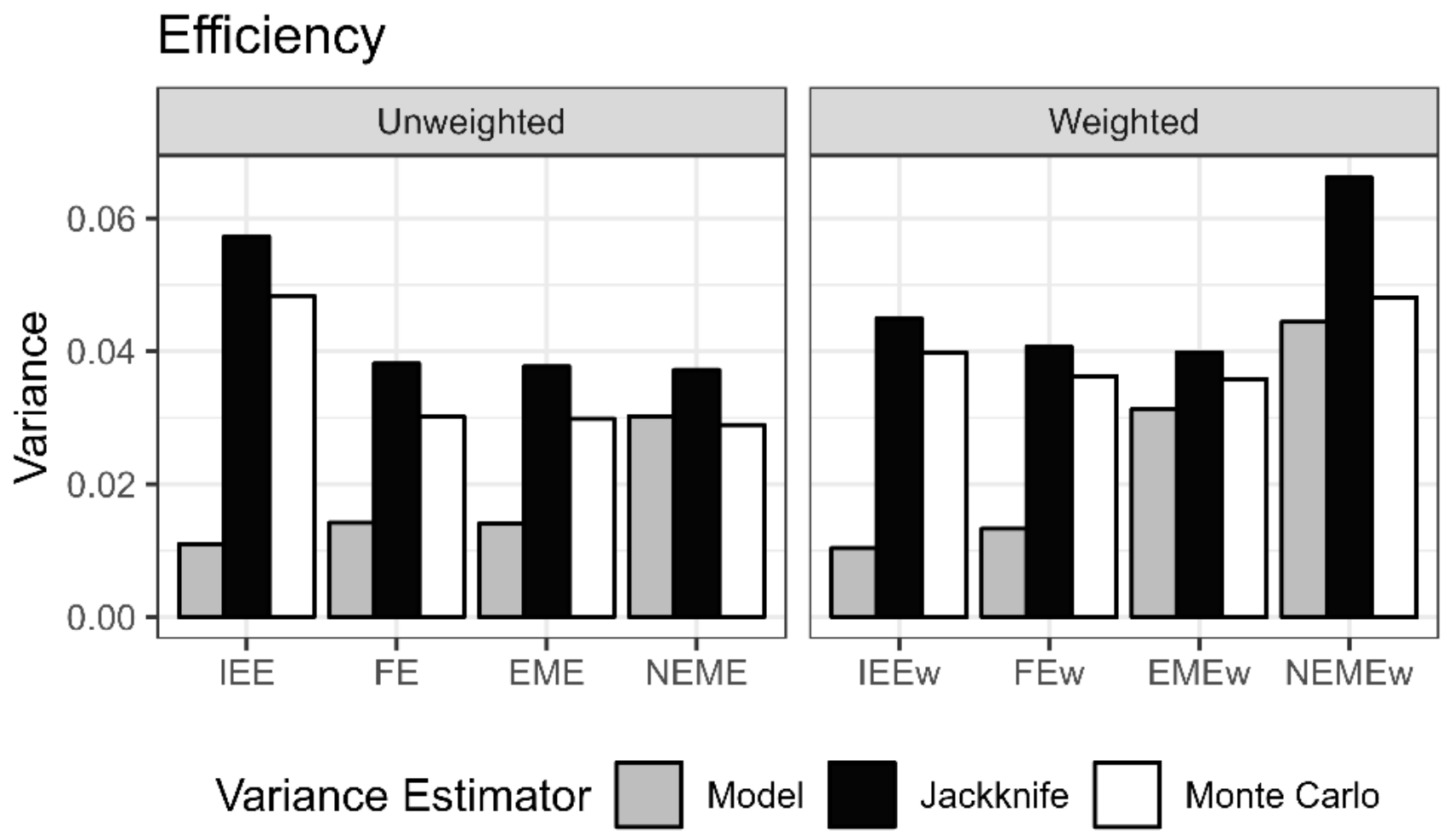

Across the unweighted analyses in our simulations, the IEE model had the largest jackknife variance estimates and Monte Carlo variances. In contrast, the FE, EME, and NEME models notably all had smaller and similar jackknife variance estimates and Monte Carlo variances. Based on these empirical observations of the Monte Carlo variances, the IEE treatment effect estimator is the least efficient of the described estimators. This occurs because the IEE estimator only makes "vertical" comparisons within periods (as described in Section 3), whereas the other described estimators use information both within and between periods.

Across the weighted analyses in our simulations, the NEMEw model has much larger model-based and jackknife variance estimates than the corresponding estimates in the other weighted analyses. Furthermore, the NEMEw model had the largest Monte Carlo variances, that were poorly approximated by the jackknife variance estimators. In contrast, the IEEw, FEw, and EMEw models had lower Monte Carlo variances, which were more closely approximated by the jackknife variance estimators. Overall, the FEw and EMEw models had the lowest Monte Carlo and jackknife variances across the weighted analyses.

We compared the weighted models against their unweighted counterparts and observe that modelling with inverse cluster-period size weights may or may not lead to worse efficiency (Figure 5). The NEMEw model had very inflated model-based, jackknife, and Monte Carlo variances in comparison to the unweighted NEME model (Figure 5). Conversely, the IEEw model yielded comparatively lower jackknife and Monte Carlo variances than the unweighted IEE model. In contrast, including inverse cluster-period size weights in the FEw and EMEw models had little effect on the jackknife variance estimates and Monte Carlo variances compared to the corresponding unweighted models (FE and EME models) (Figure 5). Overall, the FE and EME models are empirically comparable in efficiency, as are the FEw and EMEw models.

With the true underlying DGP having a nested exchangeable correlation structure, all analyses, regardless of their specified correlation structure, had proper coverage of the 95% confidence intervals with the jackknife variance estimator (Figure 6). This corresponds with previous work that demonstrated the jackknife variance estimator can yield robust inference with correlation structure misspecification in SW-CRTs (Ouyang et al., 2024). Subsequently, the power across the different analyses are only slightly reduced when using the jackknife variance estimator compared to the power using the model-based variance estimator in the correctly specified NEME and NEMEw models.

Figure 6. The coverage probability of the 95% confidence interval and the power of the different unweighted and weighted models using the model-based and leave-one-cluster jackknife variance estimators in scenarios with a homogeneous treatment effect (non-informative cluster sizes). Dashed lines show a coverage probability of 0.95.

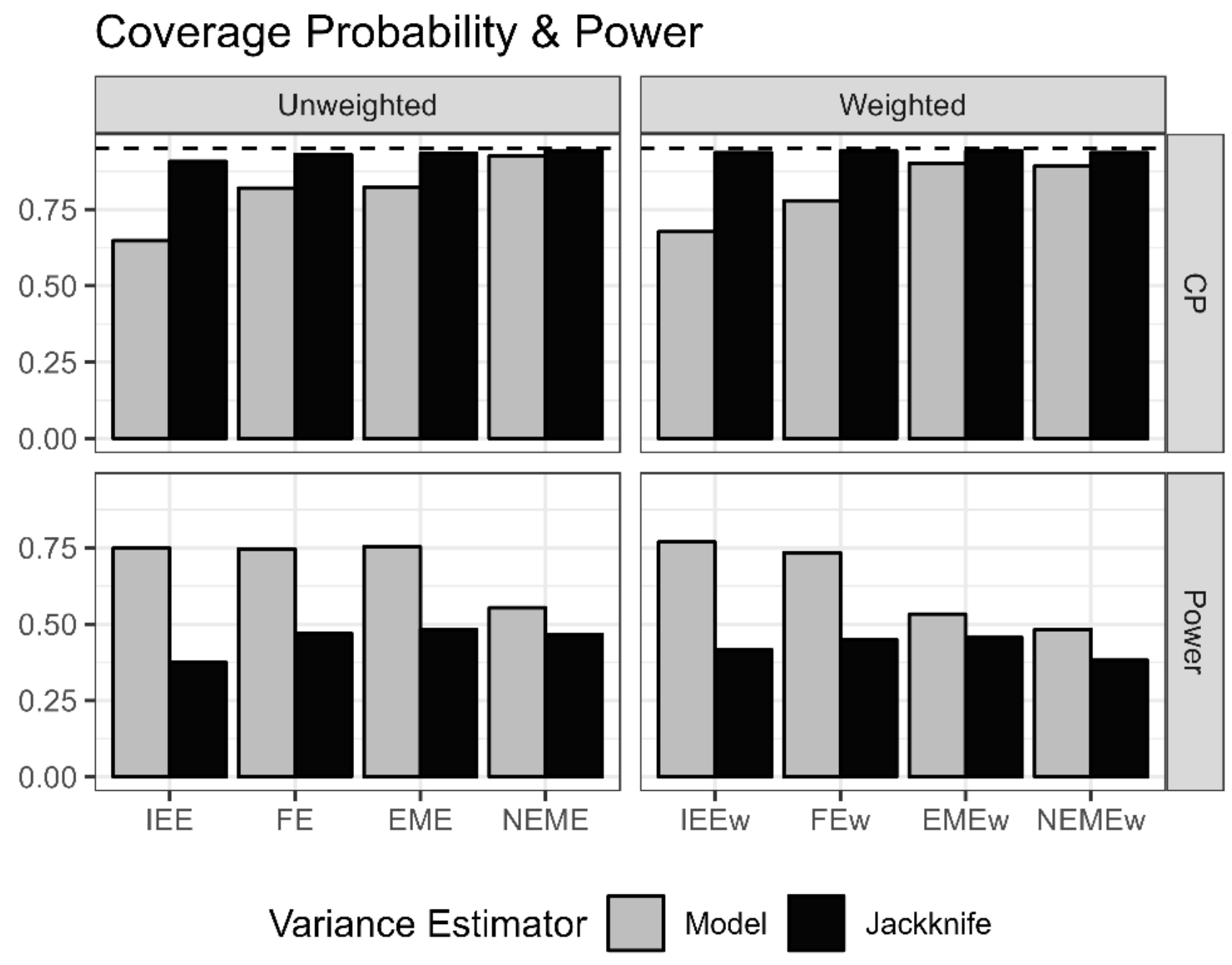

Overall, we recommend using the leave-one-cluster-out jackknife variance estimator for inference due to its robustness despite misspecification of the true correlation structure. We empirically observe that the jackknife variance estimates roughly approximate the Monte Carlo variances. Furthermore, it's simple to manually program, which makes it easily compatible with different modelling R packages, including *WeMix*. The bias-reduced linearization variance estimator yielded similar results to the jackknife variance estimator, but with slightly smaller variance estimators, resulting in slight under-coverage of the 95% confidence interval and slightly better power (Appendix A.4.1). When using the FEw model, the bias-reduced linearization variances also performs well in terms of coverage (Appendix A.4.1).

## 6.2 Simulation results with informative cluster sizes

The simulation results in scenarios with heterogeneous outcomes (informative cluster sizes) are shown below. As expected, the IEE and FE treatment effect estimators were unbiased for the true pATE estimand, and the IEEw and FEw treatment effect estimators were unbiased for the true cATE estimand (Figure 7). Surprisingly, the EME and EMEw treatment effect estimators were also empirically unbiased across our simulation replicates for the pATE and cATE, respectively (Figure 7). This was discussed earlier in Section 5. This is in contrast to the results presented by Wang et al. for a corresponding estimator in P-CRTs (X. Wang et al., 2022). In contrast, the NEME and NEMEw treatment effect estimators were biased (accepting over 10% relative bias) for the pATE and cATE, respectively, in scenarios with informative cluster sizes (Figure 7).

Figure 7. Simulation bias results for the pATE and cATE estimands in scenarios with heterogeneous treatment effects (informative cluster sizes). Dashed lines show a relative bias (%) of 5% and -5%.

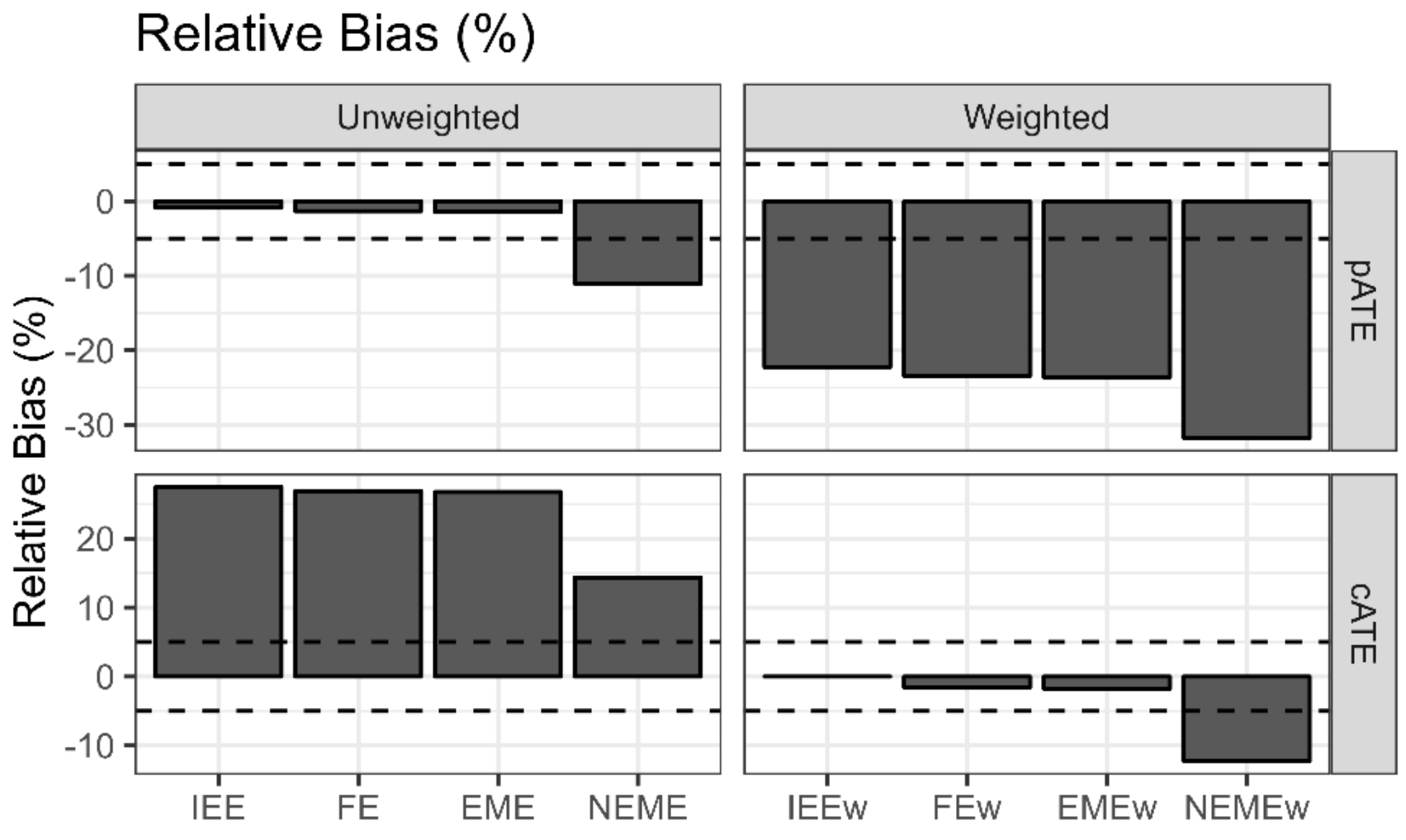

Subsequent inference using the leave-one-cluster-out jackknife variance estimator with the weighted analyses produced good coverage probability of the 95% confidence interval for unbiased estimators of the cATE (IEEw, FEw, and EMEw models), but had slight under-coverage for unbiased estimators of the pATE (IEE, FE, EME models) (Figure 8) across our simulation scenarios (Figure 8). As expected, this improved coverage probability with the jackknife variance estimator came at the cost of power (Figure 9).

Figure 8. The coverage probability of the 95% confidence interval of the different unweighted and weighted models using the model-based and leave-one-cluster jackknife variance estimators in scenarios with a heterogeneous treatment effect (informative cluster sizes). Dashed lines show a coverage probability of 0.95.

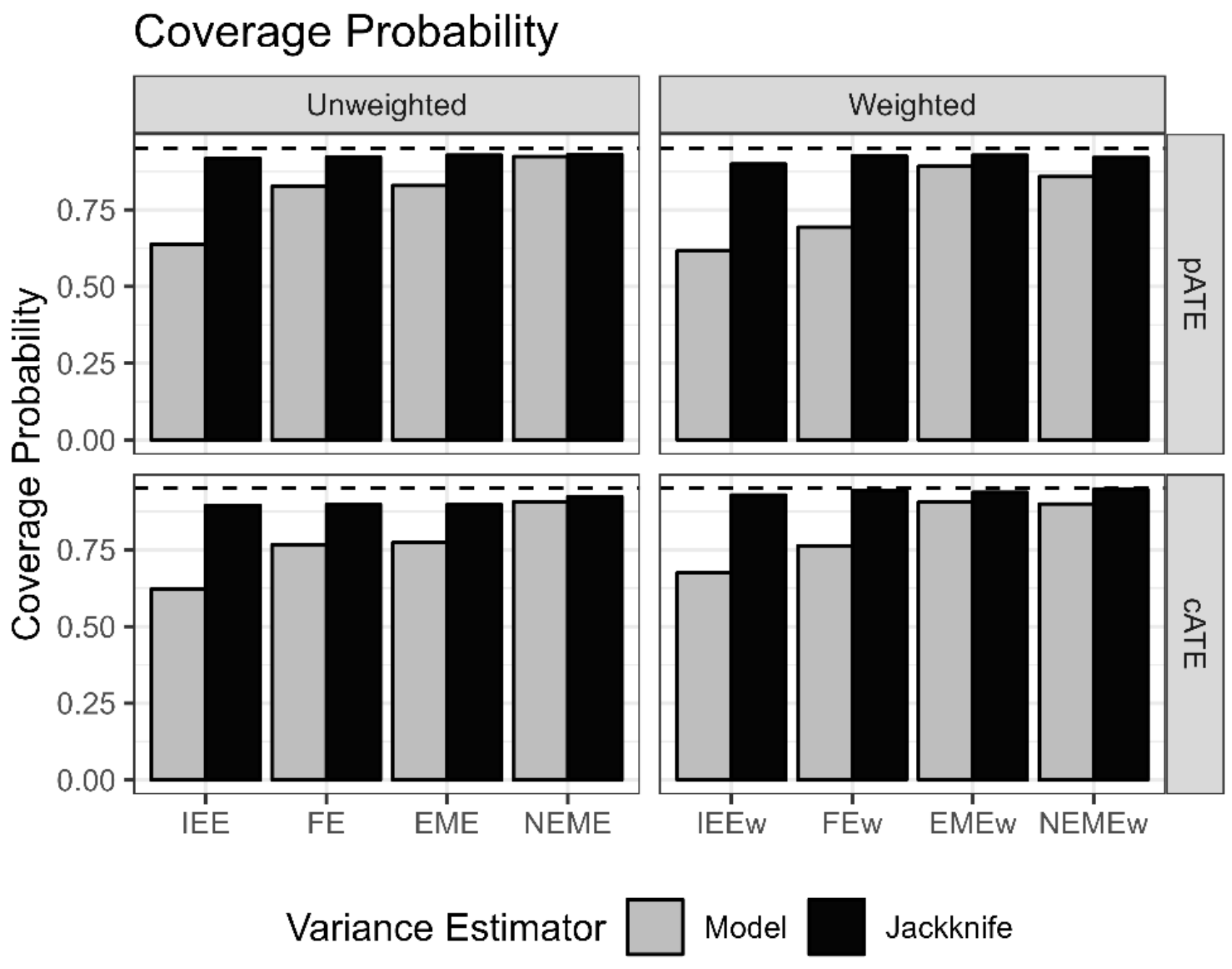

Figure 9. The power of the different unweighted and weighted models using the model-based and leave-one-cluster jackknife variance estimators in scenarios with a heterogeneous treatment effect (informative cluster sizes).

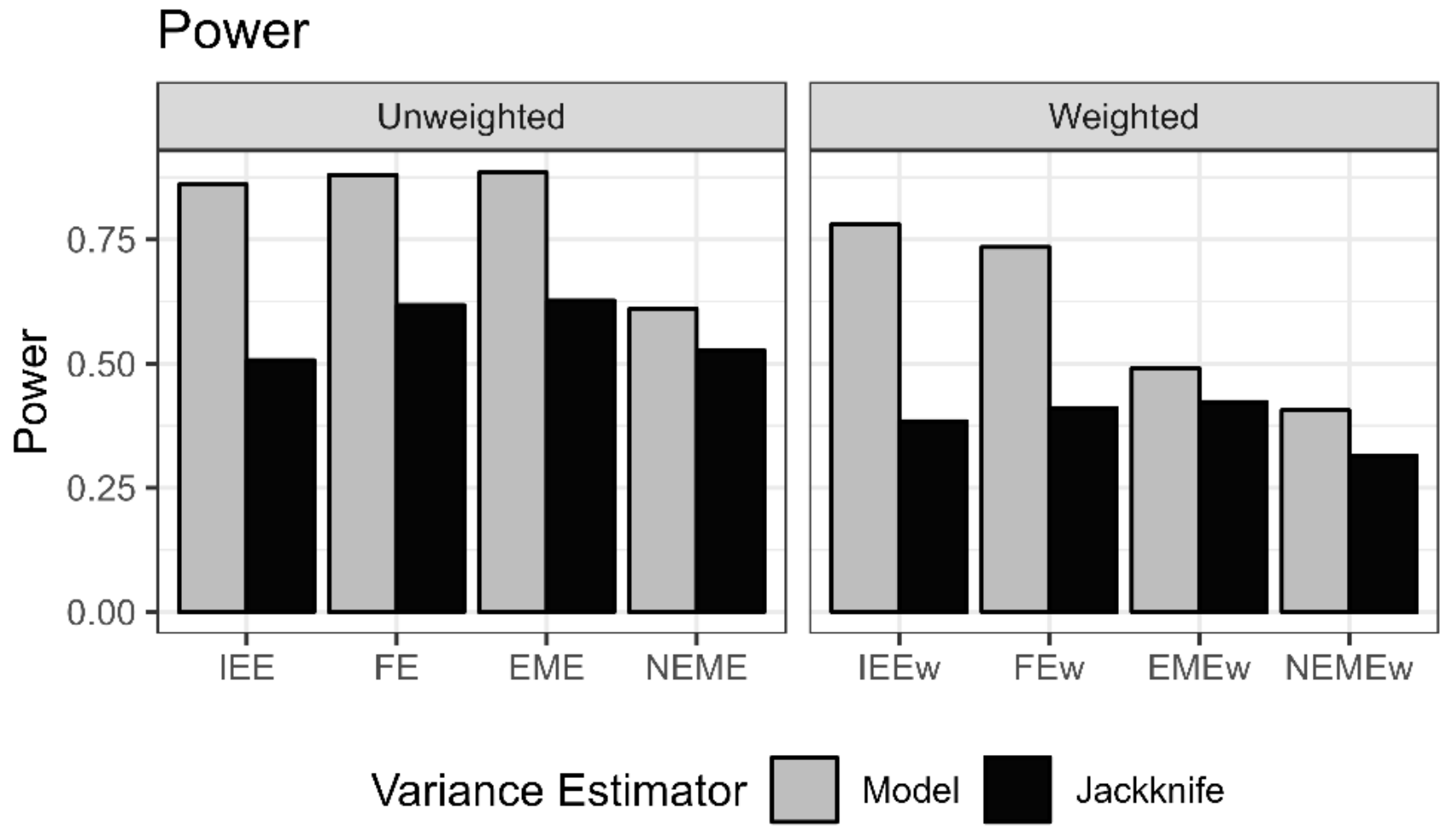

In the Appendix (A.4.2), we included bias results from two additional simulation scenarios with informative cluster sizes, testing the amount of bias yielded by the EME and EMEw estimators for the pATE and cATE estimands. Simulation parameters were set to maximize bias in the EMEw estimator, with small values of $\zeta$ and the ICC set to the corresponding value $\rho$ to optimize the bias in the EMEw estimator for the cATE estimand (as previously discussed in Section 5). Despite these scenarios being unrealistically tailored to yield biased EMEw estimates for the cATE estimand, the EMEw estimator still remained relatively unbiased. Overall, we observe that the EMEw estimator generally yields minimal bias for the cATE estimand across many scenarios, despite not being a theoretically consistent estimator for the cATE in PB-CRTs with informative cluster sizes.

# 7. Case Study

We also reanalyzed publicly available PB-CRT case study data from the JIAH trial, exploring the effects of community youth teams on a variety of outcomes among adolescent girls in rural eastern India (Bhatia et al., 2023). The JIAH trial had 28 clusters with observations from both periods of the trial. We re-analyzed the data collected from those clusters, with reported mental health scores as the continuous outcome of interest.

The cluster-period cell sizes in this trial varied greatly between periods within clusters (Appendix A.5). For weighted analyses, we used the inverse cluster-period cell size weights to generate the IEEw and FEw treatment effect estimators, as described in Section 4. To our knowledge, the extension of inverse cluster-period size weighting to analyses with variable cluster-period size and correlated errors (EMEw and NEMEw) is not as clear, and deserves additional future work. Furthermore, it is not even desirable to use the NEMEw estimator as seen in Section 6.

Figure 10. Estimates and the corresponding 95% confidence intervals as estimated using the model-based and leave-one-cluster out jackknife variance estimators from the different unweighted and weighted models in the re-analysis of the JIAH trial.

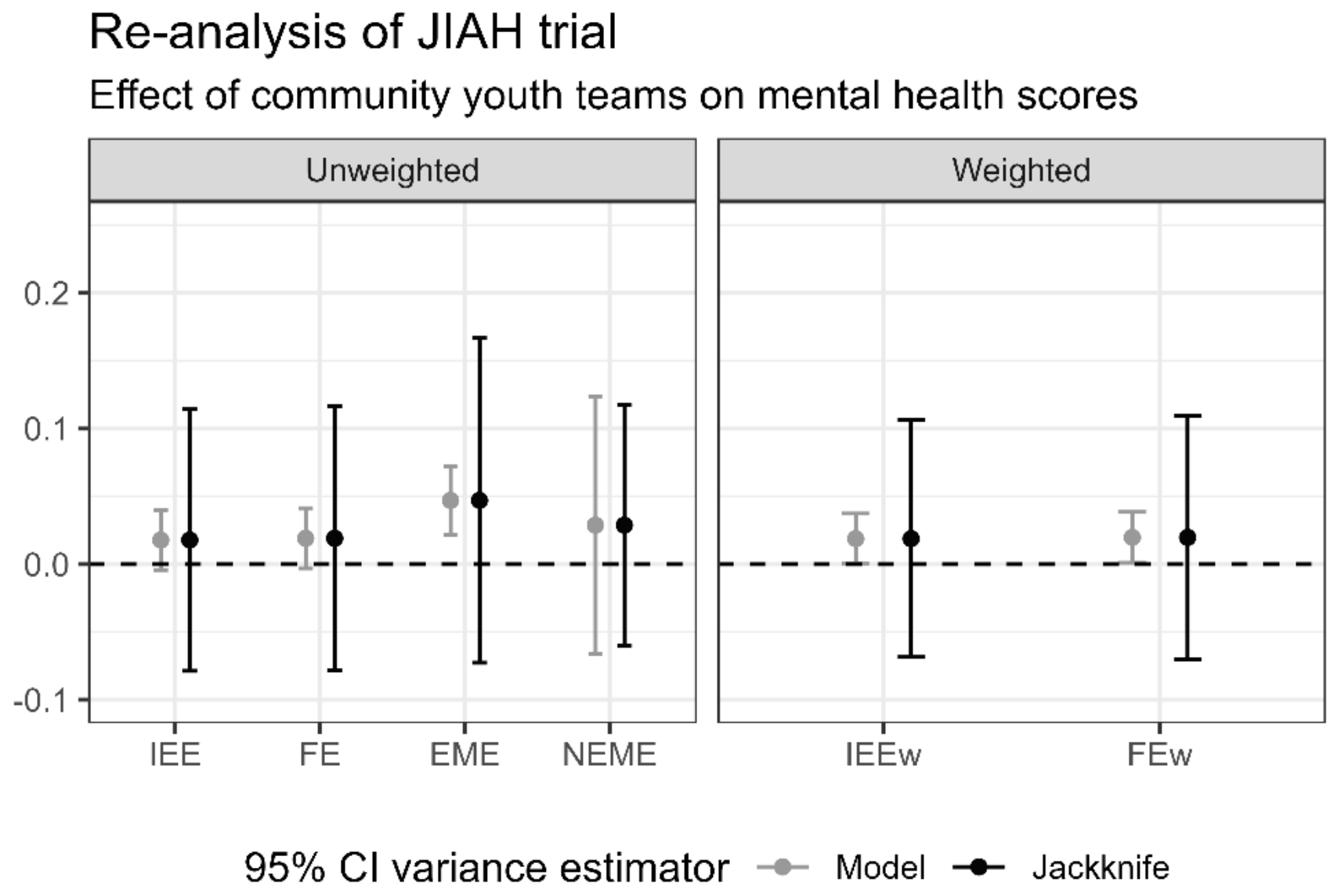

The JIAH case study estimates with corresponding 95% confidence intervals, as generated with the model-based and leave-one-cluster-out jackknife variance estimators, are presented in Figure 10. As we previously observed in our simulation results (Figure 6), the model-based variances can yield overly narrow confidence intervals which may overstate the significance of the results. Here, we similarly observe that the model-based variances are considerably smaller than the jackknife variance estimates in all models except for the NEME model. In contrast, the jackknife variance estimates across the different models roughly approximate the model-based variance estimates from the NEME model. Accordingly, we suggest using the jackknife variance estimator for inference.

The results between the weighted and unweighted analyses are qualitatively similar, which may indicate an equivalence between the pATE and cATE estimands and imply this trial has non-informative cluster sizes and a homogeneous treatment effect on mental health scores. However, the subtle difference between the IEE and FE estimates as compared to the EME and NEME estimates for the pATE estimand may suggest the opposite, as the latter two estimators are not theoretically consistent for the pATE estimand. Based on this re-analysis, we do not make a judgement on whether informative cluster sizes were present in the JIAH trial.

# 8. Discussion

Under the potential outcomes framework, we focus on two natural estimands of interest in the context of PB-CRTs with informative cluster sizes, the participant-average treatment effect (pATE) and cluster-average treatment effect (cATE). Notably, in scenarios with non-informative cluster sizes, the pATE and cATE will be equivalent and all of the described treatment effect estimators are expected to be consistent and unbiased. With informative cluster sizes, we demonstrate that the independence estimating equation (IEE) and fixed effects (FE) model point estimators are theoretically consistent estimators of the pATE estimand. However, unlike the IEE estimator, the FE estimator is only consistent for the pATE estimand in the presence of informative cluster sizes when cluster-period sizes do not vary between periods within clusters. Additionally, these two estimators with inverse cluster-period size weights, the IEEw and FEw model point estimators, are theoretically consistent estimators of the cATE estimand. We then used a simulation study with informative cluster sizes to compare the performance of the different estimators in terms of bias. As expected, the IEEw and FEw point estimators yielded empirically unbiased estimates for the cATE estimand.

Across our simulation replicates, we empirically observe through the Monte Carlo variances that the FE and FEw treatment effect estimators were more efficient than the IEE and IEEw treatment effect estimators. This observation was not previously discussed in the context of CRTs, and highlights the potential for FE and FEw estimators in estimand-driven analysis of PB-CRTs. Despite misspecification of the correlation structure in these analyses, with the true underlying simulation DGP having a nested exchangeable correlation structure, the jackknife variance estimator roughly approximated the Monte Carlo variances and yielded proper coverage of the 95% confidence interval.

We observe that the exchangeable (EME) and nested-exchangeable (NEME) mixed effects models, and their inverse cluster-period size weighted counterparts (EMEw and NEMEw) are not theoretically consistent estimators for the pATE and cATE estimands, respectively. Instead, these unweighted and weighted treatment effect estimators converge to estimands that are difficult to interpret, with cluster-specific weights that depend on both the cluster-period size as well as the probability limit of the ICC. Therefore, in general, the ICC will dictate the estimands and therefore the analysis of a different outcome may inadvertently change the estimand, which can be undesirable  However, we surprisingly demonstrate that unlike previously reported results in a P-CRT (X. Wang et al., 2022), the unweighted and weighted treatment effect estimators with an exchangeable correlation structure (EME and EMEw) typically yield minimally biased estimates of the pATE and cATE estimands in a PB-CRT. We provide in Table *2* a concise summary of the potential advantages and disadvantages of the different models explored in this paper.

Table 2. Potential advantages and disadvantages of several commonly used models for estimating the population average treatment effect (pATE) and cluster average treatment effect (cATE) estimands in PB-CRTs when informative cluster sizes (ICS) are present.

| Target | Methods | Potential advantages | Potential disadvantages |
|---|---|---|---|
| pATE | IEE | • Theoretically consistent under ICS. | • Only uses information from the follow-up period and may be inefficient.<br>• Doesn't automatically estimate ICC. |
| | FE | • Theoretically consistent under ICS when cluster-period sizes don't vary.<br>• Uses information from both baseline and follow-up periods which improves efficiency over IEE. | • Not theoretically consistent under ICS when cluster-period sizes vary.<br>• Doesn't automatically estimate ICC. |
| | EME | • Minimally biased under ICS.<br>• Automatically estimates the ICC. | • Not theoretically consistent under ICS.<br>• Doesn't automatically estimate wp-ICC and bp-ICC. |
| | NEME | • Automatically estimates wp-ICC and bp-ICC. | • Not theoretically consistent and potentially biased under ICS. |
| cATE | IEEw | • Theoretically consistent under ICS.<br>• Weights are easy to program in standard statistical software, even when cluster-period sizes vary. | • Only uses information from the follow-up period and may be inefficient.<br>• Doesn't automatically estimate ICC. |
| | FEw | • Theoretically consistent under ICS.<br>• Weights are easy to program in standard statistical software, even when cluster-period sizes vary.<br>• Uses information from both baseline and follow-up periods which improves efficiency over IEEw. | • Doesn't automatically estimate ICC. |
| | EMEw | • Minimally biased under ICS.<br>• Automatically estimates ICC. | • Not theoretically consistent under ICS.<br>• More complicated to program in standard statistical software (requires *WeMix* in R).<br>• Unclear how to specify the weights when cluster-period sizes vary. |
| | NEMEw | • Automatically estimates wp-ICC and bp-ICC. | • Not theoretically consistent and potentially biased under ICS.<br>• More complicated to program in standard statistical software (requires *WeMix* in R).<br>• Unclear how to specify the weights when cluster-period sizes vary. |

In many of our derivations for the convergence probability limit of the treatment effect estimators, we assume that cluster-period sizes are equivalent between periods within clusters. However, it is realistic for multi-period CRT studies to have variable cluster-period cell sizes (although on many occasions the within-cluster variation in cluster-period size may not be as substantial compared to between-cluster variation). It is straightforward to specify variable inverse cluster-period size weights in analyses with an independence correlation structure, as is the case with the IEEw and FEw estimators which we have demonstrated are theoretically consistent estimators for the cATE estimand. In R, such weights can be easily be specified with the "weights" option in lm(). However, to our knowledge, the extension of inverse cluster-period size weighting to analyses with variable cluster-period size and correlated errors (EMEw and NEMEw) is relatively unclear. Notably, specifying such weights deviate from the weighted estimating equation described by Williamson et al. (Williamson et al., 2003), which specifies weights on the cluster-level rather than the cluster-period-level. Nonetheless, the FEw estimator has similar properties to the EMEw estimator, is a theoretically consistent estimator for the cATE, and can be easily implemented in most statistical software. In general, if cluster-period sizes vary between periods within clusters, the IEEw and FEw estimators can be used instead to estimate the cATE in the analysis of PB-CRTs with informative cluster sizes.

Many of the models explored in this paper have been previously described for the analysis of PB-CRTs by Hooper et al. (Hooper et al., 2018). Specifically, the FE, EME, and NEME estimators have been referred to as the "difference of differences," "constrained baseline analysis with a less flexible correlation structure," and "constrained baseline analysis with a realistic correlation structure," respectively (Hooper et al., 2018). Under the assumption of non-informative cluster sizes, Hooper et al. previously discouraged the use of the FE model, citing that it can fail to achieve the statistical power that was expected when using the model-based variance estimator (Hooper et al., 2018). However, mixed-effects models can also have an increased risk of false positives when using the model-based variance estimator, as pointed out in the same paper (Hooper et al., 2018). Regardless, we suggest using the jackknife variance estimator over the model-based variance estimator due to its ability to maintain valid inference even under informative cluster size and/or with misspecification of the correlation structure.

Hooper et al. (Hooper et al., 2018) also describe and recommend an "ANCOVA" analysis, which only analyzes the follow-up period (in this article, $J = 1$) with an exchangeable correlation structure, and includes additional covariate-adjustment for the cluster-specific average baseline outcome (in this article, $\bar{Y}_{i0} = \sum_k Y_{i0k}/K_{i0}$). Notably, this analysis is equivalent to a covariate-adjusted EME analysis of a P-CRT, and is always robust under arbitrary misspecification of the analysis, assuming non-informative cluster sizes (B. Wang et al., 2023). Accordingly, the point estimator of such an "ANCOVA" analysis is equivalent to the expected EME estimator for a P-CRT, which as described by Wang et al. (referred to by them as the EEE estimator) will be vulnerable to accepting additional bias in weighted analyses in the presence of informative cluster sizes (X. Wang et al., 2022).

## 8.1 Conclusions

Previous work by Wang et al. demonstrated that out of their explored estimators, only their corresponding IEEw estimator is theoretically consistent for estimating the cATE estimand in a P-CRT design (X. Wang et al., 2022). Furthermore, they demonstrate that the commonly used exchangeable mixed-effects model with inverse cluster-size weights (corresponding to the EMEw) can yield unacceptably biased results in the presence of informative cluster sizes (X. Wang et al., 2022).

In extending their work to a PB-CRT design, we demonstrate that in addition to the IEE and IEEw estimators, the FE and FEw estimators are also theoretically consistent for estimating the pATE and cATE estimands, respectively. Additionally, the FE and FEw estimators are more efficient than the IEE and IEEw estimators in our empirical observations and are thus recommended for practice. This is especially true for the FEw estimator, which is theoretically consistent for the cATE estimand, regardless of whether cluster-period sizes vary between periods within clusters. In contrast, the FE estimator is not theoretically consistent for the pATE estimand when cluster-period sizes vary between periods within clusters. In such scenarios, the IEE may be more appropriate. Furthermore, we observe that despite being theoretically inconsistent for the pATE and cATE estimands, the EME and EMEw estimators can have surprisingly minimal bias even in the presence of informative cluster sizes.

Overall, the IEE, FE, and EME treatment effect estimators (and their weighted counterparts) are generally trustworthy estimators for the pATE (and cATE) in PB-CRTs with informative cluster sizes. In contrast, the NEME and NEMEw estimators can yield unacceptably biased estimates for the pATE and cATE in the presence of informative cluster sizes. It is likely that these conclusions would extend to other multi-period CRT designs with informative cluster-period cell sizes, including the popular stepped-wedge cluster randomized trials (SW-CRT). This will be pursued in our future work.

**Disclosure statement**

The authors declare no potential conflicts of interest with respect to the research, authorship, and/or publication of this article.

**Funding**

Research in this article was partially supported by a Patient-Centered Outcomes Research Institute Award (PCORI Award ME-2022C2-27676).

**Disclaimer**

The statements presented in this article are solely the responsibility of the authors and do not necessarily represent the official views of PCORI, its Board of Governors, or the Methodology Committee.

**ORCID**

Kenneth Menglin Lee https://orcid.org/0000-0002-0454-4537

Fan Li https://orcid.org/0000-0001-6183-1893

**Data availability statement**

Data sharing is not applicable to this article as no new data were created or analyzed in this study. R codes for the simulations will be deposited to figshare by Wiley.

# A Appendix

## A.1 Estimand derivations

Here, we derive the participant-Average Treatment Effect (pATE) and cluster-Average Treatment Effect (cATE) estimands in a PB-CRT under an infinite superpopulation framework, where $I \rightarrow \infty$.

The pATE under a finite population framework is $pATE = \frac{\sum_{i=1}^{I}\sum_{k=1}^{K_{i1}}(Y_{i1k}(1)-Y_{i1k}(0))}{\sum_{i=1}^{I}K_{i1}}$ (Kahan, Blette, et al., 2023; X. Wang et al., 2022). Accordingly, the pATE estimand under an infinite superpopulation, where $I \rightarrow \infty$, is:

$$lim_{I\rightarrow\infty} pATE = lim_{I\rightarrow\infty} \frac{\sum_{i=1}^{I}\sum_{k=1}^{K_{i1}}\big(Y_{i1k}(1) - Y_{i1k}(0)\big)}{\sum_{i=1}^{I}K_{i1}}$$

$$= \frac{lim_{I\rightarrow\infty}\sum_{i=1}^{I}\sum_{k=1}^{K_{i1}}\big(Y_{i1k}(1) - Y_{i1k}(0)\big)}{lim_{I\rightarrow\infty}\sum_{i=1}^{I}K_{i1}}$$

And:

$$pATE \overset{P}{\rightarrow} \frac{E\big[\sum_{k=1}^{K_{i1}}\big(Y_{i1k}(1) - Y_{i1k}(0)\big)\big]}{E[K_{i1}]}$$

$$= E\left[\frac{1}{E[K_{i1}]}\sum_{k=1}^{K_{i1}}[Y_{i1k}(1) - Y_{i1k}(0)]\right].$$

Crucially, if $K_{i1}$ is independent of potential outcomes $\big(Y_{i1k}(1), Y_{i1k}(0)\big)$, as is the case with non-informative cluster sizes, then we can further demonstrate that:

$$\frac{E\big[\sum_{k=1}^{K_{i1}}\big(Y_{i1k}(1) - Y_{i1k}(0)\big)\big]}{E[K_{i1}]} = \frac{E\Big[E\big[\sum_{k=1}^{K_{i1}}\big(Y_{i1k}(1) - Y_{i1k}(0)\big)\,|K_{ij}\big]\Big]}{E[K_{i1}]}$$

$$= \frac{E\big[\sum_{k=1}^{K_{i1}}E\big[Y_{i1k}(1) - Y_{i1k}(0)|K_{ij}\big]\big]}{E[K_{i1}]} = \frac{E\Big[K_{i1}E\big[Y_{i1k}(1) - Y_{i1k}(0)|K_{ij}\big]\Big]}{E[K_{i1}]}$$

$$= \frac{E\big[K_{i1}E[Y_{i1k}(1) - Y_{i1k}(0)]\big]}{E[K_{i1}]} = \frac{E[K_{i1}]E[Y_{i1k}(1) - Y_{i1k}(0)]}{E[K_{i1}]}$$

$$= E[Y_{i1k}(1) - Y_{i1k}(0)].$$

The cATE under a finite population framework is $cATE = \frac{1}{I}\sum_{i=1}^{I}\frac{\sum_{k=1}^{K_{i1}}(Y_{i1k}(1)-Y_{i1k}(0))}{K_{i1}}$ (Kahan, Blette, et al., 2023; X. Wang et al., 2022). Accordingly, the cATE estimand under an infinite superpopulation, where $I \rightarrow \infty$, is:

$$lim_{I\rightarrow\infty} cATE = lim_{I\rightarrow\infty}\frac{1}{I}\sum_{i=1}^{I}\frac{\sum_{k=1}^{K_{i1}}\big(Y_{i1k}(1) - Y_{i1k}(0)\big)}{K_{i1}}$$

and:

$$cATE \xrightarrow{P} E\left[\frac{\sum_{k=1}^{K_{i1}} Y_{i1k}(1)}{K_{i1}} - \frac{\sum_{k=1}^{K_{i1}} Y_{i1k}(0)}{K_{i1}}\right]$$

$$= E\left[\frac{1}{K_{i1}}\sum_{k=1}^{K_{i1}}[Y_{i1k}(1) - Y_{i1k}(0)]\right].$$

If $K_{i1}$ is independent of potential outcomes $\left(Y_{i1k}(1), Y_{i1k}(0)\right)$, as is the case with non-informative cluster sizes, then we can similarly demonstrate that:

$$E\left[\frac{\sum_{k=1}^{K_{i1}}(Y_{i1k}(1)-Y_{i1k}(0))}{K_{i1}}\right] = E[Y_{i1k}(1) - Y_{i1k}(0)]\,.$$

## A.2 Complete estimator derivations

### A.2.1 Independence estimating equation (IEE)

Generalized, the IEE estimator can be written with potential outcomes $\left(Y_{ijk}(0), Y_{i1k}(1)\right)$ for participant $k \in (1, \ldots, K_{i-})$ in period $j \in (0,1)$ of cluster $i \in (1, \ldots, I)$. Let $S_i$ be an indicator for whether individuals are assigned to cluster sequence $S_i = 1$:

$$\hat{\delta}_{IEE} = \left[\frac{\sum_i S_i \sum_j \sum_k Y_{i1k}(1)}{\sum_i S_i K_{i1}}\right] - \left[\frac{\sum_i (1 - S_i) \sum_j \sum_k Y_{i1k}(0)}{\sum_i (1 - S_i) K_{i1}}\right]$$

$$= \left[\frac{\sum_i S_i \sum_k Y_{i1k}(1)}{\sum_i S_i K_{i1}}\right] - \left[\frac{\sum_i (1 - S_i) \sum_k Y_{i1k}(0)}{\sum_i (1 - S_i) K_{i1}}\right].$$

We can demonstrate that this estimator is consistent and asymptotically unbiased for the pATE:

$$lim_{I\to\infty} \hat{\delta}_{IEE} = lim_{I\to\infty} \left[\frac{\sum_i S_i \sum_k Y_{i1k}(1)}{\sum_i S_i K_{i1}}\right] - lim_{I\to\infty} \left[\frac{\sum_i (1 - S_i) \sum_k Y_{i1k}(0)}{\sum_i (1 - S_i) K_{i1}}\right]$$

$$= \left[\frac{lim_{I\to\infty} \sum_i S_i \sum_k Y_{i1k}(1)}{lim_{I\to\infty} \sum_i S_i K_{i1}}\right] - \left[\frac{lim_{I\to\infty} \sum_i (1 - S_i) \sum_k Y_{i1k}(0)}{lim_{I\to\infty} \sum_i (1 - S_i) K_{i1}}\right]$$

as the number of sampled clusters $I \to \infty$, by the law of large numbers, we can demonstrate that the IEE estimator converges in probability to:

$$\hat{\delta}_{IEE} \xrightarrow{P} \left[\frac{E[\sum_k Y_{i1k}(1) \,|S_i]}{E[K_{i1}|S_i]}\right] - \left[\frac{E[\sum_k Y_{i1k}(0) \,|1 - S_i]}{E[K_{i1}|1 - S_i]}\right]$$

where with randomization, the sequence variable $S_i$ is independent of the potential outcomes and cluster-period sizes $S_i \perp\!\!\!\perp \Omega$, and $\Omega = \{Y_{i1k}(1), Y_{i1k}(0), K_{i1}\}_{i=1,k=1}^{I,K_{i1}}$:

$$= \left[\frac{E[\sum_k Y_{i1k}(1)]}{E[K_{i1}]}\right] - \left[\frac{E[\sum_k Y_{i1k}(0)]}{E[K_{i1}]}\right]$$

$$= E\left[\frac{1}{E[K_{i1}]}\sum_{k=1}^{K_{i1}} [Y_{i1k}(1) - Y_{i1k}(0)]\right].$$

### A.2.2 Independence estimating equation with inverse cluster-period size weighting (IEEw)

The weighted IEEw estimator can then be written as:

$$\hat{\delta}_{IEEw} = \left[\frac{\sum_i S_i \sum_j \frac{1}{K_{i1}} \sum_k Y_{i1k}(1)}{\sum_i S_i}\right] - \left[\frac{\sum_i (1 - S_i) \sum_j \frac{1}{K_{i1}} \sum_k Y_{i1k}(0)}{\sum_i (1 - S_i)}\right]$$

$$= \left[\frac{\sum_i S_i \frac{1}{K_{i1}} \sum_k Y_{i1k}(1)}{\sum_i S_i}\right] - \left[\frac{\sum_i (1 - S_i) \frac{1}{K_{i1}} \sum_k Y_{i1k}(0)}{\sum_i (1 - S_i)}\right]$$

where with equal allocation:

$$\hat{\delta}_{IEEw} = \left(\frac{1}{I/2}\right)\left[\sum_i S_i \frac{1}{K_{i1}} \sum_k Y_{i1k}(1) - \sum_i (1 - S_i) \frac{1}{K_{i1}} \sum_k Y_{i1k}(0)\right]$$

and as the number of sampled clusters $I \to \infty$, by the law of large numbers, we can demonstrate that the IEEw estimator converges in probability to:

$$lim_{I\to\infty} \hat{\delta}_{IEEw} = \left(\frac{1}{I/2}\right) lim_{I\to\infty} \left[\sum_i S_i \frac{1}{K_{i1}} \sum_k Y_{i1k}(1) - \sum_i (1 - S_i) \frac{1}{K_{i1}} \sum_k Y_{i1k}(0)\right]$$

$$= E\left[\frac{\sum_{k=1}^{K_{ij}} Y_{i1k}(1)}{K_{i1}} |S_i\right] - E\left[\frac{\sum_{k=1}^{K_{ij}} Y_{i1k}(0)}{K_{i1}} |1 - S_i\right]$$

where with randomization, the sequence variable $S_i$ is independent of the potential outcomes and cluster-period sizes $S_i \perp\!\!\!\perp \Omega$, and $\Omega = \{Y_{i1k}(1), Y_{i1k}(0), K_{i1}\}_{i=1,k=1}^{I,K_{i1}}$:

$$\hat{\delta}_{IEE} \xrightarrow{P} E\left[\frac{\sum_{k=1}^{K_{ij}} Y_{i1k}(1)}{K_{i1}} - \frac{\sum_{k=1}^{K_{ij}} Y_{i1k}(0)}{K_{i1}}\right].$$

In addition to being consistent and asymptotically unbiased for the cATE, we can also demonstrate that the IEEw estimator is unbiased for the cATE in expectation over the sampling distribution. We can formally define the set of potential outcomes for all $K_{i1}$ participants in the sampled clusters $i = 1, \ldots, I$ as $\Omega = \{Y_{i1k}(1), Y_{i1k}(0), K_{i1}\}_{i=1,k=1}^{I,K_{i1}}$. Formally, we take the expectation of $\hat{\delta}_{IEEw}$ by treating the potential outcomes of the samples as fixed quantities and the sequence assignment (equivalent to a treatment assignment in follow-up period $j = 1$) $S_i$ as random. Therefore, conditioning the expectation on the set of sampled potential outcomes $\Omega$:

$$E\left[\hat{\delta}_{IEEw}|\Omega\right] = E\left[\frac{\sum_i S_i \frac{1}{K_{i1}} \sum_k Y_{i1k}(1)}{\sum_i S_i} \mid \Omega\right] - E\left[\frac{\sum_i (1 - S_i) \frac{1}{K_{i1}} \sum_k Y_{i1k}(0)}{\sum_i (1 - S_i)} \mid \Omega\right]$$

$$= \frac{1}{I/2} E\left[\sum_i S_i \frac{1}{K_{i1}} \sum_k Y_{i1k}(1) \mid \Omega\right] - \frac{1}{I/2} E\left[\sum_i (1 - S_i) \frac{1}{K_{i1}} \sum_k Y_{i1k}(0) \mid \Omega\right]$$

$$= \frac{1}{I/2} \sum_i P(S_i) \frac{1}{K_{i1}} \sum_k Y_{i1k}(1) - \frac{1}{I/2} \sum_i P(1 - S_i) \frac{1}{K_{i1}} \sum_k Y_{i1k}(0)$$

$$= \frac{1}{I/2} \sum_i \frac{I/2}{I} \frac{1}{K_{i1}} \sum_k Y_{i1k}(1) - \frac{1}{I/2} \sum_i \frac{I/2}{I} \frac{1}{K_{i1}} \sum_k Y_{i1k}(0)$$

$$= \frac{1}{I} \sum_{i} \left[ \frac{1}{K_{i1}} \sum_{k} Y_{i1k}(1) - \frac{1}{K_{i1}} \sum_{k} Y_{i1k}(0) \right]$$

We assume that the sample of clusters is a simple random sample from a superpopulation of clusters. With this superpopulation framework, we have two sources of randomness, random sampling from a superpopulation of clusters and subsequent randomization of treatment assignment. With the law of total expectation:

$$E\left[E[\hat{\delta}_{IEEw} | \Omega]\right] = E[\hat{\delta}_{IEEw}]$$

$$= E\left[ \frac{1}{I} \sum_{i} \left[ \frac{1}{K_{i1}} \sum_{k} Y_{i1k}(1) - \frac{1}{K_{i1}} \sum_{k} Y_{i1k}(0) \right] \right]$$

$$= E\left[ \frac{\sum_{k=1}^{K_{ij}} Y_{i1k}(1)}{K_{i1}} - \frac{\sum_{k=1}^{K_{ij}} Y_{i1k}(0)}{K_{i1}} \right].$$

## A.2.3 Fixed-effects model (FE)

The FE estimator can be written with potential outcomes $\left(Y_{ijk}(0), Y_{i1k}(1)\right)$ for participant $k \in (1, \ldots, K_{i-})$ in period $j \in (0,1)$ of cluster $i \in (1, \ldots, I)$. Let $S_i$ be an indicator for whether individuals are assigned to cluster sequence $S_i = 1$:

$$\hat{\delta}_{FE}$$

$$= \left[ \frac{\left( \sum_{i=1}^{I} S_i \frac{\prod_{j=0}^{1} K_{ij}}{\sum_{j=0}^{1} K_{ij}} \sum_{j=1}^{1} \frac{1}{K_{i1}} \sum_{k=1}^{K_{i1}} Y_{i1k}(1) \right) - \left( \sum_{i=1}^{I} S_i \frac{\prod_{j=0}^{1} K_{ij}}{\sum_{j=0}^{1} K_{ij}} \sum_{j=0}^{0} \frac{1}{K_{i0}} \sum_{k} Y_{i0k}(0) \right)}{\sum_{i=1}^{I} S_i \frac{\prod_{j=0}^{1} K_{ij}}{\sum_{j=0}^{1} K_{ij}}} \right]$$

$$- \left[ \frac{\left( \sum_{i=1}^{I} (1 - S_i) \frac{\prod_{j=0}^{1} K_{ij}}{\sum_{j=0}^{1} K_{ij}} \sum_{j=1}^{1} \frac{1}{K_{i1}} \sum_{k} Y_{i1k}(0) \right) - \left( \sum_{i=1}^{I} (1 - S_i) \frac{\prod_{j=0}^{1} K_{ij}}{\sum_{j=0}^{1} K_{ij}} \sum_{j=0}^{0} \frac{1}{K_{i0}} \sum_{k} Y_{i0k}(0) \right)}{\sum_{i=1}^{I} (1 - S_i) \frac{\prod_{j=0}^{1} K_{ij}}{\sum_{j=0}^{1} K_{ij}}} \right]$$

$$= \left[ \frac{\left( \sum_{i=1}^{I} S_i \frac{K_{i0}}{\sum_{j=0}^{1} K_{ij}} \sum_{k=1}^{K_{i1}} Y_{i1k}(1) \right) - \left( \sum_{i=1}^{I} S_i \frac{K_{i1}}{\sum_{j=0}^{1} K_{ij}} \sum_{k} Y_{i0k}(0) \right)}{\sum_{i=1}^{I} S_i \frac{\prod_{j=0}^{1} K_{ij}}{\sum_{j=0}^{1} K_{ij}}} \right]$$

$$- \left[ \frac{\left( \sum_{i=1}^{I} (1 - S_i) \frac{K_{i0}}{\sum_{j=0}^{1} K_{ij}} \sum_{k} Y_{i1k}(0) \right) - \left( \sum_{i=1}^{I} (1 - S_i) \frac{K_{i1}}{\sum_{j=0}^{1} K_{ij}} \sum_{k} Y_{i0k}(0) \right)}{\sum_{i=1}^{I} (1 - S_i) \frac{\prod_{j=0}^{1} K_{ij}}{\sum_{j=0}^{1} K_{ij}}} \right]$$

We can demonstrate that this estimator converges in probability to:

$$lim_{I\to\infty}\hat{\delta}_{FE}$$

$$= lim_{I\to\infty}\left[\frac{\left(\sum_{i=1}^{I} S_i \frac{K_{i0}}{\sum_{j=0}^{1} K_{ij}} \sum_{k=1}^{K_{i1}} Y_{i1k}(1)\right) - \left(\sum_{i=1}^{I} S_i \frac{K_{i1}}{\sum_{j=0}^{1} K_{ij}} \sum_k Y_{i0k}(0)\right)}{\sum_{i=1}^{I} S_i \frac{\prod_{j=0}^{1} K_{ij}}{\sum_{j=0}^{1} K_{ij}}}\right]$$

$$- lim_{I\to\infty}\left[\frac{\left(\sum_{i=1}^{I}(1-S_i) \frac{K_{i0}}{\sum_{j=0}^{1} K_{ij}} \sum_k Y_{i1k}(0)\right) - \left(\sum_{i=1}^{I}(1-S_i) \frac{K_{i1}}{\sum_{j=0}^{1} K_{ij}} \sum_k Y_{i0k}(0)\right)}{\sum_{i=1}^{I}(1-S_i) \frac{\prod_{j=0}^{1} K_{ij}}{\sum_{j=0}^{1} K_{ij}}}\right]$$

$$= \frac{1}{E\left[\frac{\prod_{j=0}^{1} K_{ij}}{\sum_{j=0}^{1} K_{ij}} |S_i\right]} E\left[\frac{K_{i0}}{\sum_{j=0}^{1} K_{ij}} \sum_{k=1}^{K_{i1}} Y_{i1k}(1) - \frac{K_{i1}}{\sum_{j=0}^{1} K_{ij}} \sum_k Y_{i0k}(0) \,|S_i\right]$$
$$- \frac{1}{E\left[\frac{\prod_{j=0}^{1} K_{ij}}{\sum_{j=0}^{1} K_{ij}} |1-S_i\right]} E\left[\frac{K_{i0}}{\sum_{j=0}^{1} K_{ij}} \sum_k Y_{i1k}(0) - \frac{K_{i1}}{\sum_{j=0}^{1} K_{ij}} \sum_k Y_{i0k}(0) \,|1-S_i\right]$$

where with randomization, the sequence variable $S_i$ is independent of the potential outcomes and cluster-period sizes $S_i \perp\!\!\!\perp \Omega$, and $\Omega = \{Y_{i1k}(1), Y_{i1k}(0), K_{i1}\}_{i=1,k=1}^{I,K_{i1}}$:

$$\hat{\delta}_{FE} \overset{P}{\to} E\left[\frac{\frac{K_{i0}}{\sum_{j=0}^{1} K_{ij}} \sum_{k=1}^{K_{i1}} Y_{i1k}(1) - \frac{K_{i1}}{\sum_{j=0}^{1} K_{ij}} \sum_k Y_{i0k}(0)}{E\left[\frac{\prod_{j=0}^{1} K_{ij}}{\sum_{j=0}^{1} K_{ij}}\right]} - \frac{\frac{K_{i0}}{\sum_{j=0}^{1} K_{ij}} \sum_k Y_{i1k}(0) - \frac{K_{i1}}{\sum_{j=0}^{1} K_{ij}} \sum_k Y_{i0k}(0)}{E\left[\frac{\prod_{j=0}^{1} K_{ij}}{\sum_{j=0}^{1} K_{ij}}\right]}\right]$$

$$= E\left[\frac{\frac{K_{i0}}{\sum_{j=0}^{1} K_{ij}} \sum_{k=1}^{K_{i1}} [Y_{i1k}(1) - Y_{i1k}(0)]}{E\left[\frac{\prod_{j=0}^{1} K_{ij}}{\sum_{j=0}^{1} K_{ij}}\right]}\right].$$

Assuming that cluster-period sizes vary between clusters but not between periods within clusters, $K_{i0} = K_{i1} = K_{i-}$, gives us:

$$= E\left[\frac{\sum_{k=1}^{K_{i1}} Y_{i1k}(1)}{E[K_{i1}]} - \frac{\sum_{k=1}^{K_{i1}} Y_{i1k}(0)}{E[K_{i1}]}\right].$$

## A.2.4 Fixed-effects model with inverse cluster-period size weighting (FEw)

When cluster-period sizes vary between periods within clusters, $K_{i0} \neq K_{i1}$,we can easily specify cluster-period-specific inverse cluster-period size weights to get the following fixed-effects model with inverse cluster-period size weights (FEw):

$$\hat{\delta}_{FEw} = \left[\frac{\left(\sum_i S_i \sum_j \frac{1}{K_{i1}} \sum_k Y_{i1k}(1)\right) - \left(\sum_i S_i \sum_j \frac{1}{K_{i0}} \sum_k Y_{i0k}(0)\right)}{\sum_i S_i}\right] - \left[\frac{\left(\sum_i (1-S_i) \sum_j \frac{1}{K_{i1}} \sum_k Y_{i1k}(0)\right) - \left(\sum_i (1-S_i) \sum_j \frac{1}{K_{i0}} \sum_k Y_{i0k}(0)\right)}{\sum_i (1-S_i)}\right]$$

$$= \left[\frac{\left(\sum_{i=1}^{I} S_i \frac{1}{K_{i1}} \sum_{k=1}^{K_{i1}} Y_{i1k}(1)\right) - \left(\sum_{i=1}^{I} S_i \frac{1}{K_{i0}} \sum_{k=1}^{K_{i0}} Y_{i0k}(0)\right)}{\sum_{i=1}^{I} S_i}\right] - \left[\frac{\left(\sum_{i=1}^{I} (1-S_i) \frac{1}{K_{i1}} \sum_{k=1}^{K_{i1}} Y_{i1k}(0)\right) - \left(\sum_{i=1}^{I} (1-S_i) \frac{1}{K_{i0}} \sum_{k=1}^{K_{i0}} Y_{i0k}(0)\right)}{\sum_{i=1}^{I} (1-S_i)}\right]$$

where with equal allocation:

$$\hat{\delta}_{FEw} = \left[\frac{\left(\sum_{i=1}^{I} S_i \frac{1}{K_{i1}} \sum_{k=1}^{K_{i1}} Y_{i1k}(1)\right) - \left(\sum_{i=1}^{I} S_i \frac{1}{K_{i0}} \sum_{k=1}^{K_{i0}} Y_{i0k}(0)\right)}{I/2}\right] - \left[\frac{\left(\sum_{i=1}^{I} (1-S_i) \frac{1}{K_{i1}} \sum_{k=1}^{K_{i1}} Y_{i1k}(0)\right) - \left(\sum_{i=1}^{I} (1-S_i) \frac{1}{K_{i0}} \sum_{k=1}^{K_{i0}} Y_{i0k}(0)\right)}{I/2}\right].$$

However, such weights deviate from Williamson's derivation with cluster-specific rather than cluster-period-specific weights. Still, inverse cluster-period size weights can be easily defined in analyses with an independence correlation structure (IEEw and FEw).

As the number of sampled clusters $I \rightarrow \infty$, by the law of large numbers, we can demonstrate that the FEw estimator converges in probability to:

$$lim_{I\rightarrow\infty} \hat{\delta}_{FEw} = lim_{I\rightarrow\infty} \left[\frac{\left(\sum_{i=1}^{I} S_i \frac{1}{K_{i1}} \sum_{k=1}^{K_{i1}} Y_{i1k}(1)\right) - \left(\sum_{i=1}^{I} S_i \frac{1}{K_{i0}} \sum_{k=1}^{K_{i0}} Y_{i0k}(0)\right)}{I/2}\right] - lim_{I\rightarrow\infty} \left[\frac{\left(\sum_{i=1}^{I} (1-S_i) \frac{1}{K_{i1}} \sum_{k=1}^{K_{i1}} Y_{i1k}(0)\right) - \left(\sum_{i=1}^{I} (1-S_i) \frac{1}{K_{i0}} \sum_{k=1}^{K_{i0}} Y_{i0k}(0)\right)}{I/2}\right]$$

$$= E\left[\frac{\sum_{k=1}^{K_{i1}} Y_{i1k}(1)}{K_{i1}} - \frac{\sum_{k=1}^{K_{i0}} Y_{i0k}(0)}{K_{i0}} |S_i\right] - E\left[\frac{\sum_{k=1}^{K_{i1}} Y_{i1k}(0)}{K_{i1}} - \frac{\sum_{k=1}^{K_{i0}} Y_{i0k}(0)}{K_{i0}} |1 - S_i\right]$$

where with randomization, the sequence variable $S_i$ is independent of the potential outcomes and cluster-period sizes $S_i \perp\!\!\!\perp \Omega$, and $\Omega = \{Y_{i1k}(1), Y_{i1k}(0), K_{i1}\}_{i=1,k=1}^{I,K_{i1}}$:

$$\hat{\delta}_{FEw} \xrightarrow{P} E\left[\frac{\sum_{k=1}^{K_{ij}} Y_{i1k}(1)}{K_{i1}} - \frac{\sum_{k=1}^{K_{ij}} Y_{i1k}(0)}{K_{i1}}\right].$$

In addition to being consistent and asymptotically unbiased for the cATE, we can also demonstrate that the FEw estimator is unbiased for the cATE in expectation over the sampling distribution. We can formally define the set of potential outcomes for all $K_{i1}$ participants in the sampled clusters $i = 1, \dots, I$ as $\Omega = \{Y_{i1k}(1), Y_{i1k}(0), K_{i1}\}_{i=1,k=1}^{I,K_{i1}}$. Formally, we take the expectation of $\hat{\delta}_{FEw}$ by treating the potential outcomes of the samples as fixed quantities and the sequence assignment (equivalent to a treatment assignment in follow-up period $j = 1$) $S_i$ as random. Therefore, conditioning the expectation on the set of sampled potential outcomes $\Omega$, when the randomization assumption is not met, the FEw is still an unbiased estimator for the cluster-average treatment effect on the treated (cATT):

$$E[\hat{\delta}_{FEw}|\Omega] = \left(\frac{1}{I/2}\right) E\left[\left[\left(\sum_i S_i \frac{1}{K_{i1}} \sum_k Y_{i1k}(1)\right) - \left(\sum_i S_i \frac{1}{K_{i0}} \sum_k Y_{i0k}(0)\right)\right]\right.$$
$$\left. - \left[\left(\sum_i (1 - S_i) \frac{1}{K_{i1}} \sum_k Y_{i1k}(0)\right) - \left(\sum_i (1 - S_i) \frac{1}{K_{i0}} \sum_k Y_{i0k}(0)\right)\right]\right]$$

$$= \left(\frac{1}{I/2}\right) E\left[\left(\sum_i S_i \frac{1}{K_{i1}} \sum_k Y_{i1k}(1)\right) - \left(\sum_i S_i \frac{1}{K_{i0}} \sum_k Y_{i0k}(0)\right)\right]$$
$$- \left(\frac{1}{I/2}\right) E\left[\left(\sum_i (1 - S_i) \frac{1}{K_{i1}} \sum_k Y_{i1k}(0)\right) - \left(\sum_i (1 - S_i) \frac{1}{K_{i0}} \sum_k Y_{i0k}(0)\right)\right]$$

with the parallel trends assumption:

$$= \left(\frac{1}{I/2}\right) E\left[\left(\sum_i S_i \frac{1}{K_{i1}} \sum_k Y_{i1k}(1)\right) - \left(\sum_i S_i \frac{1}{K_{i0}} \sum_k Y_{i0k}(0)\right)\right]$$
$$- \left(\frac{1}{I/2}\right) E\left[\left(\sum_i S_i \frac{1}{K_{i1}} \sum_k Y_{i1k}(0)\right) - \left(\sum_i S_i \frac{1}{K_{i0}} \sum_k Y_{i0k}(0)\right)\right]$$

Altogether:

$$E[\hat{\delta}_{FEw}|\Omega] = \left(\frac{1}{I/2}\right) E\left[\left(\sum_i S_i \frac{1}{K_{i1}} \sum_k Y_{i1k}(1)\right) - \left(\sum_i S_i \frac{1}{K_{i1}} \sum_k Y_{i1k}(0)\right)\right]$$

$$= \left(\frac{1}{I/2}\right) E\left[\sum_i S_i \frac{1}{K_{i1}} \sum_k [Y_{i1k}(1) - Y_{i1k}(0)]\right]$$

$$= \left(\frac{1}{I/2}\right)\sum_i E\left[S_i \frac{1}{K_{i1}}\sum_k [Y_{i1k}(1) - Y_{i1k}(0)]\right]$$

We assume that the sample of clusters is a simple random sample from a superpopulation of clusters. With this superpopulation framework, we have two sources of randomness, random sampling from a superpopulation of clusters and subsequent randomization of treatment assignment. With the law of total expectation:

$$E\left[E[\hat{\delta}_{FEw}|\Omega]\right] = E[\hat{\delta}_{FEw}]$$

$$= E\left[\left(\frac{1}{I/2}\right)\sum_i E\left[S_i \frac{1}{K_{i1}}\sum_k [Y_{i1k}(1) - Y_{i1k}(0)]\right]\right]$$

$$= \left(\frac{1}{I/2}\right)\left(\frac{I}{2}\right) E\left[S_i \frac{1}{K_{i1}}\sum_k [Y_{i1k}(1) - Y_{i1k}(0)]\right]$$

$$= E\left[S_i \frac{1}{K_{i1}}\sum_k [Y_{i1k}(1) - Y_{i1k}(0)]\right]$$

$$= E\left[\frac{1}{K_{i1}}\sum_k [Y_{i1k}(1) - Y_{i1k}(0)]\,|S_i = 1\right]$$

Therefore, regardless of randomization of clusters to the corresponding sequences, with the parallel trends assumption the FEw estimator produces the cluster-average treatment effect on the treated (cATT).

With randomization, the sequence variable $S_i$ is independent of the potential outcomes and cluster sizes $S_i \perp\!\!\!\perp \{Y_{i1k}(1), Y_{i1k}(0), K_{ij}\}$, and:

$$E[\hat{\delta}_{FEw}|\Omega] = \left(\frac{1}{I/2}\right)\left[\left(\sum_i E[S_i]\frac{1}{K_{i1}}\sum_k Y_{i1k}(1)\right) - \left(\sum_i E[S_i]\frac{1}{K_{i0}}\sum_k Y_{i0k}(0)\right)\right]$$
$$- \left[\left(\sum_i E[1 - S_i]\frac{1}{K_{i1}}\sum_k Y_{i1k}(0)\right) - \left(\sum_i E[1 - S_i]\frac{1}{K_{i0}}\sum_k Y_{i0k}(0)\right)\right]$$

$$= \left(\frac{1}{I/2}\right)\left[\left(\sum_i P(S_i)\frac{1}{K_{i1}}\sum_k Y_{i1k}(1)\right) - \left(\sum_i P(S_i)\frac{1}{K_{i0}}\sum_k Y_{i0k}(0)\right)\right]$$
$$- \left[\left(\sum_i P(1 - S_i)\frac{1}{K_{i1}}\sum_k Y_{i1k}(0)\right) - \left(\sum_i P(1 - S_i)\frac{1}{K_{i0}}\sum_k Y_{i0k}(0)\right)\right]$$

$$= \left(\frac{1}{I}\right)\left[\left(\sum_i \frac{1}{K_{i1}}\sum_k Y_{i1k}(1)\right) - \left(\sum_i \frac{1}{K_{i0}}\sum_k Y_{i0k}(0)\right)\right]$$
$$- \left[\left(\sum_i \frac{1}{K_{i1}}\sum_k Y_{i1k}(0)\right) - \left(\sum_i \frac{1}{K_{i0}}\sum_k Y_{i0k}(0)\right)\right]$$

$$= \left(\frac{1}{I}\right)\sum_i \frac{1}{K_{i1}} \sum_k [Y_{i1k}(1) - Y_{i1k}(0)]$$

Regardless of the presence of parallel trends. Accordingly:

$$E\left[E\left[\hat{\delta}_{FEw}|\Omega\right]\right] = E\left[\hat{\delta}_{FEw}\right]$$

$$= E\left[\frac{1}{K_{i1}}\sum_k [Y_{i1k}(1) - Y_{i1k}(0)]\right]$$

## A.2.5 Exchangeable Mixed-effects model (EME)

The exchangeable mixed-effects treatment effect (EME) point estimator can be specified based on Equation 3.9 where the diagonal terms $(D_i)$ and off-diagonal terms $(F_i)$ in the block matrix corresponding to the observations within cluster $i$ in $V_{ijk}^{-1}$ is:

$$D_i = \frac{1}{\sigma_w^2}\left(\frac{\sigma_w^2 + (2K_{i-} - 1)\tau_\alpha^2}{\sigma_w^2 + 2K_{i-}\tau_\alpha^2}\right)$$

$$F_i = -\frac{1}{\sigma_w^2}\left(\frac{\tau_\alpha^2}{\sigma_w^2 + 2K_{i-}\tau_\alpha^2}\right)$$

assuming that cluster-period sizes vary between clusters but not between periods within clusters, $K_{ij} = K_{i0} = K_{i1} = K_{i-}$ with $2K_{i-}$ observations within each cluster.

Accordingly, the EME estimator can be written with potential outcomes $\left(Y_{ijk}(0), Y_{i1k}(1)\right)$ for participant $k \in (1, \dots, K_{i-})$ in period $j \in (0,1)$ of cluster $i \in (1, \dots, I)$. Let $S_i$ be an indicator for whether individuals are assigned to cluster sequence $S_i = 1$.

$$\hat{\delta}_{EME} = \left\{\left(\sum_i S_i A_i\right)\left(\sum_i (1 - S_i) A_i\right)\left(\sum_i C_i\right) - \left(\sum_i S_i A_i\right)\left(\sum_i (1 - S_i) B_i\right)\left(\sum_i (1 - S_i) B_i\right) - \left(\sum_i (1 - S_i) A_i\right)\left(\sum_i S_i B_i\right)\left(\sum_i S_i B_i\right)\right\}^{-1} \times$$

$$
\left\{
\begin{array}{c}
\left[\left(\sum_i S_i B_i\right)\left(\sum_i (1-S_i) B_i\right)\right]
\begin{bmatrix}
\sum_i (1-S_i) B_i \sum_j (1-\phi_j) \frac{1}{K_{ij}} \sum_k Y_{ijk}(0) \\
+\sum_i (1-S_i) A_i \sum_j \phi_j \frac{1}{K_{ij}} \sum_k Y_{ijk}(0) \\
-\sum_i S_i B_i \sum_j (1-\phi_j) \frac{1}{K_{ij}} \sum_k Y_{ijk}(0) \\
-\sum_i S_i A_i \sum_j \phi_j \frac{1}{K_{ij}} \sum_k Y_{ijk}(1)
\end{bmatrix} \\
+\left[\left(\sum_i S_i B_i\right)\left(\sum_i S_i B_i\right)-\left(\sum_i S_i A_i\right)\left(\sum_i C_i\right)\right]
\begin{bmatrix}
\sum_i (1-S_i) B_i \sum_j (1-\phi_j) \frac{1}{K_{ij}} \sum_k Y_{ijk}(0) \\
+\sum_i (1-S_i) A_i \sum_j \phi_j \frac{1}{K_{ij}} \sum_k Y_{ijk}(0)
\end{bmatrix} \\
+\left[\left(\sum_i (1-S_i) B_i\right)\left(\sum_i (1-S_i) B_i\right)-\left(\sum_i (1-S_i) A_i\right)\left(\sum_i C_i\right)\right]
\begin{bmatrix}
-\sum_i S_i B_i \sum_j (1-\phi_j) \frac{1}{K_{ij}} \sum_k Y_{ijk}(0) \\
-\sum_i S_i A_i \sum_j \phi_j \frac{1}{K_{ij}} \sum_k Y_{ijk}(1)
\end{bmatrix} \\
-\left[\left(\sum_i (1-S_i) A_i\right)\left(\sum_i S_i B_i\right)-\left(\sum_i S_i A_i\right)\left(\sum_i (1-S_i) B_i\right)\right]
\begin{bmatrix}
\sum_i C_i \sum_j (1-\phi_j) \frac{1}{K_{ij}} \sum_k Y_{ijk}(0) \\
+\sum_i S_i B_i \sum_j \phi_j \frac{1}{K_{ij}} \sum_k Y_{ijk}(1) \\
+\sum_i (1-S_i) B_i \sum_j \phi_j \frac{1}{K_{ij}} \sum_k Y_{ijk}(0)
\end{bmatrix}
\end{array}
\right\}
$$

where:

$$
A_i = K_{i2}[D_i + (K_{i2}-1)F_i] = K_{i2}\frac{1}{\sigma_w^2}\left(\frac{\sigma_w^2 + K_{i1}\tau_\alpha^2}{\sigma_w^2 + (K_{i1}+K_{i2})\tau_\alpha^2}\right)
$$

$$
= K_{i2}\frac{1}{\sigma_w^2}\left(\frac{1+(K_{i1}-1)\rho}{1+(K_{i1}+K_{i2}-1)\rho}\right)
$$

$$
B_i = K_{i1}K_{i2}F_i = -K_{i1}K_{i2}\frac{1}{\sigma_w^2}\left(\frac{\tau_\alpha^2}{\sigma_w^2 + (K_{i1}+K_{i2})\tau_\alpha^2}\right)
$$

$$
= -K_{i1}K_{i2}\frac{1}{\sigma_w^2}\left(\frac{\rho}{1+(K_{i1}+K_{i2}-1)\rho}\right)
$$

$$
C_i = K_{i1}[D_i + (K_{i1}-1)F_i] = K_{i1}\frac{1}{\sigma_w^2}\left(\frac{\sigma_w^2 + K_{i2}\tau_\alpha^2}{\sigma_w^2 + (K_{i1}+K_{i2})\tau_\alpha^2}\right)
$$

$$
= K_{i1}\frac{1}{\sigma_w^2}\left(\frac{1+(K_{i2}-1)\rho}{1+(K_{i1}+K_{i2}-1)\rho}\right)
$$

Assuming that cluster-period sizes vary between clusters but not between periods within clusters, $K_{ij} = K_{i0} = K_{i1} = K_{i-}$, produces:

$$\hat{\delta}_{EME} = \Bigg\{\left(\sum_i S_i A_i\right)\left(\sum_i (1-S_i)A_i\right)\left(\sum_i A_i\right) - \left(\sum_i S_i A_i\right)\left(\sum_i (1-S_i)B_i\right)\left(\sum_i (1-S_i)B_i\right) - \left(\sum_i (1-S_i)A_i\right)\left(\sum_i S_i B_i\right)\left(\sum_i S_i B_i\right)\Bigg\}^{-1} \times$$

$$\left\{\begin{array}{c}
\left(\sum_i S_i B_i\right)\left(\sum_i (1-S_i)B_i\right)\begin{pmatrix} \sum_i (1-S_i)B_i \frac{1}{K_{i-}}\sum_k Y_{i0k}(0) \\ +\sum_i (1-S_i)A_i \frac{1}{K_{i-}}\sum_k Y_{i1k}(0) \\ -\sum_i S_i B_i \frac{1}{K_{i-}}\sum_k Y_{i0k}(0) \\ -\sum_i S_i A_i \frac{1}{K_{i-}}\sum_k Y_{i1k}(1) \end{pmatrix} \\
+\left(\left(\sum_i S_i B_i\right)\left(\sum_i S_i B_i\right) - \left(\sum_i S_i A_i\right)\left(\sum_i A_i\right)\right)\begin{pmatrix} \sum_i (1-S_i)B_i \frac{1}{K_{i-}}\sum_k Y_{i0k}(0) \\ +\sum_i (1-S_i)A_i \frac{1}{K_{i-}}\sum_k Y_{i1k}(0) \end{pmatrix} \\
-\left(\left(\sum_i (1-S_i)B_i\right)\left(\sum_i (1-S_i)B_i\right) - \left(\sum_i (1-S_i)A_i\right)\left(\sum_i A_i\right)\right)\begin{pmatrix} \sum_i S_i B_i \frac{1}{K_{i-}}\sum_k Y_{i0k}(0) \\ +\sum_i S_i A_i \frac{1}{K_{i-}}\sum_k Y_{i1k}(1) \end{pmatrix} \\
-\left(\left(\sum_i (1-S_i)A_i\right)\left(\sum_i S_i B_i\right) - \left(\sum_i S_i A_i\right)\left(\sum_i (1-S_i)B_i\right)\right)\begin{pmatrix} \sum_i A_i \frac{1}{K_{i-}}\sum_k Y_{i0k}(0) \\ +\sum_i S_i B_i \frac{1}{K_{i-}}\sum_k Y_{i1k}(1) \\ +\sum_i (1-S_i)B_i \frac{1}{K_{i-}}\sum_k Y_{i1k}(0) \end{pmatrix}
\end{array}\right\}$$

Where:

$$A_i = K_{i-}[D_i + (K_{i-}-1)F_i] = \frac{1}{\sigma_w^2}\left(\frac{1+(K_{i-}-1)\rho}{1+(2K_{i-}-1)\rho}\right)K_{i-}$$

$$B_i = \sigma_w^2 K_{i-}^2 F_i = -\frac{1}{\sigma_w^2}\left(\frac{\rho}{1+(2K_{i-}-1)\rho}\right)K_{i-}^2$$

with intracluster correlation $\rho = \frac{\tau_\alpha^2}{\tau_\alpha^2+\sigma_w^2}$ .

We can demonstrate that this estimator is consistent and asymptotically unbiased for the pATE with randomization where the sequence variable $S_i$ is independent of the potential outcomes and cluster-period sizes $S_i \perp\!\!\!\perp \Omega$, and $\Omega = \{Y_{i1k}(1), Y_{i1k}(0), K_{i1}\}_{i=1,k=1}^{I,K_{i1}}$:

:

$$lim_{I\to\infty}\hat{\delta}_{EME} =$$

$$\frac{(E[A_i]E[A_i] - E[B_i]E[B_i])E\left[A_i \frac{1}{K_{i-}} \sum_k [Y_{i1k}(1) - Y_{i1k}(0)]\right]}{E[A_i]E[A_i]E[A_i] - \left(\frac{1}{2}\right)E[A_i]E[B_i]E[B_i] - \left(\frac{1}{2}\right)E[A_i]E[B_i]E[B_i]}$$

$$= \frac{E\left[A_i \frac{1}{K_{i-}} \sum_k [Y_{i1k}(1) - Y_{i1k}(0)]\right]}{E[A_i]}$$

$$= E\left[\frac{A_i/K_{i-}}{E[A_i]} \sum_k [Y_{i1k}(1) - Y_{i1k}(0)]\right].$$

Recall:

$$A_i = K_{i-}[D_i + (K_{i-} - 1)F_i] = \frac{1}{\sigma_w^2}\left(\frac{1 + (K_{i-} - 1)\rho}{1 + (2K_{i-} - 1)\rho}\right)K_{i-}$$

Altogether, we demonstrate that the EME estimator converges in probability to the following difficult to interpret estimand:

$$\hat{\delta}_{EME} \xrightarrow{P} \frac{E\left[\left(\frac{1 + (K_{i-} - 1)\rho}{1 + (2K_{i-} - 1)\rho}\right)\sum_k [Y_{i1k}(1) - Y_{i1k}(0)]\right]}{E\left[\frac{1 + (K_{i-} - 1)\rho}{1 + (2K_{i-} - 1)\rho}\right]}.$$

In general, we can demonstrate:

$$\hat{\delta}_{ME} \xrightarrow{P} E\left[\frac{A_i/K_{i-}}{E[A_i]} \sum_k [Y_{i1k}(1) - Y_{i1k}(0)]\right]$$

but with different model-specific values of $A_i$ for any of the mixed-effects models explored here (EME, EMEw, NEME, and NEMEw).

### A.2.6 Exchangeable Mixed-effects model with inverse cluster-period size weighting (EMEw)

The weighted EMEw estimator can be specified similarly to the unweighted EME estimator, but with the diagonal terms ($D_i$) and off-diagonal terms ($F_i$) in the block matrices of the inverted weighted correlation structure $\boldsymbol{W}_i^{-1}$ corresponding to the observations within cluster $i$ specified as:

$$D_i = \frac{1}{(K_{i-})\sigma_w^2}\left(\frac{(K_{i-})\sigma_w^2 + (2K_{i-} - 1)(K_{i-})\tau_\alpha^2}{(K_{i-})\sigma_w^2 + 2K_{i-}(K_{i-})\tau_\alpha^2}\right) = \frac{1}{(K_{i-})\sigma_w^2}\left(\frac{\sigma_w^2 + (2K_{i-} - 1)\tau_\alpha^2}{\sigma_w^2 + 2K_{i-}\tau_\alpha^2}\right)$$

$$F_i = -\frac{1}{(K_{i-})\sigma_w^2}\left(\frac{(K_{i-})\tau_\alpha^2}{(K_{i-})\sigma_w^2 + 2K_{i-}\tau_\alpha^2(K_{i-})}\right) = -\frac{1}{(K_{i-})\sigma_w^2}\left(\frac{\tau_\alpha^2}{\sigma_w^2 + 2K_{i-}\tau_\alpha^2}\right)$$

As a result, exchangeable mixed-effects model with inverse cluster-period size weighting (EMEw) estimator has an identical estimator to **Error! Reference source not found.**, but with a different weighted specification of $A_i$ and $B_i$:

$$A_i = K_{i-}\left[\frac{1}{(K_{i-})\sigma_w^2}\left(\frac{\sigma_w^2 + (2K_{i-} - 1)\tau_\alpha^2}{\sigma_w^2 + 2K_{i-}\tau_\alpha^2}\right) - (K_{i-} - 1)\frac{1}{(K_{i-})\sigma_w^2}\left(\frac{\tau_\alpha^2}{\sigma_w^2 + 2K_{i-}\tau_\alpha^2}\right)\right]$$

$$= \frac{1}{\sigma_w^2}\left(\frac{\sigma_w^2 + (2K_{i-} - 1)\tau_\alpha^2 - (K_{i-} - 1)\tau_\alpha^2}{\sigma_w^2 + 2K_{i-}\tau_\alpha^2}\right)$$

$$= \frac{1}{\sigma_w^2}\left(\frac{\sigma_w^2 + K_{i-}\tau_\alpha^2}{\sigma_w^2 + 2K_{i-}\tau_\alpha^2}\right)$$

$$= \frac{1}{\sigma_w^2}\left(\frac{\sigma_w^2 + K_{i-}\tau_\alpha^2}{\sigma_w^2 + 2K_{i-}\tau_\alpha^2}\right)$$

$$= \frac{1}{\sigma_w^2}\left(\frac{\sigma_w^2 + \tau_\alpha^2 + (K_{i-} - 1)\tau_\alpha^2}{\sigma_w^2 + \tau_\alpha^2 + (2K_{i-} - 1)\tau_\alpha^2}\right)$$

$$= \frac{1}{\sigma_w^2}\left(\frac{1 + (K_{i-} - 1)\left(\frac{\tau_\alpha^2}{\sigma_w^2 + \tau_\alpha^2}\right)}{1 + (2K_{i-} - 1)\left(\frac{\tau_\alpha^2}{\sigma_w^2 + \tau_\alpha^2}\right)}\right)$$

$$= \frac{1}{\sigma_w^2}\left(\frac{1 + (K_{i-} - 1)\rho}{1 + (2K_{i-} - 1)\rho}\right)$$

and

$$B_i = -\frac{K_{i-}^2}{(K_{i-})\sigma_w^2}\left(\frac{\tau_\alpha^2}{\sigma_w^2 + 2K_{i-}\tau_\alpha^2}\right)$$

$$= -\frac{K_{i-}}{\sigma_w^2}\left(\frac{\tau_\alpha^2}{\sigma_w^2 + 2K_{i-}\tau_\alpha^2}\right)$$

$$= -K_{i-}\frac{1}{\sigma_w^2}\left(\frac{\rho}{1 + (2K_{i-} - 1)\rho}\right)$$

assuming that cluster-period sizes vary between clusters but not between periods within clusters, $K_{ij} = K_{i0} = K_{i1} = K_{i-}$, with intracluster correlation $\rho = \frac{\tau_\alpha^2}{\tau_\alpha^2 + \sigma_w^2}$.

Therefore, the EMEw estimator converges in probability to the following difficult to interpret estimand:

$$\hat{\delta}_{EMEw} \xrightarrow{P} E\left[\frac{\left(\frac{1 + (K_{i-} - 1)\rho}{1 + (2K_{i-} - 1)\rho}\right)}{E\left[\left(\frac{1 + (K_{i-} - 1)\rho}{1 + (2K_{i-} - 1)\rho}\right)\right]}\left(\frac{1}{K_{i-}}\sum_k [Y_{i1k}(1) - Y_{i1k}(0)]\right)\right]$$

## A.2.7 Nested Exchangeable Mixed-effects model (NEME)

The nested exchangeable mixed-effects treatment effect (NEME) point estimator can be specified based on Equation 3.12. Assuming that cluster-period sizes vary between clusters but not between periods within clusters, $K_{ij} = K_{i0} = K_{i1} = K_{i-}$ with $2K_{i-}$ observations within each cluster, the correlation structure is then $\check{V}_{ijk} = \mathrm{I}_I \otimes R_i$, where:

$$R_i = \begin{pmatrix} R_{i1}^{NME} & R_{i2}^{NME} \\ R_{i3}^{NME} & R_{i4}^{NME} \end{pmatrix},$$

with block matrices:

$$R_{i1}^{NME} = R_{i4}^{NME} = \left(\mathrm{I}_{K_{i-}}\sigma_w^2 + \mathrm{J}_{K_{i-}}\left(\tau_\alpha^2 + \tau_\gamma^2\right)\right)$$
$$= \begin{pmatrix} \sigma_w^2 + \tau_\alpha^2 + \tau_\gamma^2 & \tau_\alpha^2 + \tau_\gamma^2 & \cdots & \tau_\alpha^2 + \tau_\gamma^2 \\ \tau_\alpha^2 + \tau_\gamma^2 & \sigma_w^2 + \tau_\alpha^2 + \tau_\gamma^2 & \cdots & \tau_\alpha^2 + \tau_\gamma^2 \\ \vdots & \vdots & \ddots & \vdots \\ \tau_\alpha^2 + \tau_\gamma^2 & \tau_\alpha^2 + \tau_\gamma^2 & \cdots & \sigma_w^2 + \tau_\alpha^2 + \tau_\gamma^2 \end{pmatrix},$$

$$R_{i2}^{NME} = R_{i3}^{NME} = \left(\mathrm{J}_{K_{i-}}\tau_\alpha^2\right) = \begin{pmatrix} \tau_\alpha^2 & \tau_\alpha^2 & \cdots & \tau_\alpha^2 \\ \tau_\alpha^2 & \tau_\alpha^2 & \cdots & \tau_\alpha^2 \\ \vdots & \vdots & \ddots & \vdots \\ \tau_\alpha^2 & \tau_\alpha^2 & \cdots & \tau_\alpha^2 \end{pmatrix}$$

where $\mathrm{I}_{K_{i-}}$ is a $K_{i-}$ by $K_{i-}$ dimension identity matrix and $\mathrm{J}_{K_{i-}}$ is a $K_{i-}$ by $K_{i-}$ dimension matrix of ones. Accordingly, $\tilde{V}_{ijk}^{-1} = \mathrm{I}_I \otimes R_i^{-1}$, where:

$$R_i^{-1} = \begin{pmatrix} [R_{i1}^{NME} - R_{i2}^{NME}(R_{i1}^{NME})^{-1}R_{i2}^{NME}]^{-1} & -[R_{i1}^{NME} - R_{i2}^{NME}(R_{i1}^{NME})^{-1}R_{i2}^{NME}]^{-1}R_{i2}^{NME}(R_{i1}^{NME})^{-1} \\ -[R_{i1}^{NME} - R_{i2}^{NME}(R_{i1}^{NME})^{-1}R_{i2}^{NME}]^{-1}R_{i2}^{NME}(R_{i1}^{NME})^{-1} & [R_{i1}^{NME} - R_{i2}^{NME}(R_{i1}^{NME})^{-1}R_{i2}^{NME}]^{-1} \end{pmatrix}$$

We define:

$$[R_{i1}^{NME} - R_{i2}^{NME}(R_{i1}^{NME})^{-1}R_{i2}^{NME}]^{-1} = \left(\mathrm{I}_{K_{i-}}(D_i - F_i) + \mathrm{J}_{K_{i-}}(F_i)\right)$$

where the diagonal terms $(D_i)$ and off-diagonal terms $(F_i)$ are:

$$D_i = \frac{1}{\sigma_w^2}\left(\frac{\sigma_w^2 + (K_{i-} - 1)\left[\left(\tau_\alpha^2 + \tau_\gamma^2\right) - \dfrac{(K_{i-})(\tau_\alpha^2)^2}{\sigma_w^2 + (K_{i-})\left(\tau_\alpha^2 + \tau_\gamma^2\right)}\right]}{\sigma_w^2 + (K_{i-})\left[\left(\tau_\alpha^2 + \tau_\gamma^2\right) - \dfrac{(K_{i-})(\tau_\alpha^2)^2}{\sigma_w^2 + (K_{i-})\left(\tau_\alpha^2 + \tau_\gamma^2\right)}\right]}\right)$$

$$F_i = -\frac{1}{\sigma_w^2}\left(\frac{\left(\tau_\alpha^2 + \tau_\gamma^2\right) - \dfrac{(K_{i-})(\tau_\alpha^2)^2}{\sigma_w^2 + (K_{i-})\left(\tau_\alpha^2 + \tau_\gamma^2\right)}}{\sigma_w^2 + (K_{i-})\left[\left(\tau_\alpha^2 + \tau_\gamma^2\right) - \dfrac{(K_{i-})(\tau_\alpha^2)^2}{\sigma_w^2 + (K_{i-})\left(\tau_\alpha^2 + \tau_\gamma^2\right)}\right]}\right).$$

Subsequently, we define:

$$-[R_{i1}^{NME} - R_{i2}^{NME}(R_{i1}^{NME})^{-1}R_{i2}^{NME}]^{-1}R_{i2}^{NME}(R_{i1}^{NME})^{-1} = \mathrm{J}_{K_{i-}}(G_i)$$

where:

$$G_i = -\left(\frac{\tau_\alpha^2}{\sigma_w^2 + (K_{i-})(\tau_\alpha^2 + \tau_\gamma^2)}\right)\left(\frac{1}{\sigma_w^2 + (K_{i-})\left[(\tau_\alpha^2 + \tau_\gamma^2) - \frac{(K_{i-})(\tau_\alpha^2)^2}{\sigma_w^2 + (K_{i-})(\tau_\alpha^2 + \tau_\gamma^2)}\right]}\right)$$

$$= -\left(\frac{\tau_\alpha^2}{\sigma_w^2 + (K_{i-})(\tau_\alpha^2 + \tau_\gamma^2)}\right)\left(\frac{\sigma_w^2 + (K_{i-})(\tau_\alpha^2 + \tau_\gamma^2)}{\left(\sigma_w^2 + (K_{i-})(\tau_\alpha^2 + \tau_\gamma^2)\right)^2 - (K_{i-})^2(\tau_\alpha^2)^2}\right)$$

$$= -\left(\frac{\tau_\alpha^2}{\left(\sigma_w^2 + (K_{i-})(\tau_\alpha^2 + \tau_\gamma^2)\right)^2 - (K_{i-})^2(\tau_\alpha^2)^2}\right)$$

Accordingly, the NEME estimator can be written with potential outcomes $\left(Y_{ijk}(0), Y_{i1k}(1)\right)$ for participant $k \in (1, \dots, K_{i-})$ in period $j \in (0,1)$ of cluster $i \in (1, \dots, I)$. Let $S_i$ be an indicator for whether individuals are assigned to cluster sequence $S_i = 1$. The resulting, nested exchangeable mixed-effects model (NEME) estimator has an identical estimator to **Error! Reference source not found.**, but with a different weighted specification of $A_i$ and $B_i$:

$$A_i = K_{i-}[D_i + (K_{i-} - 1)F_i]$$

$$= K_{i-}\left[\frac{1}{\sigma_w^2}\left(\frac{\sigma_w^2 + (K_{i-} - 1)\left[(\tau_\alpha^2 + \tau_\gamma^2) - \frac{(K_{i-})(\tau_\alpha^2)^2}{\sigma_w^2 + (K_{i-})(\tau_\alpha^2 + \tau_\gamma^2)}\right]}{\sigma_w^2 + (K_{i-})\left[(\tau_\alpha^2 + \tau_\gamma^2) - \frac{(K_{i-})(\tau_\alpha^2)^2}{\sigma_w^2 + (K_{i-})(\tau_\alpha^2 + \tau_\gamma^2)}\right]}\right) - \frac{1}{\sigma_w^2}\left(\frac{(K_{i-} - 1)\left[(\tau_\alpha^2 + \tau_\gamma^2) - \frac{(K_{i-})(\tau_\alpha^2)^2}{\sigma_w^2 + (K_{i-})(\tau_\alpha^2 + \tau_\gamma^2)}\right]}{\sigma_w^2 + (K_{i-})\left[(\tau_\alpha^2 + \tau_\gamma^2) - \frac{(K_{i-})(\tau_\alpha^2)^2}{\sigma_w^2 + (K_{i-})(\tau_\alpha^2 + \tau_\gamma^2)}\right]}\right)\right]$$

$$= K_{i-}\left(\frac{1}{\sigma_w^2 + (K_{i-})\left[(\tau_\alpha^2 + \tau_\gamma^2) - \frac{(K_{i-})(\tau_\alpha^2)^2}{\sigma_w^2 + (K_{i-})(\tau_\alpha^2 + \tau_\gamma^2)}\right]}\right)$$

$$= K_{i-}\left(\frac{\sigma_w^2 + (K_{i-})(\tau_\alpha^2 + \tau_\gamma^2)}{\left(\sigma_w^2 + (K_{i-})(\tau_\alpha^2 + \tau_\gamma^2)\right)^2 - (K_{i-})^2(\tau_\alpha^2)^2}\right)$$

which can be further written in terms of the within-period ICC $\left(\rho_{wp} = \frac{\tau_\alpha^2 + \tau_\gamma^2}{\tau_\alpha^2 + \tau_\gamma^2 + \sigma_w^2}\right)$ and between-period ICC $\left(\rho_{bp} = \frac{\tau_\alpha^2}{\tau_\alpha^2 + \tau_\gamma^2 + \sigma_w^2}\right)$:

$$= K_{i-}\left(\frac{1+(K_{i-}-1)\rho_{wp}}{\left[\left(1+(K_{i-}-1)\rho_{wp}\right)^2-(K_{i-})^2\rho_{bp}^2\right]\left(\tau_\alpha^2+\tau_\gamma^2+\sigma_w^2\right)}\right)$$

and:

$$B_i = K_{i-}^2 G_i$$

$$= -K_{i-}^2\left(\frac{\tau_\alpha^2}{\left(\sigma_w^2+(K_{i-})\left(\tau_\alpha^2+\tau_\gamma^2\right)\right)^2-(K_{i-})^2(\tau_\alpha^2)^2}\right)$$

Therefore, the NEME estimator converges in probability to the following difficult to interpret estimand:

$$\hat{\delta}_{NEME} \xrightarrow{P} E\left[\frac{\left(\frac{1+(K_{i-}-1)\rho_{wp}}{\left(1+(K_{i-}-1)\rho_{wp}\right)^2-(K_{i-})^2\rho_{bp}^2}\right)}{E\left[\left(\frac{1+(K_{i-}-1)\rho_{wp}}{\left(1+(K_{i-}-1)\rho_{wp}\right)^2-(K_{i-})^2\rho_{bp}^2}\right)K_{i-}\right]}\sum_k\left[Y_{i1k}(1)-Y_{i1k}(0)\right]\right]$$

### A.2.8 Nested Exchangeable Mixed-effects model with inverse cluster-period size weighting (NEMEw)

The weighted NEMEw estimator can be specified similarly to the unweighted NEME estimator, but with the following terms

$$D_i = \frac{1}{(K_{i-})\sigma_w^2}\left(\frac{\sigma_w^2+(K_{i-}-1)\left[\left(\tau_\alpha^2+\tau_\gamma^2\right)-\frac{(K_{i-})(\tau_\alpha^2)^2}{\sigma_w^2+(K_{i-})\left(\tau_\alpha^2+\tau_\gamma^2\right)}\right]}{\sigma_w^2+(K_{i-})\left[\left(\tau_\alpha^2+\tau_\gamma^2\right)-\frac{(K_{i-})(\tau_\alpha^2)^2}{\sigma_w^2+(K_{i-})\left(\tau_\alpha^2+\tau_\gamma^2\right)}\right]}\right)$$

$$F_i = -\frac{1}{(K_{i-})\sigma_w^2}\left(\frac{\left(\tau_\alpha^2+\tau_\gamma^2\right)-\frac{(K_{i-})(\tau_\alpha^2)^2}{\sigma_w^2+(K_{i-})\left(\tau_\alpha^2+\tau_\gamma^2\right)}}{\sigma_w^2+(K_{i-})\left[\left(\tau_\alpha^2+\tau_\gamma^2\right)-\frac{(K_{i-})(\tau_\alpha^2)^2}{\sigma_w^2+(K_{i-})\left(\tau_\alpha^2+\tau_\gamma^2\right)}\right]}\right)$$

$$G_i = -\frac{1}{(K_{i-})}\left(\frac{\tau_\alpha^2}{\sigma_w^2+(K_{i-})\left(\tau_\alpha^2+\tau_\gamma^2\right)}\right)\left(\frac{1}{\sigma_w^2+(K_{i-})\left[\left(\tau_\alpha^2+\tau_\gamma^2\right)-\frac{(K_{i-})(\tau_\alpha^2)^2}{\sigma_w^2+(K_{i-})\left(\tau_\alpha^2+\tau_\gamma^2\right)}\right]}\right)$$

$$= -\frac{1}{(K_{i-})}\left(\frac{\tau_\alpha^2}{\left(\sigma_w^2+(K_{i-})\left(\tau_\alpha^2+\tau_\gamma^2\right)\right)^2-(K_{i-})^2(\tau_\alpha^2)^2}\right)$$

As a result, nested exchangeable mixed-effects model with inverse cluster-period size weighting (EMEw) estimator has an identical estimator to **Error! Reference source not found.**, but with a different weighted specification of $A_i$ and $B_i$:

$$A_i = K_{i-}[D_i + (K_{i-} - 1)F_i]$$

$$= \frac{1}{\sigma_w^2 + (K_{i-})\left[(\tau_\alpha^2 + \tau_\gamma^2) - \frac{(K_{i-})(\tau_\alpha^2)^2}{\sigma_w^2 + (K_{i-})(\tau_\alpha^2 + \tau_\gamma^2)}\right]}$$

$$= \frac{\sigma_w^2 + (K_{i-})(\tau_\alpha^2 + \tau_\gamma^2)}{\left(\sigma_w^2 + (K_{i-})(\tau_\alpha^2 + \tau_\gamma^2)\right)^2 - (K_{i-})^2(\tau_\alpha^2)^2}$$

which can be further written in terms of the within-period ICC $\left(\rho_{wp} = \frac{\tau_\alpha^2 + \tau_\gamma^2}{\tau_\alpha^2 + \tau_\gamma^2 + \sigma_w^2}\right)$ and between-period ICC $\left(\rho_{bp} = \frac{\tau_\alpha^2}{\tau_\alpha^2 + \tau_\gamma^2 + \sigma_w^2}\right)$:

$$= \frac{1 + (K_{i-} - 1)\rho_{wp}}{\left[\left(1 + (K_{i-} - 1)\rho_{wp}\right)^2 - (K_{i-})^2\rho_{bp}^2\right](\tau_\alpha^2 + \tau_\gamma^2 + \sigma_w^2)}$$

and:

$$B_i = K_{i-}^2 G_i$$

$$= -K_{i-}\left(\frac{\tau_\alpha^2}{\sigma_w^2 + (K_{i-})(\tau_\alpha^2 + \tau_\gamma^2)}\right)\left(\frac{1}{\sigma_w^2 + (K_{i-})\left[(\tau_\alpha^2 + \tau_\gamma^2) - \frac{(K_{i-})(\tau_\alpha^2)^2}{\sigma_w^2 + (K_{i-})(\tau_\alpha^2 + \tau_\gamma^2)}\right]}\right)$$

$$= -K_{i-}\left(\frac{\tau_\alpha^2}{\left(\sigma_w^2 + (K_{i-})(\tau_\alpha^2 + \tau_\gamma^2)\right)^2 - (K_{i-})^2(\tau_\alpha^2)^2}\right)$$

assuming that cluster-period sizes vary between clusters but not between periods within clusters, $K_{ij} = K_{i0} = K_{i1} = K_{i-}$ .

Therefore, the NEMEw estimator converges in probability to the following difficult to interpret estimand:

$$\hat{\delta}_{NEMEw} \overset{P}{\rightarrow} E\left[\frac{\left(\frac{1 + (K_{i-} - 1)\rho_{wp}}{\left(1 + (K_{i-} - 1)\rho_{wp}\right)^2 - (K_{i-})^2\rho_{bp}^2}\right)}{E\left[\frac{1 + (K_{i-} - 1)\rho_{wp}}{\left(1 + (K_{i-} - 1)\rho_{wp}\right)^2 - (K_{i-})^2\rho_{bp}^2}\right]}\left(\frac{1}{K_{i-}}\sum_k [Y_{i1k}(1) - Y_{i1k}(0)]\right)\right]$$

## A.3 Evaluation of the EMEw estimator

### A.3.1 Derivation of ICC for maximum bias

Here, we include the derivations for the ICC value that yields maximum bias in the EMEw estimator. As stated in Section 5, the EMEw estimator can be interpreted as a weighted average of the cluster-specific cATE estimands, with weights $\lambda_{EMEw}$. We assume there are two subpopulations $u = 1,2$ with fixed cluster-period sizes $K_{i-,1}$ and $K_{i-,2}$ and corresponding weights $\lambda_{EMEw,1}$ and $\lambda_{EMEw,2}$.

The maximum amount of bias in the EMEw estimator for cATE estimand relative to the ICC $\rho$ occurs when there is the maximum amount of difference between the estimand weights of the two subpopulations in the EMEw estimator $(\lambda_{EMEw,1} - \lambda_{EMEw,2})$. Accordingly:

$$\lambda_{EMEw,1} - \lambda_{EMEw,2} = \frac{\left(\frac{1+(K_{i-,1}-1)\rho}{1+(2K_{i-,1}-1)\rho} - \frac{1+(K_{i-,2}-1)\rho}{1+(2K_{i-,2}-1)\rho}\right)}{E\left[\left(\frac{1+(K_{i-}-1)\rho}{1+(2K_{i-}-1)\rho}\right)\right]}$$

This difference $(\lambda_{EMEw,1} - \lambda_{EMEw,2})$ is then maximized for values of $\rho$ where $\frac{d(\lambda_{EMEw,1}-\lambda_{EMEw,2})}{d\rho} = 0$. Accordingly:

$$\frac{d(\lambda_{EMEw,1} - \lambda_{EMEw,2})}{d\rho}$$

$$= \frac{d}{d\rho}\left(E\left[\left(\frac{1+(K_{i-}-1)\rho}{1+(2K_{i-}-1)\rho}\right)\right]\right)^{-1}\left(\frac{1+(K_{i-,1}-1)\rho}{1+(2K_{i-,1}-1)\rho} - \frac{1+(K_{i-,2}-1)\rho}{1+(2K_{i-,2}-1)\rho}\right)$$
$$+\left(E\left[\left(\frac{1+(K_{i-}-1)\rho}{1+(2K_{i-}-1)\rho}\right)\right]\right)^{-1}\frac{d}{d\rho}\left(\frac{1+(K_{i-,1}-1)\rho}{1+(2K_{i-,1}-1)\rho} - \frac{1+(K_{i-,2}-1)\rho}{1+(2K_{i-,2}-1)\rho}\right)$$

$$= -\left(E\left[\left(\frac{1+(K_{i-}-1)\rho}{1+(2K_{i-}-1)\rho}\right)\right]\right)^{-2} E\left[\frac{-K_{i-}}{(1+(2K_{i-}-1)\rho)^2}\right]\left(\frac{1+(K_{i-,1}-1)\rho}{1+(2K_{i-,1}-1)\rho} - \frac{1+(K_{i-,2}-1)\rho}{1+(2K_{i-,2}-1)\rho}\right)$$
$$+\left(E\left[\left(\frac{1+(K_{i-}-1)\rho}{1+(2K_{i-}-1)\rho}\right)\right]\right)^{-1}\left(\frac{-K_{i-.1}}{(1+(2K_{i-.1}-1)\rho)^2} - \frac{-K_{i-,2}}{(1+(2K_{i-,2}-1)\rho)^2}\right)$$

which we set $= 0$. Solving for $\rho$:

$$\left(E\left[\left(\frac{1+(K_{i-}-1)\rho}{1+(2K_{i-}-1)\rho}\right)\right]\right)^{-1} E\left[\frac{K_{i-}}{(1+(2K_{i-}-1)\rho)^2}\right]\left(\frac{1+\left(K_{i-,1}-1\right)\rho}{1+\left(2K_{i-,1}-1\right)\rho} - \frac{1+\left(K_{i-,2}-1\right)\rho}{1+\left(2K_{i-,2}-1\right)\rho}\right) = \left(\frac{K_{i-.1}}{(1+(2K_{i-.1}-1)\rho)^2} - \frac{K_{i-,2}}{\left(1+\left(2K_{i-,2}-1\right)\rho\right)^2}\right)$$

$$E\left[\frac{K_{i-}}{(1+(2K_{i-}-1)\rho)^2}\right]\left(\frac{1+\left(K_{i-,1}-1\right)\rho}{1+\left(2K_{i-,1}-1\right)\rho} - \frac{1+\left(K_{i-,2}-1\right)\rho}{1+\left(2K_{i-,2}-1\right)\rho}\right) = E\left[\left(\frac{1+(K_{i-}-1)\rho}{1+(2K_{i-}-1)\rho}\right)\right]\left(\frac{K_{i-.1}}{(1+(2K_{i-.1}-1)\rho)^2} - \frac{K_{i-,2}}{\left(1+\left(2K_{i-,2}-1\right)\rho\right)^2}\right)$$

$$\left(P(u=1)\frac{K_{i-.1}}{(1+(2K_{i-.1}-1)\rho)^2} + P(u=2)\frac{K_{i-,2}}{\left(1+\left(2K_{i-,2}-1\right)\rho\right)^2}\right)\left(\frac{1+\left(K_{i-,1}-1\right)\rho}{1+\left(2K_{i-,1}-1\right)\rho} - \frac{1+\left(K_{i-,2}-1\right)\rho}{1+\left(2K_{i-,2}-1\right)\rho}\right)$$
$$= \left(P(u=1)\frac{1+\left(K_{i-,1}-1\right)\rho}{1+\left(2K_{i-,1}-1\right)\rho} + P(u=2)\frac{1+\left(K_{i-,2}-1\right)\rho}{1+\left(2K_{i-,2}-1\right)\rho}\right)\left(\frac{K_{i-.1}}{(1+(2K_{i-.1}-1)\rho)^2} - \frac{K_{i-,2}}{\left(1+\left(2K_{i-,2}-1\right)\rho\right)^2}\right)$$

$$\left(\frac{K_{i-,2}}{\left(1+\left(2K_{i-,2}-1\right)\rho\right)^2}\right)\left(\frac{1+\left(K_{i-,1}-1\right)\rho}{1+\left(2K_{i-,1}-1\right)\rho}\right) = \left(\frac{K_{i-.1}}{(1+(2K_{i-.1}-1)\rho)^2}\right)\left(\frac{1+\left(K_{i-,2}-1\right)\rho}{1+\left(2K_{i-,2}-1\right)\rho}\right)$$

$$\frac{K_{i-,2}\left(1+\left(K_{i-,1}-1\right)\rho\right)}{1+\left(2K_{i-,2}-1\right)\rho} = \frac{K_{i-,1}\left(1+\left(K_{i-,2}-1\right)\rho\right)}{1+\left(2K_{i-,1}-1\right)\rho}$$

$$K_{i-,2}\left(1+\left(K_{i-,1}-1\right)\rho\right)\left(1+\left(2K_{i-,1}-1\right)\rho\right) = K_{i-,1}\left(1+\left(K_{i-,2}-1\right)\rho\right)\left(1+\left(2K_{i-,2}-1\right)\rho\right)$$

$$K_{i-,2}\left(2K_{i-,1}^2\rho^2 + (1-\rho)^2\right) = K_{i-,1}\left(2K_{i-,2}^2\rho^2 + (1-\rho)^2\right)$$

$$(1-\rho)^2\left(K_{i-,2} - K_{i-,1}\right) = \rho^2\left(2K_{i-,1}K_{i-,2}^2 - 2K_{i-,1}^2K_{i-,2}\right)$$

$$(1-\rho)\sqrt{K_{i-,2} - K_{i-,1}} = \rho\sqrt{2\left(K_{i-,2} - K_{i-,1}\right)K_{i-,1}K_{i-,2}}$$

where $\rho \leq 1$ and we set $K_{i-,1} \leq K_{i-,2}$. Finally, solving for $\rho$, we get:

$$\rho = \frac{\sqrt{K_{i-,2} - K_{i-,1}}}{\sqrt{K_{i-,2} - K_{i-,1}} + \sqrt{2\left(K_{i-,2} - K_{i-,1}\right)K_{i-,1}K_{i-,2}}} = \frac{1}{1+\sqrt{2K_{i-,1}K_{i-,2}}}.$$

In conclusion, we prove that the maximum difference between the estimand weights, and accordingly the maximum amount of bias, occurs when the ICC is equivalent to:

$$\rho = \frac{1}{1 + \sqrt{2K_{i-,1}K_{i-,2}}}.$$

### A.3.2 Derivation of $P(u = 1)$ for maximum bias

Here, we include the derivations for the $P(u = 1)$ value that yields maximum bias in the EMEw estimator. As stated in Section 5, the EMEw estimator can be interpreted as a weighted average of the cluster-specific cATE estimands, with weights $\lambda_{EMEw}$. We assume there are two subpopulations $u = 1,2$ with fixed cluster-period sizes $K_{i-,1}$ and $K_{i-,2}$ and corresponding weights $\lambda_{EMEw,1}$ and $\lambda_{EMEw,2}$.

Assuming that subpopulations $u = 1$ and 2 have corresponding treatment effects $\delta_1$ and $\delta_2$, the cATE is then equivalent to $P(u = 1)\delta_1 + P(u = 2)\delta_2$. Accordingly, the bias is the difference between the estimator and the estimand:

$$Bias = \big(P(u = 1)\lambda_{EMEw,1}\delta_1 + P(u = 2)\lambda_{EMEW,2}\delta_2\big) - (P(u = 1)\delta_1 + P(u = 2)\delta_2)$$

$$= P(u = 1)\big(\lambda_{EMEw,1} - 1\big)\delta_1 + \big(1 - P(u = 1)\big)\big(\lambda_{EMEW,2} - 1\big)\delta_2$$

where:

$$\big(\lambda_{EMEw,1} - 1\big) = \frac{\big(1 - P(u = 1)\big)\left(\frac{1 + (K_{i-,1} - 1)\rho}{1 + (2K_{i-,1} - 1)\rho} - \frac{1 + (K_{i-,2} - 1)\rho}{1 + (2K_{i-,2} - 1)\rho}\right)}{P(u = 1)\left(\frac{1 + (K_{i-,1} - 1)\rho}{1 + (2K_{i-,1} - 1)\rho}\right) + \big(1 - P(u = 1)\big)\left(\frac{1 + (K_{i-,2} - 1)\rho}{1 + (2K_{i-,2} - 1)\rho}\right)}$$

and:

$$\big(\lambda_{EMEW,2} - 1\big) = \frac{P(u = 1)\left(\frac{1 + (K_{i-,2} - 1)\rho}{1 + (2K_{i-,2} - 1)\rho} - \frac{1 + (K_{i-,1} - 1)\rho}{1 + (2K_{i-,1} - 1)\rho}\right)}{P(u = 1)\left(\frac{1 + (K_{i-,1} - 1)\rho}{1 + (2K_{i-,1} - 1)\rho}\right) + \big(1 - P(u = 1)\big)\left(\frac{1 + (K_{i-,2} - 1)\rho}{1 + (2K_{i-,2} - 1)\rho}\right)}$$

Altogether, the bias can be written as:

$$Bias = \frac{P(u = 1)\big(1 - P(u = 1)\big)\left(\frac{1 + (K_{i-,2} - 1)\rho}{1 + (2K_{i-,2} - 1)\rho} - \frac{1 + (K_{i-,1} - 1)\rho}{1 + (2K_{i-,1} - 1)\rho}\right)}{P(u = 1)\left(\frac{1 + (K_{i-,1} - 1)\rho}{1 + (2K_{i-,1} - 1)\rho}\right) + \big(1 - P(u = 1)\big)\left(\frac{1 + (K_{i-,2} - 1)\rho}{1 + (2K_{i-,2} - 1)\rho}\right)}(\delta_1 - \delta_2).$$

The bias is then maximized for values of $P(u = 1)$ where $\frac{dln(Bias)}{dP(u=1)} = 0$. We maximize the log bias $\ln(Bias)$ to make calculations more straightforward. Accordingly:

$$\ln(Bias) = \ln\left(P(u=1)\right) + \ln\left(1 - P(u=1)\right) + \ln\left(\frac{1+\left(K_{i-,2}-1\right)\rho}{1+\left(2K_{i-,2}-1\right)\rho} - \frac{1+\left(K_{i-,1}-1\right)\rho}{1+\left(2K_{i-,1}-1\right)\rho}\right) + \ln(\delta_1 - \delta_2) - \ln\left(P(u=1)\left(\frac{1+\left(K_{i-,1}-1\right)\rho}{1+\left(2K_{i-,1}-1\right)\rho}\right) + \left(1-P(u=1)\right)\left(\frac{1+\left(K_{i-,2}-1\right)\rho}{1+\left(2K_{i-,2}-1\right)\rho}\right)\right)$$

and:

$$\frac{d\ln(Bias)}{dP(u=1)} = \frac{1}{P(u=1)} - \frac{1}{1-P(u=1)} - \frac{\left(\frac{1+\left(K_{i-,1}-1\right)\rho}{1+\left(2K_{i-,1}-1\right)\rho} - \frac{1+\left(K_{i-,2}-1\right)\rho}{1+\left(2K_{i-,2}-1\right)\rho}\right)}{P(u=1)\left(\frac{1+\left(K_{i-,1}-1\right)\rho}{1+\left(2K_{i-,1}-1\right)\rho}\right) + \left(1-P(u=1)\right)\left(\frac{1+\left(K_{i-,2}-1\right)\rho}{1+\left(2K_{i-,2}-1\right)\rho}\right)}$$

which we set $= 0$. Solving for $P(u=1)$:

$$\frac{1}{P(u=1)} - \frac{1}{1-P(u=1)} = \frac{\left(\frac{1+\left(K_{i-,1}-1\right)\rho}{1+\left(2K_{i-,1}-1\right)\rho} - \frac{1+\left(K_{i-,2}-1\right)\rho}{1+\left(2K_{i-,2}-1\right)\rho}\right)}{P(u=1)\left(\frac{1+\left(K_{i-,1}-1\right)\rho}{1+\left(2K_{i-,1}-1\right)\rho}\right) + \left(1-P(u=1)\right)\left(\frac{1+\left(K_{i-,2}-1\right)\rho}{1+\left(2K_{i-,2}-1\right)\rho}\right)}$$

$$\frac{1-2P(u=1)}{P(u=1)\left(1-P(u=1)\right)} = \frac{\left(\frac{1+\left(K_{i-,1}-1\right)\rho}{1+\left(2K_{i-,1}-1\right)\rho} - \frac{1+\left(K_{i-,2}-1\right)\rho}{1+\left(2K_{i-,2}-1\right)\rho}\right)}{P(u=1)\left(\frac{1+\left(K_{i-,1}-1\right)\rho}{1+\left(2K_{i-,1}-1\right)\rho}\right) + \left(1-P(u=1)\right)\left(\frac{1+\left(K_{i-,2}-1\right)\rho}{1+\left(2K_{i-,2}-1\right)\rho}\right)}$$

$$P(u=1)\left(1-P(u=1)\right)\left(\frac{1+\left(K_{i-,1}-1\right)\rho}{1+\left(2K_{i-,1}-1\right)\rho} - \frac{1+\left(K_{i-,2}-1\right)\rho}{1+\left(2K_{i-,2}-1\right)\rho}\right) - \left(1-2P(u=1)\right)\left(P(u=1)\left(\frac{1+\left(K_{i-,1}-1\right)\rho}{1+\left(2K_{i-,1}-1\right)\rho}\right) + \left(1-P(u=1)\right)\left(\frac{1+\left(K_{i-,2}-1\right)\rho}{1+\left(2K_{i-,2}-1\right)\rho}\right)\right) = 0$$

$$P(u=1)^2\left(\frac{1+(K_{i-,1}-1)\rho}{1+(2K_{i-,1}-1)\rho}\right)-\left(\frac{1+(K_{i-,2}-1)\rho}{1+(2K_{i-,2}-1)\rho}\right)+2P(u=1)\left(\frac{1+(K_{i-,2}-1)\rho}{1+(2K_{i-,2}-1)\rho}\right)$$
$$-P(u=1)^2\left(\frac{1+(K_{i-,2}-1)\rho}{1+(2K_{i-,2}-1)\rho}\right)=0$$

$$P(u=1)^2\left(\frac{1+(K_{i-,1}-1)\rho}{1+(2K_{i-,1}-1)\rho}\right)-\left(1-P(u=1)\right)^2\left(\frac{1+(K_{i-,2}-1)\rho}{1+(2K_{i-,2}-1)\rho}\right)=0$$

$$P(u=1)^2\left(\frac{1+(K_{i-,1}-1)\rho}{1+(2K_{i-,1}-1)\rho}\right)=\left(1-P(u=1)\right)^2\left(\frac{1+(K_{i-,2}-1)\rho}{1+(2K_{i-,2}-1)\rho}\right)$$

$$P(u=1)\sqrt{\frac{1+(K_{i-,1}-1)\rho}{1+(2K_{i-,1}-1)\rho}}=\left(1-P(u=1)\right)\sqrt{\frac{1+(K_{i-,2}-1)\rho}{1+(2K_{i-,2}-1)\rho}}$$

Finally, solving for $P(u=1)$, we get:

$$P(u=1)=\frac{\sqrt{\frac{1+(K_{i-,2}-1)\rho}{1+(2K_{i-,2}-1)\rho}}}{\sqrt{\frac{1+(K_{i-,1}-1)\rho}{1+(2K_{i-,1}-1)\rho}}+\sqrt{\frac{1+(K_{i-,2}-1)\rho}{1+(2K_{i-,2}-1)\rho}}}=\frac{\sqrt{\lambda_{EMEw,2}}}{\sqrt{\lambda_{EMEw,1}}+\sqrt{\lambda_{EMEw,2}}}.$$

In conclusion, we prove that the maximum bias occurs when the probability of belonging to subpopulation $u=1$ is equivalent to:

$$P(s=1)=\frac{\sqrt{\lambda_{EMEw,2}}}{\sqrt{\lambda_{EMEw,1}}+\sqrt{\lambda_{EMEw,2}}}$$

with $P(u=2)=1-P(u=1)$. Overall this term tends to be approximately $P(u=1)\approx 0.5$ for most values of $K_{i-,1}, K_{i-,2}$, and $\rho$.

## A.3.3 Additional estimand weight figures

EMEw weights are plotted for different values of $\zeta=\frac{K_{i-,1}}{K_{i-,2}}$ and $P(u=1)=0.5, 0.9$, or the "optimal" value $\left(P(u=1)=\frac{\sqrt{\lambda_{EMEw,2}}}{\sqrt{\lambda_{EMEw,1}}+\sqrt{\lambda_{EMEw,2}}}\right)$ using the optimal ICC $\left(\rho=\frac{1}{1+\sqrt{2K_{i-,1}K_{i-,2}}}\right)$ to yield the maximum amount of bias. The optimal ICC is plotted in a gray dashed line.

The "optimal" $P(u=1)$ values were $0.42, 0.42, 0.43, 0.46, 0.47$, and $0.49$ for conditions with $\zeta=0.001, 0.002, 0.005, 0.1, 0.2$, and $0.5$, respectively. Notably, the "optimal" $P(u=1)$ are all $\approx 0.5$ and the graph of their corresponding estimand weights $\lambda_{EMEw,s}$ are qualitatively similar to the graphs with $P(u=1)=0.5$.

Intuitively, the maximum amount of bias occurs around scenarios where $P(u=1)=0.5$, since both subpopulations $u=1$ and 2 are equally overweighted and underweighted, respectively. In contrast, in the graphed scenario where $P(u = 1) = 0.9$, outcomes from subpopulations $s = 1$ are correctly weighted near 1, whereas outcomes from subpopulation $u = 2$ are more drastically underweighted. However, since subpopulation $u = 1$ makes up 90% of the sampled clusters, the estimate ends up being more correctly weighted and less biased for the cATE estimand as a result.

Overall, we observe that the maximum difference between the estimand weights $\lambda_{EMEw,u}$ increases with larger values of $\zeta$. However, graphing up to an extreme value of $\zeta = 0.001$ with the optimal $P(u = 1)$ and $\rho$ to maximize the bias still yielded fairly conservative values for the estimand weights.

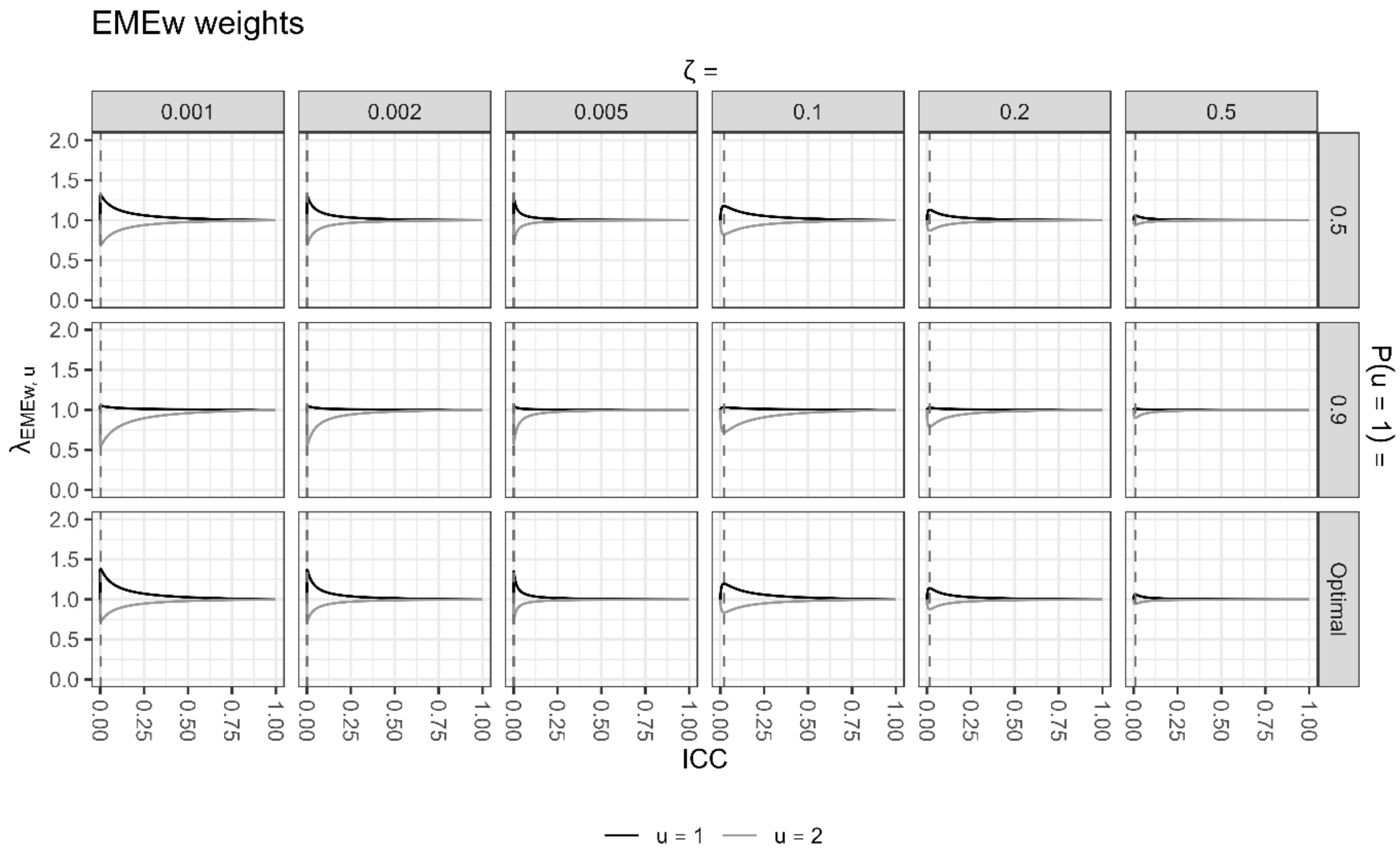

# A.4 Simulation

## A.4.1 Simulation BRL variance estimator results

Simulation Results with bias-reduced linearization (BRL) variance estimators

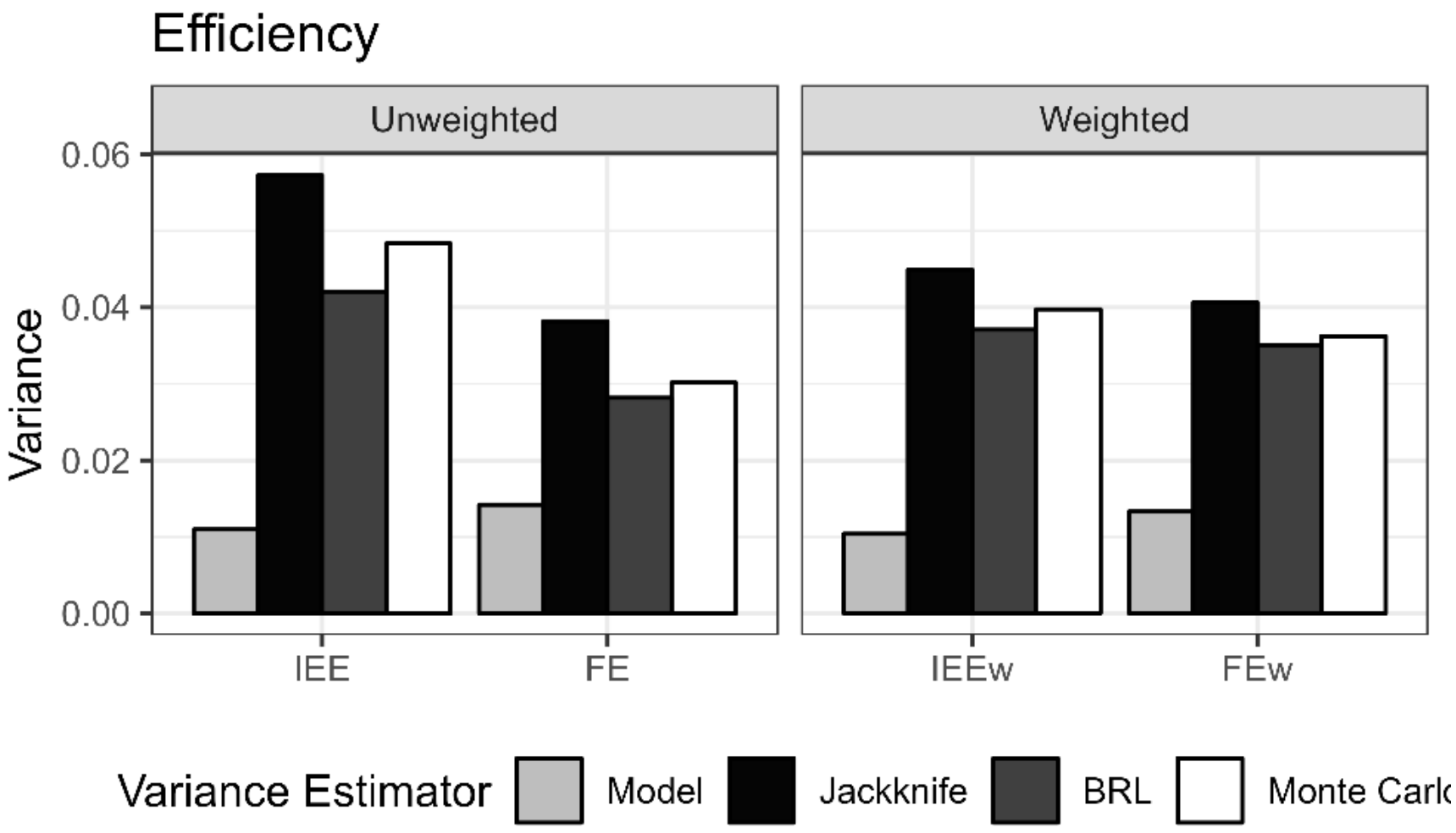

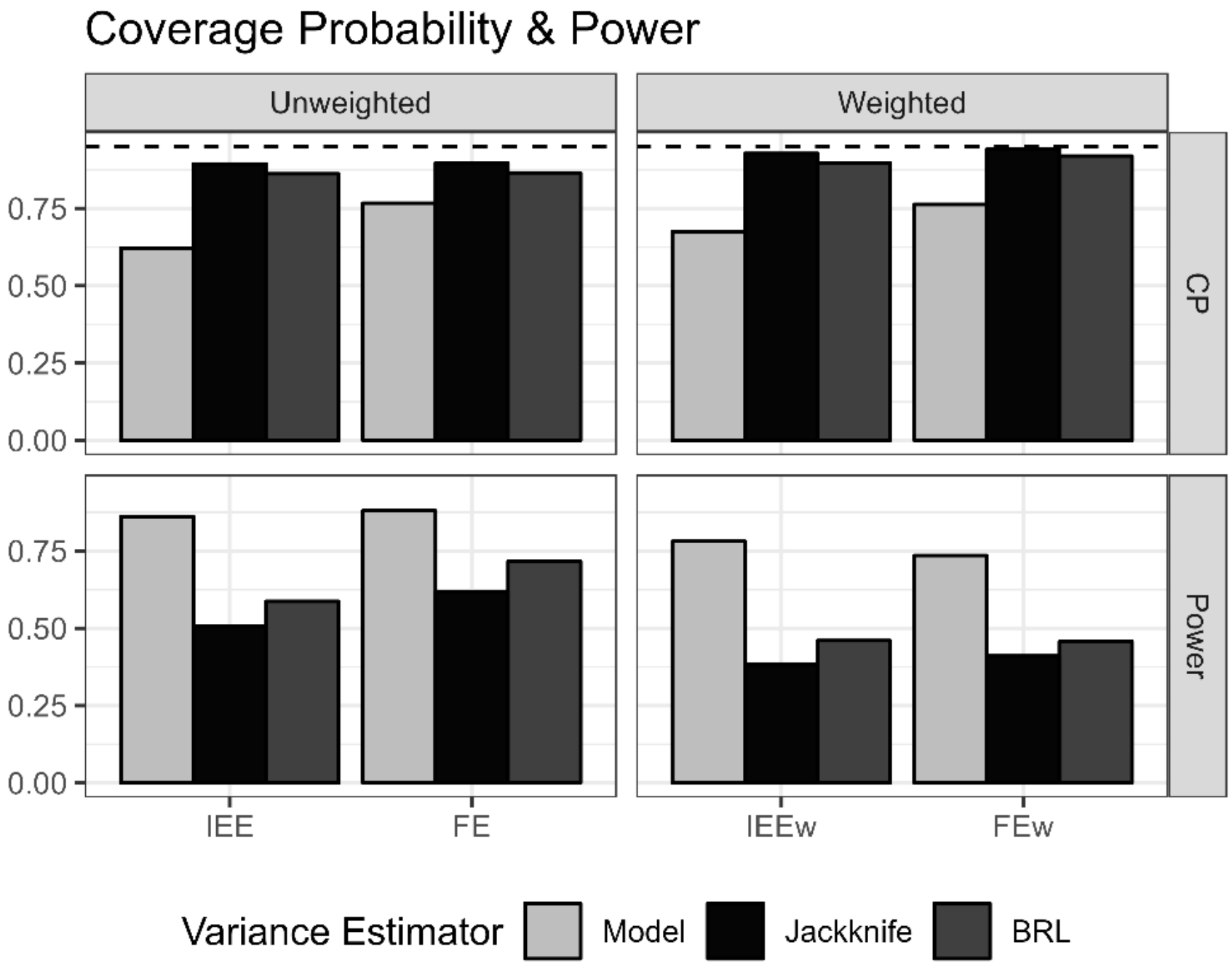

## A.4.2 Additional simulation scenario bias results

Relative bias results from two additional simulation scenarios. Simulation parameters were set to maximize bias in the EMEw estimator. The two simulation scenarios were simulated with a similar data generating process as described in Section 6 with the same 10 cluster PB-CRT design, but with $E\left[K_{i-,1}\right] = 20, E\left[K_{i-,2}\right] = 200, \zeta = \frac{E\left[K_{i-,1}\right]}{E\left[K_{i-,2}\right]} = 0.1$ or $E\left[K_{i-,1}\right] = 50, E\left[K_{i-,2}\right] = 1000, \zeta = \frac{E\left[K_{i-,1}\right]}{E\left[K_{i-,2}\right]} = 0.05$. Both scenarios are simulated with $P(u = 1) = 0.5$, CAC=1, and the corresponding $\rho = \frac{1}{1+\sqrt{2E[K_{i-,1}]E[K_{i-,2}]}}$ to optimize the bias in the EMEw estimator for the cATE estimand.

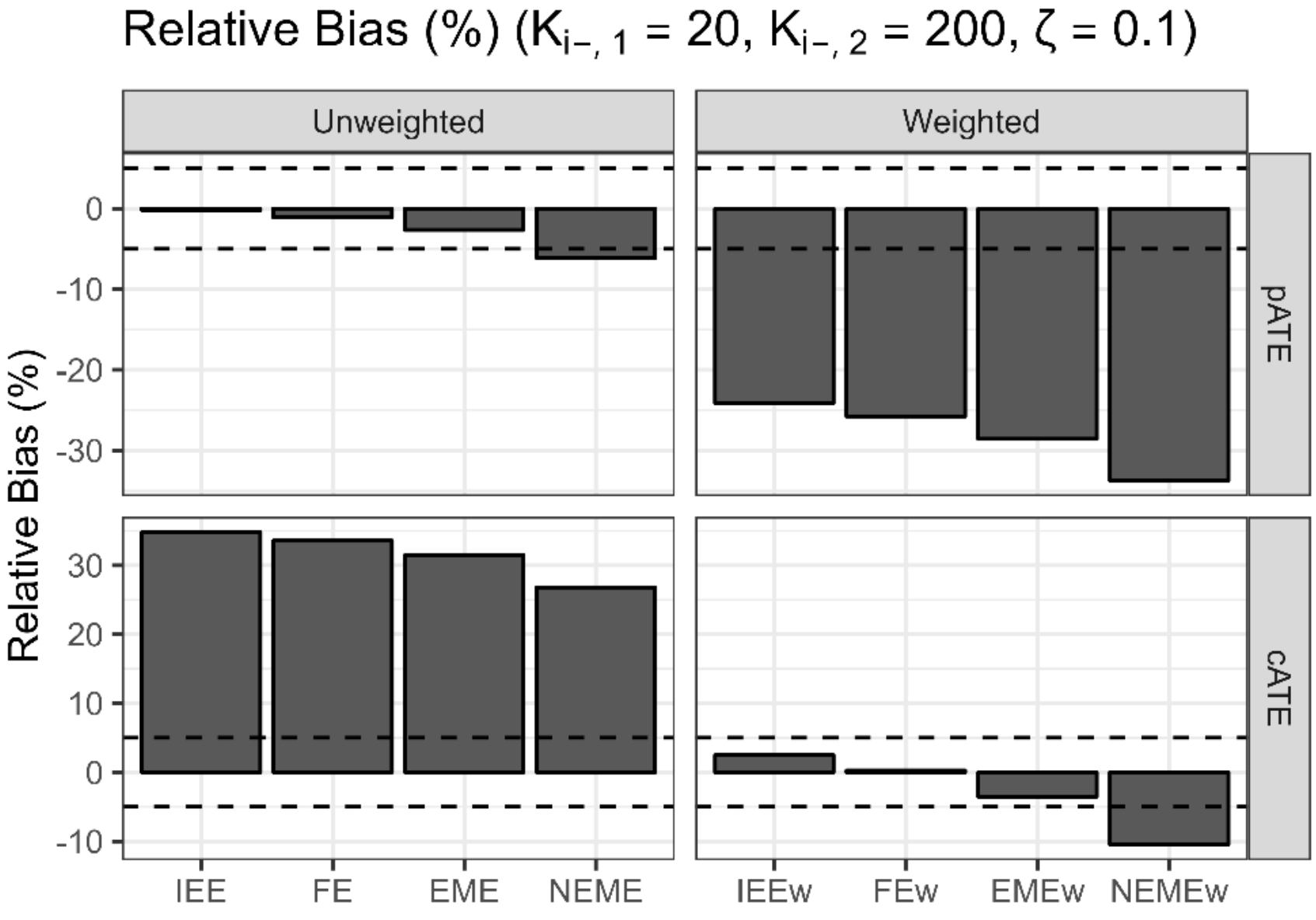

In the first scenario with $\zeta = 0.1$, the EME and EMEw estimator still maintained less than 5% relative bias for the pATE and cATE estimands, respectively. In the second scenario with $\zeta = 0.05$, the EME estimator still maintained less than 5% relative bias for the pATE estimand. The EMEw estimator had slightly over 5% relative bias for the cATE estimand over the 1000 simulation replicates.

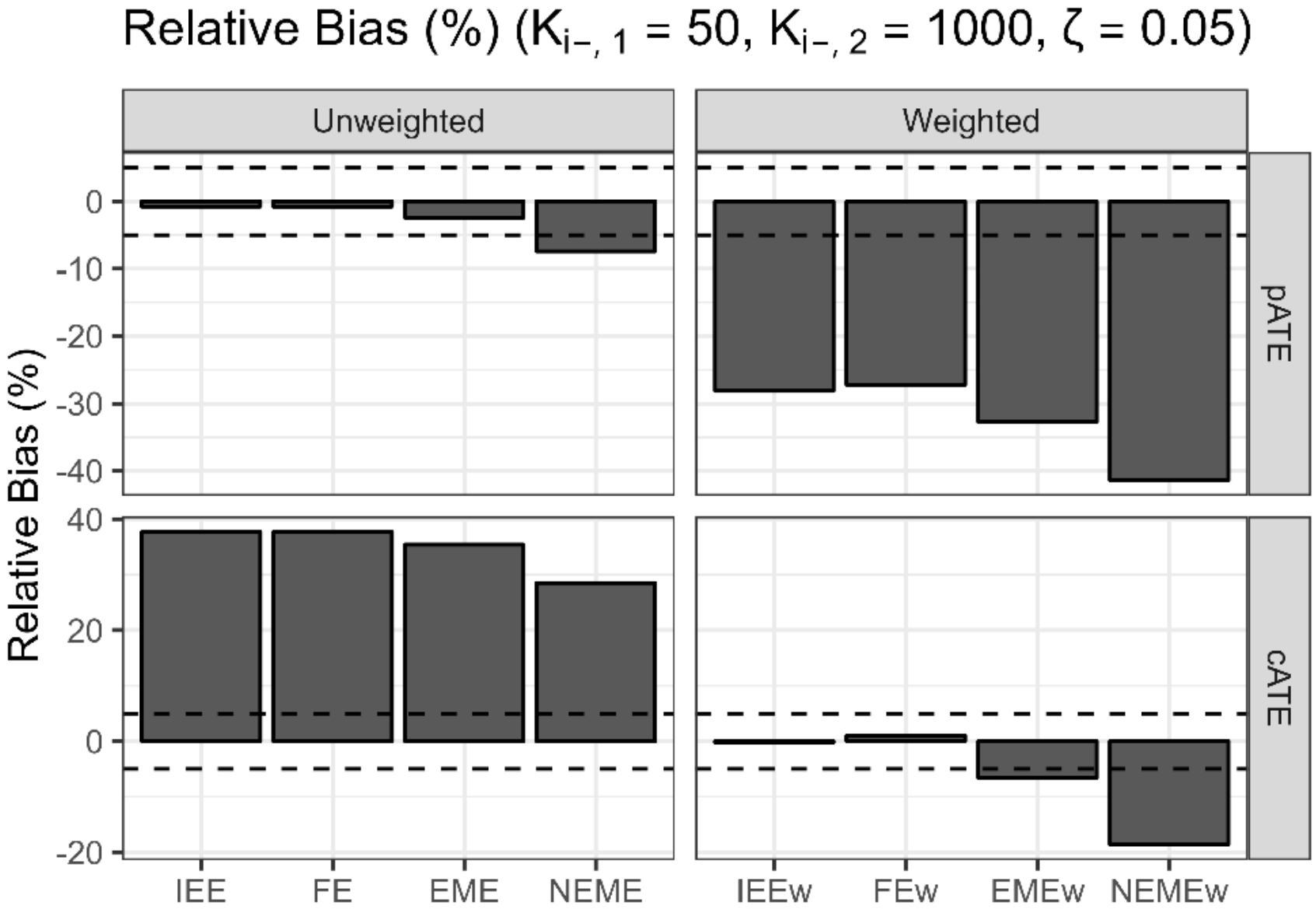

Despite these scenarios being unrealistically tailored to yield biased EMEw estimates for the cATE estimand, the EMEw estimator still remained relatively unbiased. Overall, the EMEw bias is generally unbiased for the cATE estimand across the explored scenarios, despite not being a theoretically consistent estimator for the cATE in PB-CRTs with informative cluster sizes.

## A.5 Additional case study information

Cluster-period cell sizes from the JIAH trial. Cells receiving the intervention are shaded in gray.

| P0 | P1 |
|---|---|
| 48 | 70 |
| 60 | 76 |
| 114 | 81 |
| 47 | 74 |
| 51 | 48 |
| 54 | 17 |
| 75 | 37 |
| 105 | 85 |
| 119 | 73 |
| 69 | 3 |
| 87 | 3 |
| 97 | 92 |
| 132 | 15 |
| 68 | 62 |
| 116 | 94 |
| 92 | 77 |
| 83 | 40 |
| 78 | 22 |
| 101 | 81 |
| 46 | 23 |
| 104 | 93 |
| 138 | 12 |
| 69 | 37 |
| 109 | 52 |
| 50 | 43 |
| 108 | 101 |
| 101 | 5 |
| 69 | 61 |